\documentclass[a4paper,fleqn,usenatbib]{mnras}

\usepackage{newtxtext,newtxmath}

\usepackage[T1]{fontenc}
\usepackage{ae,aecompl}

\usepackage{graphicx}	
\usepackage{amsmath}	
\usepackage{amssymb}	
\usepackage{hyperref}
\usepackage{longtable}
\usepackage{caption}
\usepackage{pdflscape}

\newcommand{\sextractor}{\textsc{SExtractor}}

\defcitealias{2011ApJS..196...11S}{S11}
\defcitealias{2014ApJ...794..120A}{A14}

\title[Stripe 82 deep image morphologies]{Bulge plus disc and S\'ersic decomposition catalogues for 16,908 galaxies in the SDSS Stripe 82 co-adds: A detailed study of the $ugriz$ structural measurements}

\author[Bottrell et al.]{
Connor Bottrell,$^{1}$\thanks{E-mail: cbottrel@uvic.ca}
Luc Simard,$^{2,1}$
J. Trevor Mendel,$^{3,4}$
Sara L. Ellison$^{1}$
\\
$^{1}$Department of Physics and Astronomy, University of Victoria, Victoria, British Columbia V8P 1A1, Canada\\
$^{2}$National Research Council of Canada, 5071 West Saanich Road, Victoria, British Columbia V9E 2E7, Canada\\
$^{3}$Research School of Astronomy and Astrophysics, Australian National University, Canberra, ACT 2611, Australia\\
$^{4}$ARC Centre of Excellence for All Sky Astrophysics in 3 Dimensions (ASTRO 3D)
}

\date{Accepted XXX. Received YYY; in original form ZZZ}

\pubyear{2018}

\hypersetup{draft}
\begin{document}
\label{firstpage}
\pagerange{\pageref{firstpage}--\pageref{lastpage}}
\maketitle

\begin{abstract}
Quantitative characterization of galaxy morphology is vital in enabling comparison of observations to predictions from galaxy formation theory. However, without significant overlap between the observational footprints of deep and shallow galaxy surveys, the extent to which structural measurements for large galaxy samples are robust to image quality (e.g., depth, spatial resolution) cannot be established. Deep images from the Sloan Digital Sky Survey (SDSS) Stripe 82 co-adds provide a unique solution to this problem - offering $1.6-1.8$ magnitudes improvement in depth with respect to SDSS Legacy images. Having similar spatial resolution to Legacy, the co-adds make it possible to examine the sensitivity of parametric morphologies to depth alone. Using the \textsc{gim2d} surface-brightness decomposition software, we provide public morphology catalogs for 16,908 galaxies in the Stripe 82 $ugriz$ co-adds. Our methods and selection are completely consistent with the \cite{2011ApJS..196...11S} and \cite{2014ApJS..210....3M} photometric decompositions. We rigorously compare measurements in the deep and shallow images. We find no systematics in total magnitudes and sizes except for faint galaxies in the $u$-band and the brightest galaxies in each band. However, characterization of bulge-to-total fractions is significantly improved in the deep images. Furthermore, statistics used to determine whether single-S\'ersic or two-component (e.g., bulge+disc) models are required become more bimodal in the deep images. Lastly, we show that asymmetries are enhanced in the deep images and that the enhancement is positively correlated with the asymmetries measured in Legacy images.
\end{abstract}




\section{Introduction}

Morphology is the visual connection to the formation and evolutionary histories of galaxies. From a theoretical perspective, observed stellar structure is tied intricately to these histories. Stellar discs are built-up from gas that either collapsed gravitationally from a protogalactic cloud in a rotating dark matter halo \citep{1978MNRAS.183..341W,1980MNRAS.193..189F} or was accreted at least semi-coherently with the rotation of a pre-existing galaxy \citep{2005MNRAS.363....2K,2012MNRAS.423.1544S,2013ApJ...769...74S,2015MNRAS.449.2087D}. Amongst stellar bulges, the observed agreement between classical dispersion-supported bulges and elliptical galaxies across a number of key scaling relations (e.g., \citealt{1977ApJ...217..406K,1976ApJ...204..668F,1987ApJ...313...59D,1987ApJ...313...42D,2008AJ....136..773F,2009MNRAS.393.1531G,2010ApJ...716..942F,2012ApJS..198....2K}), the black-hole mass vs. bulge velocity dispersion relation \citep{2011Natur.469..374K,2013ARA&A..51..511K}, and the similarity of their stellar populations \citep{2006MNRAS.371..583M} suggest a similar and coordinated formation mechanism. Mergers are widely regarded as key suspects due to broad theoretical compatibility with these observed similarities \citep{1967MNRAS.136..101L,1972ApJ...178..623T,1977egsp.conf..401T,1983MNRAS.205.1009N,1988ApJ...331..699B,1992ApJ...400..460H,2008ApJ...679..156H,2009ApJ...691.1168H,2013ApJ...778...61T,2017MNRAS.467.3083R,2018MNRAS.480.2266M}. Meanwhile, a class of morphologically and kinematically disc-like pseudo-bulges which do not neatly subscribe to trends shared by classical bulges and ellipticals (see \citealt{2004ARA&A..42..603K} for a review) are observed as downward dips in the stellar velocity dispersion at the centres of galaxies and are often linked to secular formation through disc instabilities \citep{2001A&A...368...52E,2003A&A...409..459M,2006MNRAS.369..529F,2007MNRAS.379..445P}. Accurate characterization of morphology is therefore of enormous astrophysical importance because the allocation of stars to the bulge and disc and their structures encode the dominant physical processes in a galaxy's formation (e.g., \citealt{2014MNRAS.440L..66B,2019MNRAS.tmp..363B}).

The dominance of a stellar disc or bulge forms the basis of most visual morphological classification schemes while additional stellar components can motivate splitting of the main track into parallel sequences \citep{1926ApJ....64..321H,1936rene.book.....H,1959HDP....53..275D,1961hag..book.....S}. However, purely visual morphology classification has limited capacity for comparison with theoretical predictions. Intrinsic subjectivity in visual classification combined with the enormous scale of modern galaxy surveys demands robust and quantitative characterization of galaxy morphology. This demand has been met with a wealth of tools for accurately characterizing the global and/or component structures based on galaxy surface-brightness distributions \citep{2002AJ....124..266P,2002ApJS..142....1S,2004ApJS..153..411D,2004AJ....128..163L,2006MNRAS.373.1389C,2008A&A...478..353M,2013MNRAS.435..623V,2015ApJ...810..120C,2017MNRAS.466.1513R}. These tools enable routine and reliable separation of bulge and disc light in tens to hundreds of thousands of galaxies (e.g., \citealt{2006MNRAS.371....2A,2007MNRAS.379..841B,2011ApJS..196...11S,2012MNRAS.421.2277L,2012MNRAS.421.1007K,2015MNRAS.446.3943M}). Indeed, marriage of visual and quantitative morphologies on such large scales may provide even more accurate descriptions of galaxy structures that go beyond the bulge and disc. Recent work by \cite{2018MNRAS.473.4731K} has shown that it is possible to use statistics for visual morphologies harvested from the Galaxy Zoo citizen-science project \citep{2008MNRAS.389.1179L,2015MNRAS.449..820W} as priors on the forms of models used in quantitative photometric decompositions. Given that the same quantitative and visual methods can now be employed on realistic mock-observations of galaxies produced in large-volume cosmological hydrodynamical simulations (e.g., \citealt{2014MNRAS.444.1518V,2015MNRAS.446..521S,2018MNRAS.473.4077P}), the capacity with which observed structures may be compared with theoretical predictions has never been better (e.g., \citealt{2015MNRAS.454.1886S,2017MNRAS.467.1033B,2017MNRAS.467.2879B,2018ApJ...853..194D,2019MNRAS.483.4140R}).

But just how robust are these quantitative measurements? If the bulge-to-disc $(B/D)$ or bulge-to-total $(B/T)$ light ratios encode the above-mentioned formation histories, then accurate characterization of these quantities and their limitations within large galaxy surveys is vital. A classic test of the sensitivity of measured parameters to various metrics for image quality is to insert analytic galaxy bulge+disc models into real survey fields and to evaluate their recovery (e.g., \citealt{2002ApJS..142....1S}). However, these tests do not capture the systematic and random errors that arise from additional substructure in the bulge and disc components or, indeed, structures that are neither bulge nor disc. Instead, comparisons of structural measurements obtained for real galaxies in images of varying quality do capture these biases and enable a more critical evaluation of the sensitivity of measured quantities (e.g., \citealt{2004AJ....128..163L,2006ApJ...636..592L,2008ApJS..179..319L}). Thus far, no such image-quality comparison has been performed to evaluate the limitations for the measured properties of galaxies in the bulge/disc decomposition analysis of 1.12 million galaxies by \cite{2011ApJS..196...11S} (hereafter \citetalias{2011ApJS..196...11S}) using Sloan Digital Sky Survey (SDSS) Legacy photometry \citep{2009ApJS..182..543A}. The deep co-added images offered for the SDSS Stripe 82 \citep[][hereafter \citetalias{2014ApJ...794..120A}]{2014ApJ...794..120A} provide the ideal basis for such a test -- extending 1.6-1.8 magnitudes deeper than the Legacy single-pass images. Having been combined from images taken with the same instrument and resulting in images with similar spatial resolution, the deep co-adds offer a direct look at how the measured properties of galaxies respond to depth alone. Furthermore, we may expect that faint substructures that were previously inaccessible in Legacy are revealed in the deep images. 

In this paper, we characterize the extent to which structural measurements for galaxies derived from parametric two-dimensional surface brightness decompositions are robust to improved depth. To do so, we compare structural measurements taken from existing catalogs using SDSS Legacy photometry \citep{2011ApJS..196...11S,2014ApJS..210....3M} to those in a new set of publicly available catalogs for galaxies in the SDSS Stripe 82 deep co-add images. Surface-brightness decompositions are performed using the \textsc{gim2d} IRAF software \citep{2002ApJS..142....1S}. As such, our comparison of $ugriz$ structural measurements in deep and shallow images is entirely self-consistent. This paper is structured as follows. Section \ref{sec:data} provides a description of the salient features of the SDSS image acquisition, features, and definitions that are relevant to creation of the co-adds. We also describe the galaxy sample selection and provide a characterization of typical spatial resolution in the co-adds. Section \ref{sec:catalogs} describes the decomposition models, methods, and catalogs. Section \ref{sec:comparison} compares the global and component structures, statistics, and asymmetries of galaxies in the deep and shallow images. Our main results are summarized in section \ref{sec:summary}. All rest-frame quantities and distance measures computed for the catalogs and for this paper assume a $(H_0,\Omega_m,\Omega_{\Lambda}) = (70$ km/s/Mpc, 0.3, 0.7) cosmology and magnitudes are quoted in the AB magnitude system \citep{1983ApJ...266..713O}.

\section{Data}\label{sec:data}
In this section we describe the imaging methods, construction of the co-add images, and galaxy selection criteria. We also provide a brief characterization of image quality in the co-add images with respect to the single-epoch images.

\subsection{The Sloan Digital Sky Survey}\label{sec:sdss}

SDSS Legacy imaging was performed with the dedicated 2.5-m wide-field telescope at Apache Point Observatory (APO) in New Mexico \citep{2009ApJS..182..543A}. A total of around 45,000 square degrees of science quality images across five broad bands ($ugriz$) were produced during operation and calibrated to the AB magnitude system \citep{1983ApJ...266..713O}. Taken together, the imaging covers a unique footprint of 8,423 square degrees of sky. The telescope and distinctive terminology used to describe its imaging are presented in \citet{2006AJ....131.2332G} and \citet{2002AJ....123..485S}, respectively. Imaging \textit{strips} are integrated in drift-scan mode along great circles to form six parallel \textit{scanlines} (one for each camera column) -- each $13.5$ arcmin wide. Two interleaving strips, filling in the gaps and overlapping slightly between scanlines, make a single \textit{stripe} that is 2.5 degrees in width \citep{2000AJ....120.1579Y}.

DR7 imaging is complimented by a 640-fiber-fed pair of multi-object double spectrographs with coverage from $3800-9200$\AA\ at a resolution of $\lambda/\Delta\lambda\simeq2000$. The SDSS DR7 Main Galaxy spectroscopic sample consists of a magnitude-limited sample of 928,567 galaxies complete to a Galactic extinction-corrected Petrosian magnitude limit $m_{\mathrm{r,Petro}}=17.77$ \citep{1998ApJ...500..525S,2002AJ....124.1810S}.

\subsection{Stripe 82}\label{sec:stripe82}

The SDSS Stripe 82 resides at $-1\overset{\circ}{.}25<\delta<+1\overset{\circ}{.}25$, $-50^{\circ}<\alpha<+60^{\circ}$ along the celestial equator in the Southern Galactic Cap. Repeated imaging runs of Stripe 82 were carried out to enable stacking of images and reach fainter magnitudes. Images from Stripe 82 runs $125\leq\texttt{run}\leq5924$ were used in the \cite{2014ApJ...794..120A} stacks -- numbering 123 runs in total. This selection rejects a large number of the original 303 total runs that were taken under poor seeing, bright/moonlit sky, and/or non-photometric conditions. Individual fields of poor quality were then rejected based on quantitative cuts to the $r-$band seeing, sky brightness, atmospheric transparency, and number of standard calibrating stars. The matching $ugiz$ frames which did not satisfy the corresponding $r-$band quality cuts were also rejected. As such, the \emph{effective} number of images used in the co-add stacks varies along the Northern and Southern strips from 15 to 34. The depth is therefore not perfectly homogeneous and varies at most by around 0.4 mag between the shallowest and deepest co-add fields. On average, the stacks go $\Delta(u, g, r, i, z) = (1.6, 1.8, 1.6, 1.6, 1.8)$ magnitudes deeper than the single-epoch SDSS Legacy images based on 50\% completeness limits for point sources. For reference, the average 50\% completeness limits for point sources in the single-pass Legacy images are $(u, g, r, i, z)_{\mathrm{Legacy}} = (22.3, 23.2, 23.0, 22.6, 21.1)$ magnitudes. These limits are determined by inserting and measuring recovery of artificial point-source models into the respective single-epoch and stacked image fields \citep{2016MNRAS.456.1359F}.

Stacks were also constructed by \cite{2014ApJS..213...12J} and \cite{2016MNRAS.456.1359F} (hereafter IAC) which each achieve 1.7-1.9 mag and 1.7-2.0 mag improved depth compared to the single-epoch images, respectively (again with stellar 50\% completion limits as the metric) -- improving slightly on the depth achieved in the \citetalias{2014ApJ...794..120A} stacks. Unlike \citetalias{2014ApJ...794..120A}, the \cite{2014ApJS..213...12J} stacks use all Stripe 82 runs with less strict cuts on input image quality and a slightly modified weighting scheme. In particular, use of images with significant moonlight or strong variations in sky brightness along the drift direction without overly degrading quality in the stacks was facilitated by their background subtraction method -- which efficiently handles such cases. \cite{2016MNRAS.456.1359F} take a completely different approach in the IAC stacks. Instead of attempting to remove the sky, they deliberately place special emphasis on preserving low surface brightness structures on all spatial and intensity scales. Rejecting about 1/3 of the Stripe 82 runs with the poorest quality, their stacks offer the depth and resolution to examine stellar streams, ultra-diffuse galaxies, intracluster light, and even the dust filamentary structure of our Galaxy via optical cirrus. 

The quality of the stacked images is sensitive to the combination procedure and therefore may be optimized for specific science objectives. In particular, the \citetalias{2014ApJ...794..120A} stacks were designed to enable processing of the co-adds using the SDSS standard measurement code, \texttt{PHOTO} \citep{2001ASPC..238..269L,2002SPIE.4836..350L,2012photolite}. For our purposes, this choice offers three main advantages over other stacking strategies for a small cost in depth: (1) the SDSS collaboration has thoroughly tested the \texttt{PHOTO} pipeline and characterized the uncertainties in the measurements made by its algorithms; (2) consistency between the methods and measurements for the single-epoch images and stacks; and (3) the data access architecture in the Catalog Archive Server (CAS) and Data Archive Server (DAS) are conveniently consistent with the single-epoch images. \citetalias{2014ApJ...794..120A} also imposed strict quality cuts and an unforgiving quality-weighting scheme for the images considered in the co-adds. For our purposes, these factors facilitate the most straight-forward comparison between quantitive morphologies of galaxies in the \citetalias{2014ApJ...794..120A} stacks and the Legacy images. 

\subsection{Estimation of the sky and total variance}\label{sec:gain}

Accurate modelling and meaningful uncertainties in model parameters rely crucially on a correct interpretation of noise. As with a single-epoch image, the total variance in a calibrated co-add image pixel is well approximated by:
\begin{align}
\label{eq:noise}
\sigma_{\mathrm{tot}}^2 = \sigma^2_{\mathrm{source}} + \sigma^2_{\mathrm{sky}} = (I_{\mathrm{source}}- \langle I_{\mathrm{sky}}\rangle )/G_{\mathrm{eff}} + \sigma^2_{\mathrm{sky}}
\end{align}
where $I_{\mathrm{source}}- \langle I_{\mathrm{sky}}\rangle$ is the background-subtracted source intensity measured in the pixel (D.U.) and $G_{\mathrm{eff}}$ is the  effective gain (electrons/D.U.) of the stacked image. The variance in a co-add image pixel intensity derives from noise contributions from the individual input images which are generally non-homogeneous in $S/N$. However, the sky variance, $\sigma^2_{\mathrm{sky}}$, can be measured directly and locally around a particular galaxy in the co-add images. We use the approach of \citetalias{2011ApJS..196...11S} to estimate $\sigma^2_{\mathrm{sky}}$. First, \textsc{SExtractor} \citep{1996A&AS..117..393B} is used to delineate between sources and sky and to deblend pixels in overlapping sources. The \textsc{SExtractor} parameters were calibrated in \citetalias{2011ApJS..196...11S} using sensitive tests based on the results of subsequent bulge+disc decompositions with \textsc{gim2d}: the size luminosity relation of discs and the colour-magnitude diagrams and fiber colours of pairs in the \cite{2011MNRAS.412..591P} pair catalog. The resulting \textsc{SExtractor} segmentation image is used to measure the local sky level and noise around the target source in the co-add image.\footnote{A corrected co-add image is already subtracted by a standard estimate of the global sky. Therefore, we measure a local \emph{residual} sky level with respect to this estimate in the corrected co-add image.} A minimum of 20,000 of the nearest sky pixels to the target, excluding those within four arcseconds of any source pixel, are used to estimate the local sky level and variance. We assume that the readout noise is captured by this direct sky noise estimate. The source Poisson variance in Equation \ref{eq:noise} is then estimated in each target pixel using sky-corrected intensities and the effective gain that can be obtained from the corresponding atlas image headers.

\subsection{Spatial Resolution}

\begin{figure}
  \includegraphics[width=\linewidth]{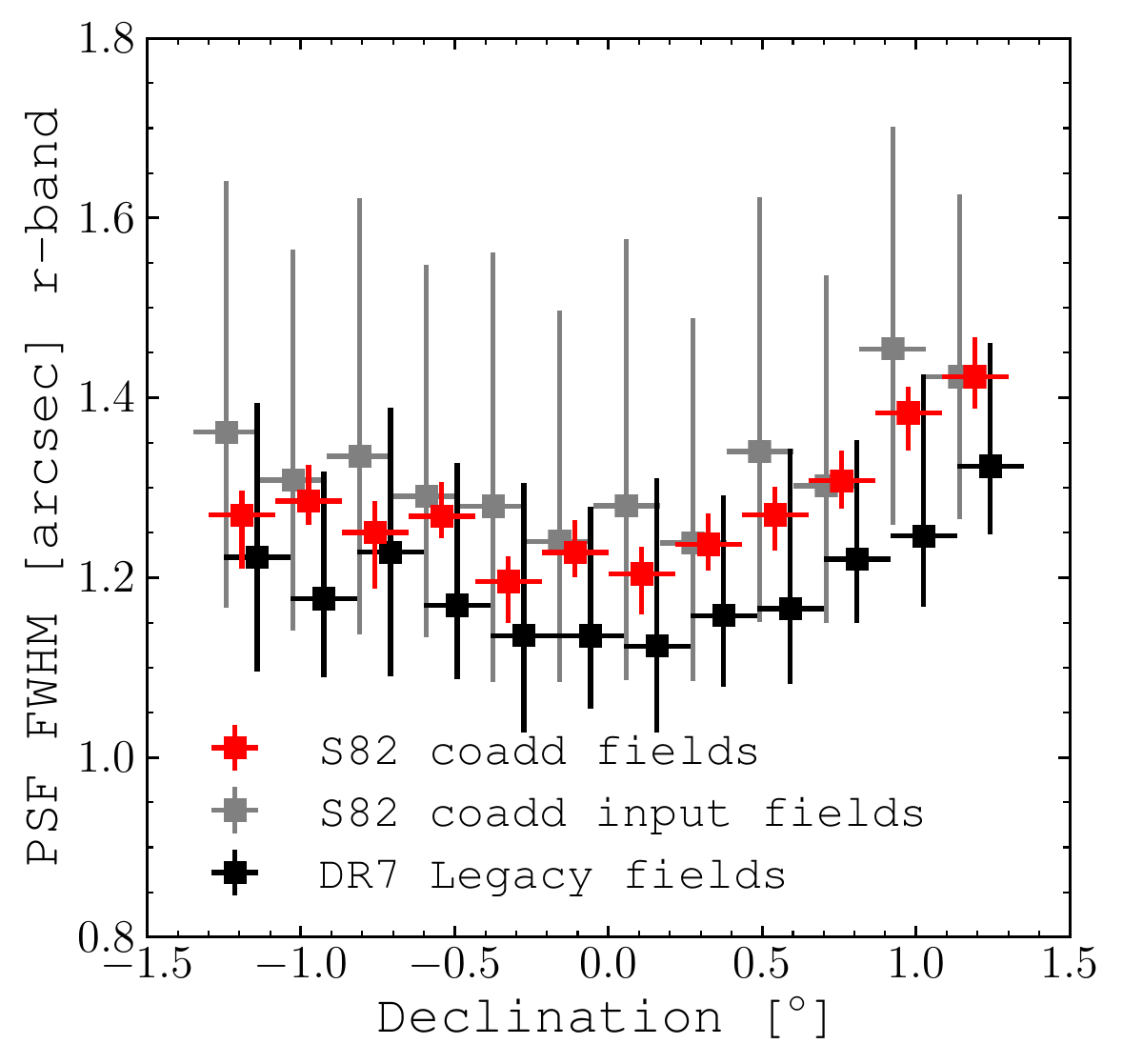}
\caption[Characterization of PSF resolution]{Characterization of spatial resolution in $r$-band co-add images. Point-spread function FWHM are obtained from the respective co-add and single epoch \textsc{field} tables and are measured from 2D gaussian fits to the point-spread function in each (co-add) image field. Points show the medians and $16^{\mathrm{th}}-84^{\mathrm{th}}$ percentiles of PSF FWHM for all co-add stacks (red), single-epoch DR7 Legacy (black) fields, and co-add input fields in each of the 12 scanlines. Black and grey markers are offset by $0.05^{\circ}$ right and left of their respective bin centres for clarity. Input co-add fields (grey) include also SDSS-II Supernovae Project runs which have poorer photometric standards standards than Legacy fields.
}
\label{fig:fwhm}
\end{figure}

The modified version of \texttt{PHOTO} that was used for the Stripe 82 co-adds models the spatial variation of the Point-Spread Function (PSF) using the suitably weighted sum of the model PSFs from the input image fields -- for which the spatial variations were already separately determined. The weighting scheme is designed such that images with the best seeing have higher weights in the co-add construction. Figure \ref{fig:fwhm} compares the median spatial resolution along each scanline in the co-add images (red markers), the DR7 Legacy images used in the \citetalias{2011ApJS..196...11S} decompositions (black markers), and the full ensemble of input images used in the co-add stacks (grey markers). Measurements for the PSF were taken from the SDSS \textsc{field} tables for Stripe 82 and DR7. The \citetalias{2011ApJS..196...11S} decompositions used Stripe 82 \texttt{runs}: (2583, 2662, 3325, 3388, 2738, 2659) from Legacy. The spatial resolution in the Stripe 82 co-adds is $\sim0.1"$ poorer than in the Legacy fields but is more homogenous along each scanline; reducing the typical scatter from $\sim0.3"$ to $\sim0.1"$. The median seeing and its scatter are improved in the co-adds with respect to the full set of input images -- as expected from the image weighting strategy. The main difference between the Legacy and co-add input fields is the co-add inputs include images from the SDSS-II SNeS project runs which were generally taken under poorer photometric conditions. Fields from these runs generally have small weights in the co-add construction but are more numerous than Legacy fields; offsetting the median seeing to poorer quality. 

\subsection{Galaxy Sample Selection}

We select galaxies from the SDSS DR7 spectroscopic galaxy catalog that are previously analyzed by \citetalias{2011ApJS..196...11S} and which reside in Stripe 82. This provides a sample of 16,908 galaxies for which decompositions in the deeper images can be directly compared with the more shallow reference decompositions from \citetalias{2011ApJS..196...11S}. We avoid misclassified non-galaxy sources in our galaxy sample by selecting targets determined to be galaxies from SDSS fiber spectroscopy. Consequently, we have spectroscopic redshifts for all galaxies in our sample. Furthermore, galaxies from the \citetalias{2011ApJS..196...11S} decompositions all have Petrosian magnitudes $m_{\mathrm{r,Petro}}\leq17.77$ mag -- the faint limit of the SDSS Legacy spectroscopic sample. Our selection, combined with $\sim1.6-1.8$ magnitudes deeper imaging for this large galaxy sample, enables at least two unique experiments using bulge+disc decompositions. First, we can investigate the impact of added depth on global and component properties. Second, components and asymmetric structures that may previously have been too faint for accurate characterization, or indeed detection, in the single-epoch images may be better characterized in the Stripe 82 co-adds.

\section{Decompositions and Catalogs}\label{sec:catalogs}
\subsection{Photometric Decompositions}
To be consistent with the analyses of \citetalias{2011ApJS..196...11S} and \cite{2014ApJS..210....3M} in shallower images, we use the same analysis pipeline for our decompositions in Stripe 82. Quantitative morphologies were obtained using the parametric surface-brightness decomposition tool \textsc{gim2d}. An important feature of the \textsc{gim2d} software is that it uses a Bayesian Metropolis-Hastings maximum-likelihood optimization algorithm (but see also \textsc{ProFit}; \citealt{2017MNRAS.466.1513R}). As such, the best-fitting model parameters are largely insensitive to the volume of parameter space in which optimization iterations begin. The nature of the Metropolis-Hastings algorithm is that there is always a finite probability of leaving a position of greater likelihood for a position of lesser likelihood. Therefore, the algorithm need not settle at a local maximum but can escape to (hopefully) converge on the global maximum. Each galaxy was fitted using three models: 
\begin{itemize}
\item (\texttt{ps}) single-component S\'ersic profile \\
\item (\texttt{n4}) two-component $n_b=4$ bulge and exponential disc \\
\item (\texttt{fn}) two-component free-$n_b$ bulge and exponential disc 
\end{itemize}
The parameters of these models are described in \citetalias{2011ApJS..196...11S}. As in \cite{2014ApJS..210....3M}, we perform $ur$, $gr$, $ir$, and $zr$ pairwise simultaneous surface-brightness decompositions. The $r-$band mask from \textsc{SExtractor} is used for all images to avoid inconsistencies in pixel allocations in each band. Similarly, following \cite{2014ApJS..210....3M}, structural parameters in each band of a pairwise simultaneous fit are forced to be the same. These tied structural parameters are the bulge effective radius, bulge ellipticity, bulge S\'ersic index, disc scale length, disc inclination, and bulge and disc position angles. Total galaxy magnitudes, bulge-to-total fractions, and centroid offsets are free to vary in each band of a pairwise fit. 

In addition, following \cite{2014ApJS..210....3M}, the structural parameters for $ur$, $ir$, and $zr$ pairwise fits were fixed to values first derived in $gr$ decompositions (with the exception of S\'ersic index in the \texttt{ps} decompositions -- which is allowed to vary freely in each pairwise fit; see Appendix A of \citealt{2014ApJS..210....3M} for discussion). As such, the only truly independent structural measurements are those derived in the $gr$ simultaneous decompositions -- the structural measurements of all other pairwise decompositions for a given model type (\texttt{ps}, \texttt{n4}, \texttt{fn}) are held fixed to the $gr$ values. Consequently, this method does not offer the flexibility required to capture the known wavelength dependence of galaxy structural parameters that has been identified using more rigorous multi-wavelength models (e.g., \textsc{Sigma}-- \citealt{2012MNRAS.421.1007K}; and \textsc{MegaMorph}-- \citealt{2013MNRAS.430..330H,2013MNRAS.435..623V}). Such a model is not employed here because our primary goal is to characterize the impact of improved imaging depth on the \cite{2014ApJS..210....3M} and \cite{2011ApJS..196...11S} decomposition measurements. Exploiting the same model and analysis pipeline is of fundamental importance to the nature of this task. So while the wavelength dependence of galaxy magnitudes and bulge-fractions in our models can be largely considered independent for each band, the structural parameters that are held fixed across bands cannot. \cite{2014ApJS..210....3M} adopted this (albeit restrictive) multi-wavelength strategy to suitably construct a robust broadband SED for each galaxy which would not be strongly affected by the covariances between model parameters for sources that are faint in any particular band. In summary, any changes to $uiz$ model structural parameters are \emph{entirely} due to changes in the $gr$ decomposition results (by construction). But total magnitudes, bulge and disc fractions, and corresponding component magnitudes are independent in each band. Any changes to these features are reflections of the improved constraints provided by the deeper images.

Our fiducial \texttt{n4} model adopts an $n_b=4$ S\'ersic profile for the bulge. The choice of S\'ersic index for bulge component models in galaxy surface-brightness decompositions has been a long-standing debate in the literature -- ultimately resolving that a continuous range of bulge profile shapes must exist (\citealt{1994MNRAS.267..283A,1995MNRAS.275..874A,1996A&A...313...45D,2000ApJ...531L.103K,2001AJ....121..820G,2001A&A...368...16M,2003ApJ...582..689M,2003AJ....125.2936G,2003ApJ...582L..79B,2004ARA&A..42..603K,2010ApJ...716..942F,2010MNRAS.405.1089L}; see also \citealt{2013pss6.book...91G} \S2 and \S4 for a review). At the heart of the remaining challenges to bulge characterization is whether S\'ersic index can be used to classify and discriminate between bulges of different formation mechanisms. The challenge is exacerbated by the presence of additional structural components such as bars, excess nuclear light, and internal lenses (rings that do not derive from gravitational lensing) which are known to affect structural measurements -- in particular the bulge-to-total fractions, $B/T$ \citep{2003AJ....125.2936G,2005MNRAS.362.1319L,2010AJ....139.2097P,2013MNRAS.430.3489L}. Without independent priors for the presence of such features (e.g., by thorough visual inspection of each galaxy's surface brightness profile), adding model components will only serve to create degeneracies between model parameters and produce meaningless results. Recently however, \cite{2018MNRAS.473.4731K} have shown that visual classification from the Galaxy Zoo citizen science project \citep{2008MNRAS.389.1179L} can serve as useful priors for using additional components in the decompositions. Calibration of such priors and their implementation is beyond the scope of the present work.

The Stripe 82 co-adds present an opportunity to investigate whether added depth enables better characterization of the bulge profile. \citetalias{2011ApJS..196...11S} showed that the majority of galaxies in the single-epoch images had insufficient $S/N$ and/or spatial resolution to discriminate between bulge profiles. Figure \ref{fig:fwhm} shows that improved characterization of the peak due to better seeing is not expected. The seeing in a co-add image is not improved in quality (indeed, it is slightly poorer) compared to a corresponding single-epoch Legacy image. As such, changes to measurements due to seeing are generally not expected. As for depth, the 1.6-1.8 magnitude improvement (factor of 4.4-5.2 in $S/N$) should increase the number of pixels assigned to objects in the deblending -- as more will meet the intensity threshold with respect to the sky noise. Larger footprints and more degrees of freedom make the fits slower, but enable better characterization of the low surface-brightness limits of galaxy profiles. With respect to the bulge, the increase in $S/N$ should improve delineation of the sky from the wings of the bulge profile -- which is also important for accurate bulge characterization.

\subsection{Tables \& Catalogs}

\begin{table*}
	\centering
	\caption[Photometric decomposition catalogs]{Photometric decomposition catalogs. Fits using each of the three models are performed simultaneously in two bandpasses -- one of which is always the $r-$band. Structural parameters (bulge effective radius, bulge ellipticity, disc scale length, disc inclination, and bulge and disc position angles) are either free or fixed to the $gr$ results as indicated. The model's (bulge) S\'ersic index is also either free or fixed to the $gr$ results as indicated (though it is always free in the single S\'ersic component model decomposition). The number of successful fits to each bandpass pair, $N_{\mathrm{success}}$, is shown in the final column. }
	\vspace{5pt}
	\label{tab:catalogs}
	\begin{tabular}{lccccr} 
		\hline
		Catalog Name & Decomposition Model & Bandpasses & Structural Parameters & (Bulge) S\'ersic index & N$_{\mathrm{success}}$  \\
		\hline
		\texttt{sdss\_s82\_morph\_gr\_n4} & $n_b=4$ bulge + exp. disc & $g$ and $r$ & Free          & Fixed $n_b=4$ & 16,822\\
		\texttt{sdss\_s82\_morph\_ur\_n4} & $n_b=4$ bulge + exp. disc & $u$ and $r$ & Fixed to $gr$ & Fixed $n_b=4$ & 16,672\\
		\texttt{sdss\_s82\_morph\_ir\_n4} & $n_b=4$ bulge + exp. disc & $i$ and $r$ & Fixed to $gr$ & Fixed $n_b=4$ & 16,663\\
		\texttt{sdss\_s82\_morph\_zr\_n4} & $n_b=4$ bulge + exp. disc & $z$ and $r$ & Fixed to $gr$ & Fixed $n_b=4$ & 16,704\\
		\texttt{sdss\_s82\_morph\_gr\_ps} & single free-$n$ S\'ersic & $g$ and $r$  & Free          & Free          & 16,892\\
		\texttt{sdss\_s82\_morph\_ur\_ps} & single free-$n$ S\'ersic & $u$ and $r$  & Fixed to $gr$ & Free          & 16,764\\
		\texttt{sdss\_s82\_morph\_ir\_ps} & single free-$n$ S\'ersic & $i$ and $r$  & Fixed to $gr$ & Free          & 16,724\\
		\texttt{sdss\_s82\_morph\_zr\_ps} & single free-$n$ S\'ersic & $z$ and $r$  & Fixed to $gr$ & Free          & 16,729\\
		\texttt{sdss\_s82\_morph\_gr\_fn} & free-$n_b$ bulge + exp. disc & $g$ and $r$ & Free          & Free          & 16,822\\
		\texttt{sdss\_s82\_morph\_ur\_fn} & free-$n_b$ bulge + exp. disc & $u$ and $r$ & Fixed to $gr$ & Fixed to $gr$ & 16,707\\
		\texttt{sdss\_s82\_morph\_ir\_fn} & free-$n_b$ bulge + exp. disc & $i$ and $r$ & Fixed to $gr$ & Fixed to $gr$ & 16,681\\
		\texttt{sdss\_s82\_morph\_zr\_fn} & free-$n_b$ bulge + exp. disc & $z$ and $r$ & Fixed to $gr$ & Fixed to $gr$ & 16,664\\
		\hline
	\end{tabular}
\end{table*}

\begin{figure*}  
\includegraphics[width=0.237\linewidth,angle=90]{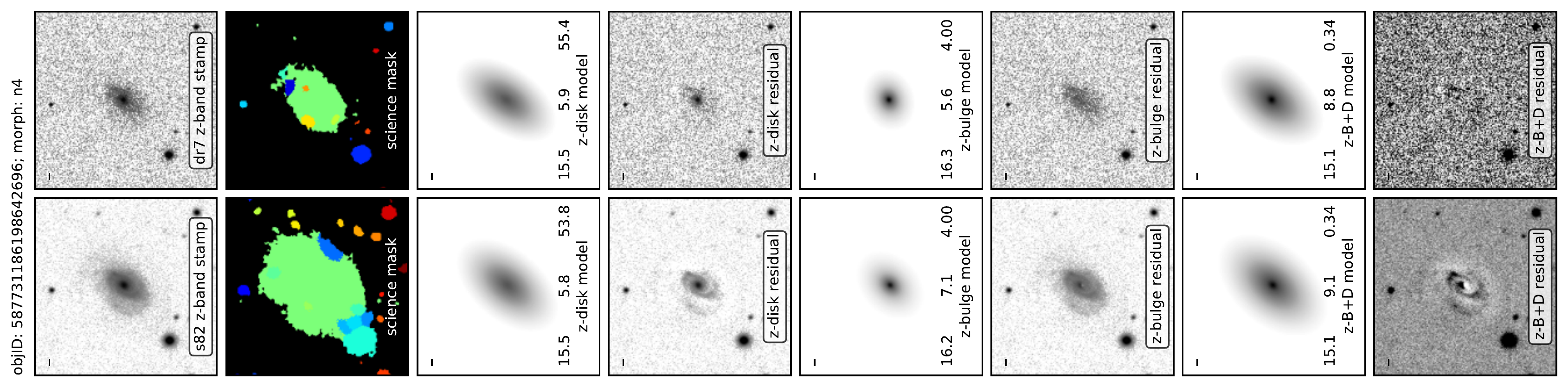}  
\includegraphics[width=0.237\linewidth,angle=90]{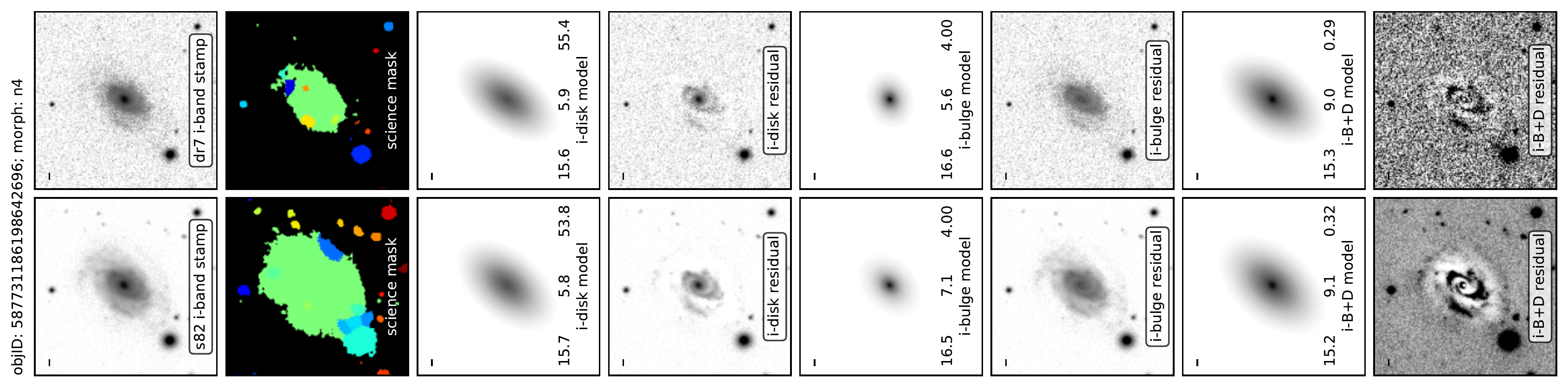}  
\includegraphics[width=0.237\linewidth,angle=90]{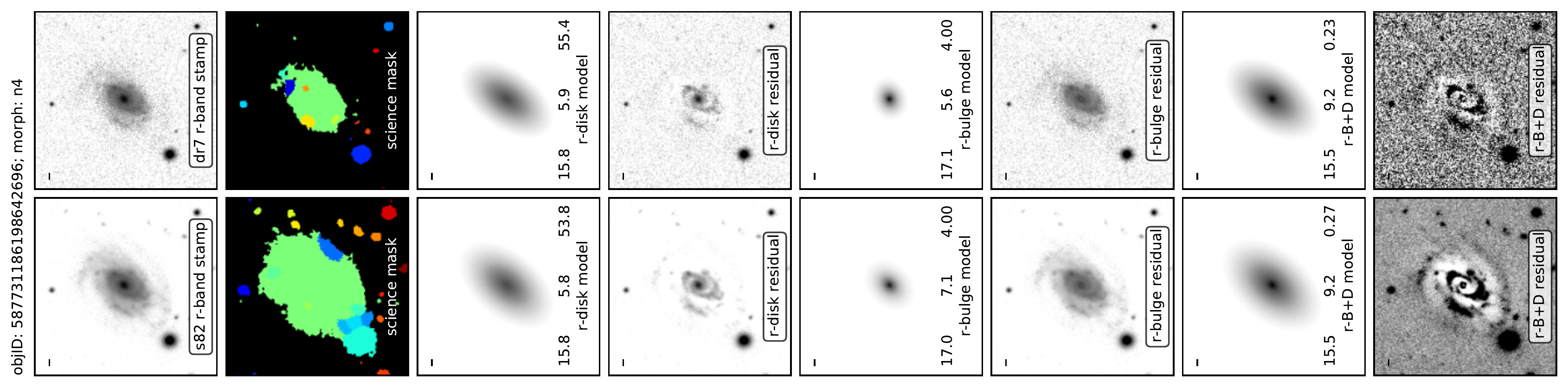}  
\includegraphics[width=0.237\linewidth,angle=90]{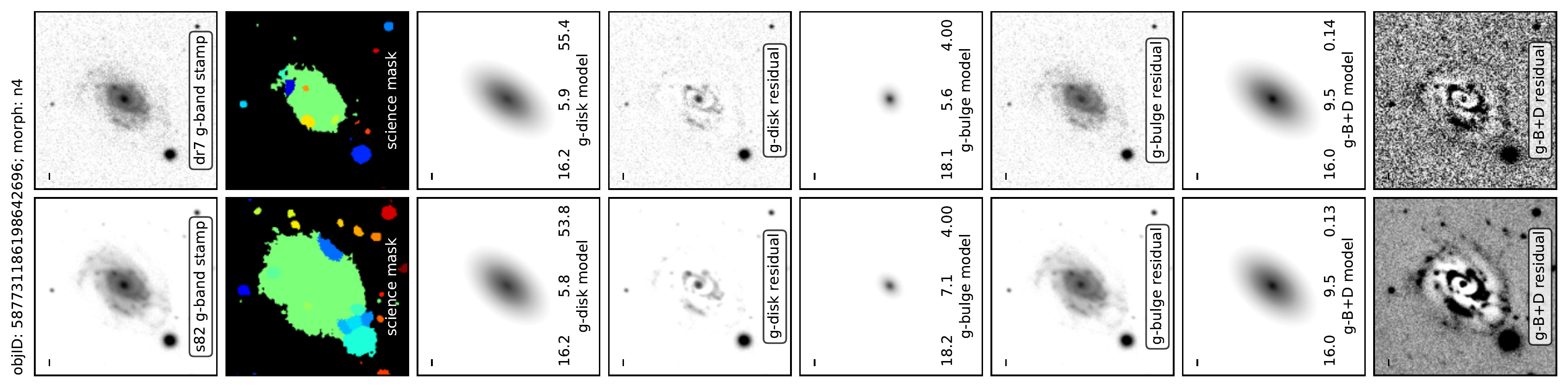}
\includegraphics[width=0.237\linewidth,angle=90]{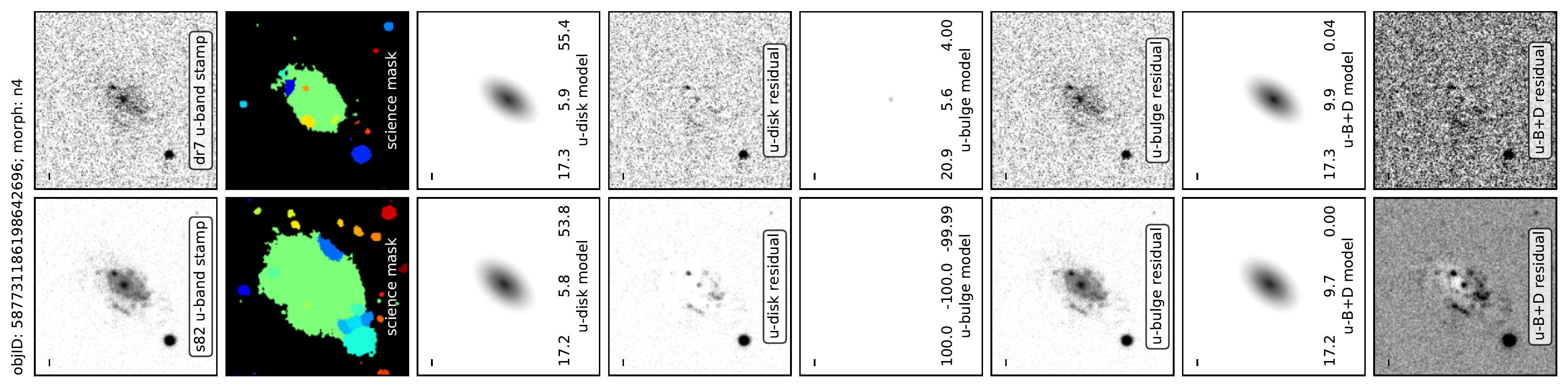}

\caption[Band mosaic]{Bulge+disc decompositions in SDSS $ugriz$ bands for a single randomly selected galaxy in our sample. Each pair of columns (reading the figure in landscape) shows the decomposition results for a corresponding band in Stripe 82 (left columns) and DR7 Legacy (right columns). The galaxy models in the right columns are constructed from the \citetalias{2011ApJS..196...11S} results. Each image, with the exception of the mask, includes a ruler on the upper left measuring 3 arcsec. \emph{First row}: science image cut-outs; \emph{Second row}: mask image generated with \textsc{SExtractor} from the $r$-band image frame (pixels with the same colour as the central pixel belong to the source, black=sky, and all other colours are deblended sources); \emph{Third row}: \texttt{n4} disc model with disc apparent magnitude, disc scale length (arcsec), and disc position angle; \emph{Fourth row}: exponential disc model residual; \emph{Fifth row}: \texttt{n4} bulge model with apparent bulge magnitude, bulge effective radius (arcsec), and bulge S\'ersic index (always 4.0 in this case); \emph{Sixth row}: bulge model residual; \emph{Seventh row}: full \texttt{n4} model with total apparent magnitude, half-light radius (arcsec, derived from the combination of the bulge and disc model), and bulge-to-total fraction; \emph{Eighth row}: \texttt{n4} model residual. The bulge+disc model residual images use a linear greyscale in which the contrast limits are set to $(-5,5)\sigma_{\mathrm{sky,S82}}$. All other images use a logarithmic greyscale in which the contrast limits are set to (0.01,$I_{10}$) nanomaggies where $I_{10}$ is the maximum intensity within 10 pixels of the target galaxy barycenter.}
\label{fig:bands}
\end{figure*}

Catalogs for the photometric decompositions are made available in plain text and Structured Query Language (SQL) format. The catalogs have the same naming convention and structures as the \citetalias{2011ApJS..196...11S} public catalog. The salient features of each table are shown in Table \ref{tab:catalogs}. The basic structure and parameters of each catalog are shown in Table \ref{tab:morph} in Appendix \ref{sec:appB}. Uncertainties on model parameters are the $16\%$ and $84\%$ confidence limits acquired by sampling the convergence region about the best-fitting model after convergence. The statistical uncertainties on observed fluxes from \textsc{gim2d} are lower limits, which do not account for additional uncertainties arising from relative and absolute SDSS photometric calibration and sky subtraction. \cite{2014ApJS..210....3M} used Monte Carlo simulations of analytic bulge+disc models in SDSS fields to confirm that the \textsc{gim2d} statistical uncertainties and the rms noise in the sky background are consistent with the total uncertainties in the recovery of simulated galaxy fluxes. Additionally, an uncertainty of $\sim3\%$ in $griz$ and $\sim5\%$ in $u$ bandpasses arises from the relative and absolute photometric calibrations. Taken together, the combination of these three sources of uncertainty account for the total uncertainties in the model fluxes. 

\section{Comparison with Legacy Results}\label{sec:comparison}

\begin{figure*}
  \includegraphics[width=0.33\linewidth]{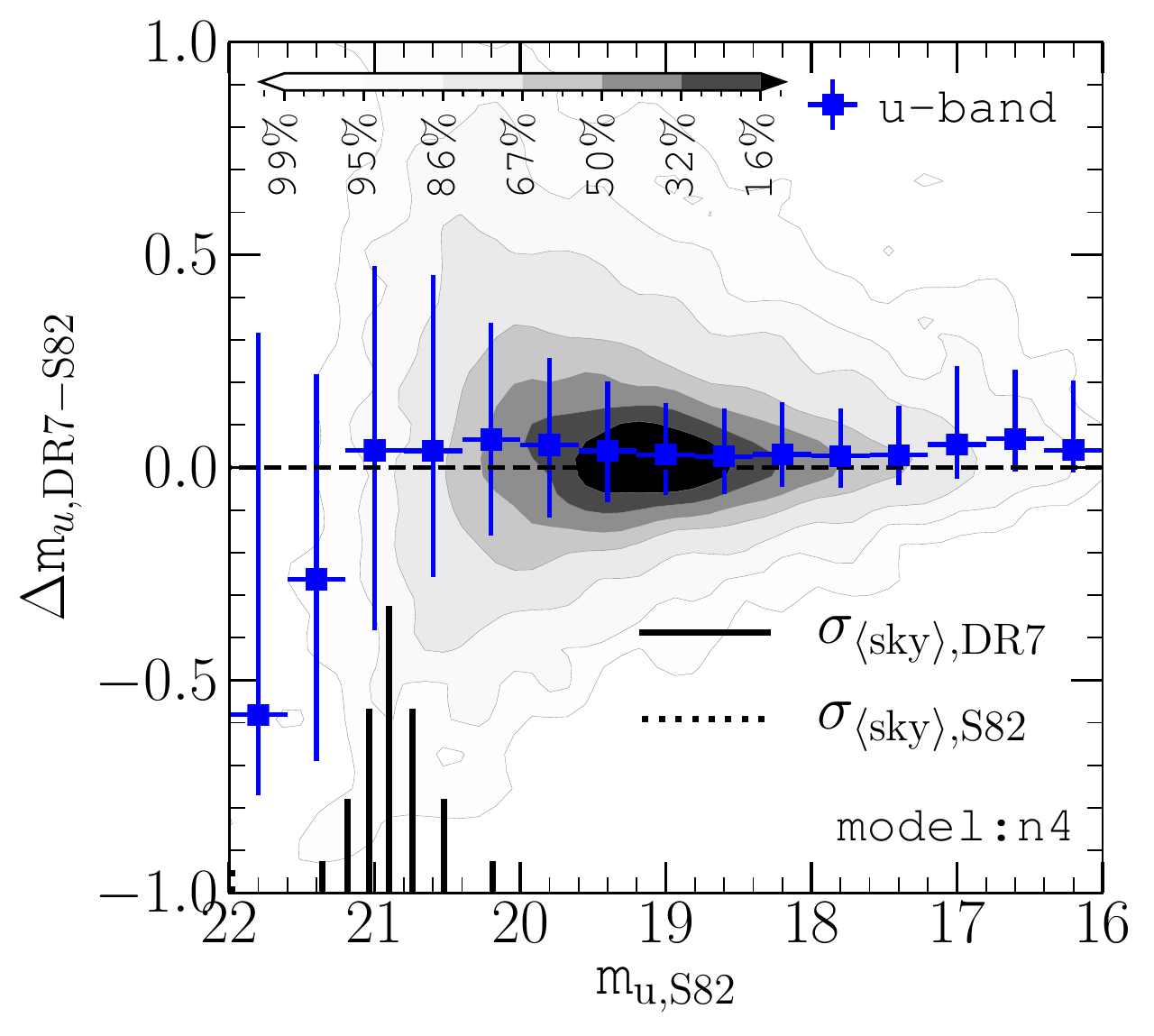}
  \includegraphics[width=0.33\linewidth]{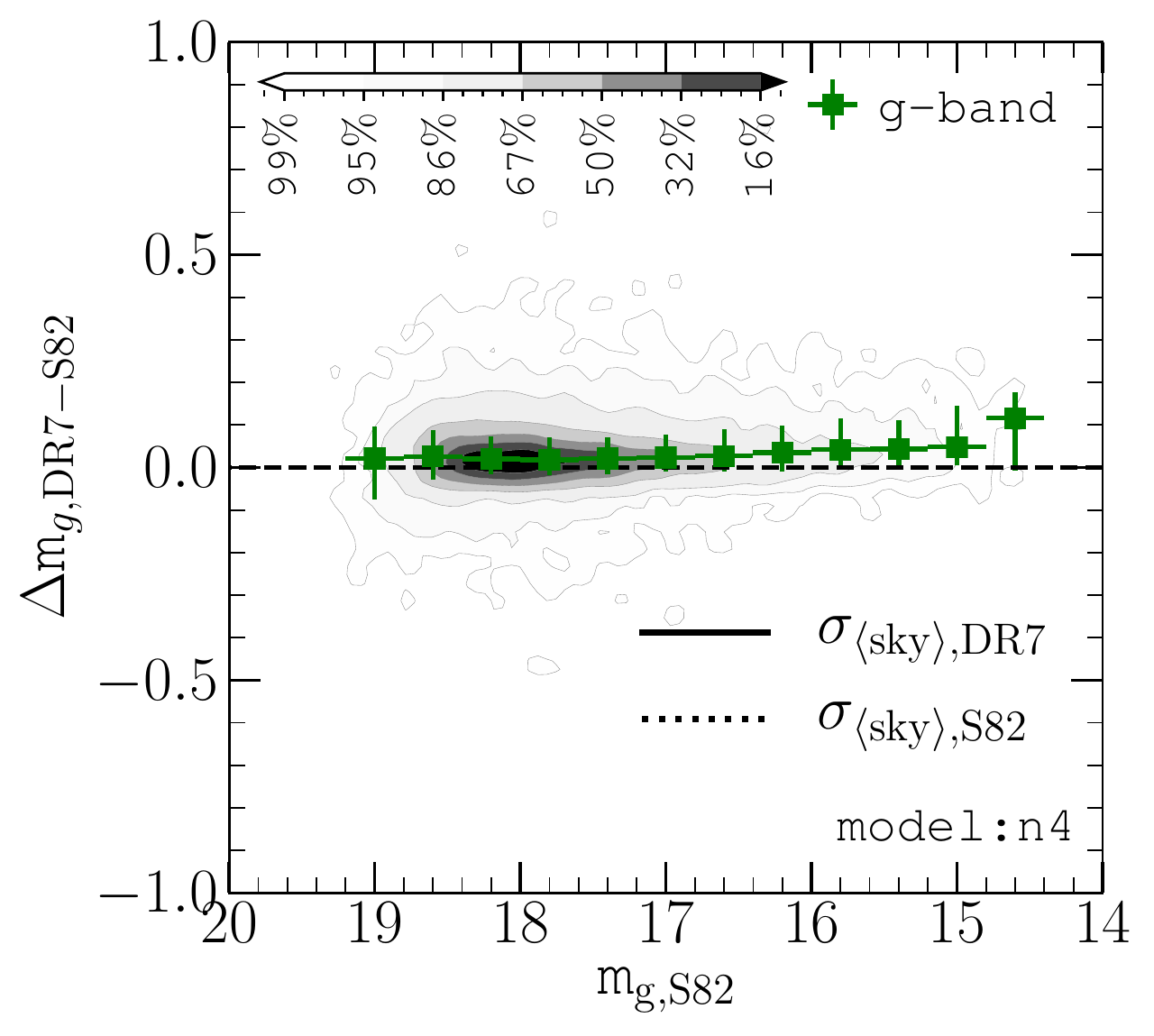}
  \includegraphics[width=0.33\linewidth]{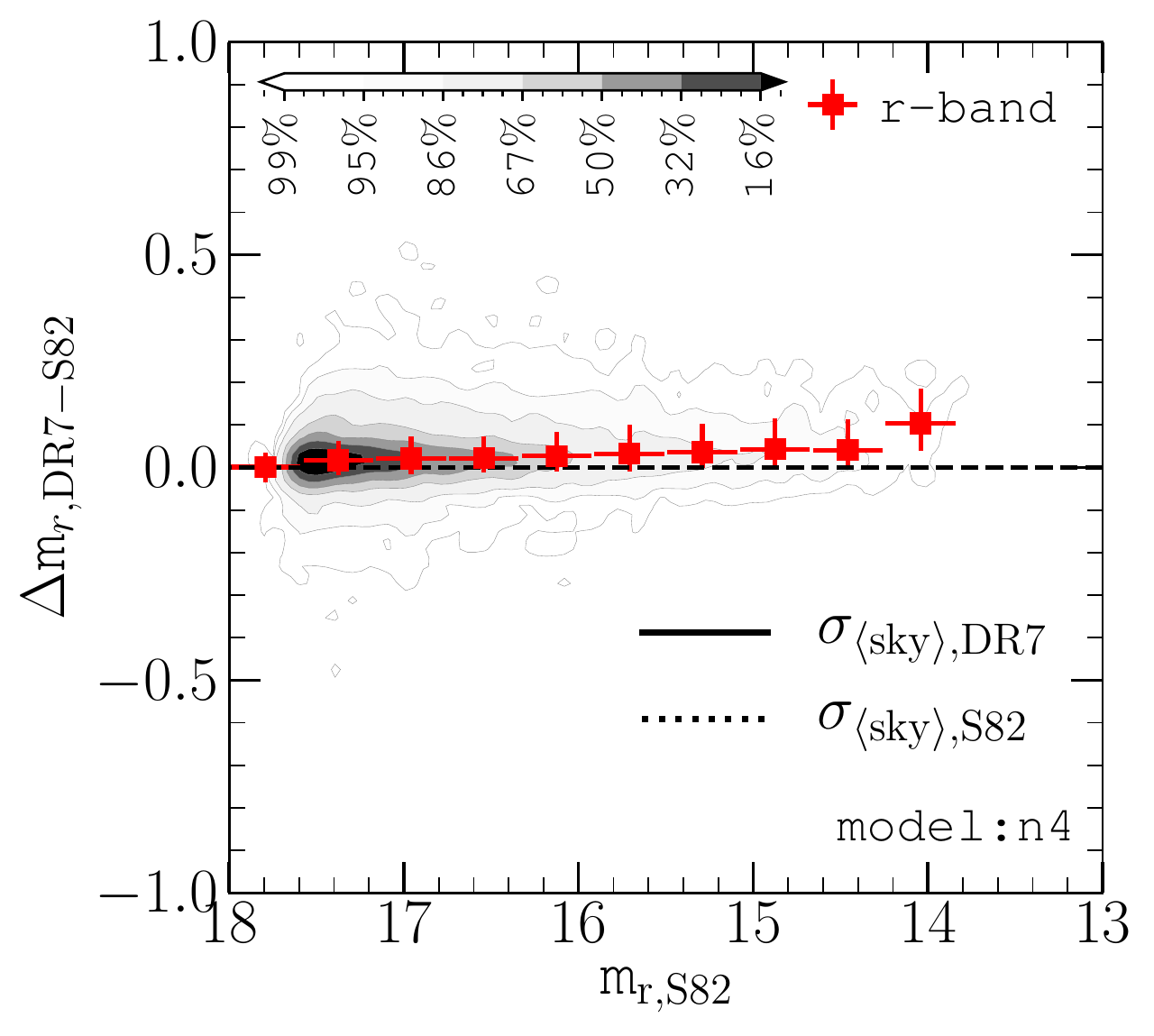}
  \includegraphics[width=0.33\linewidth]{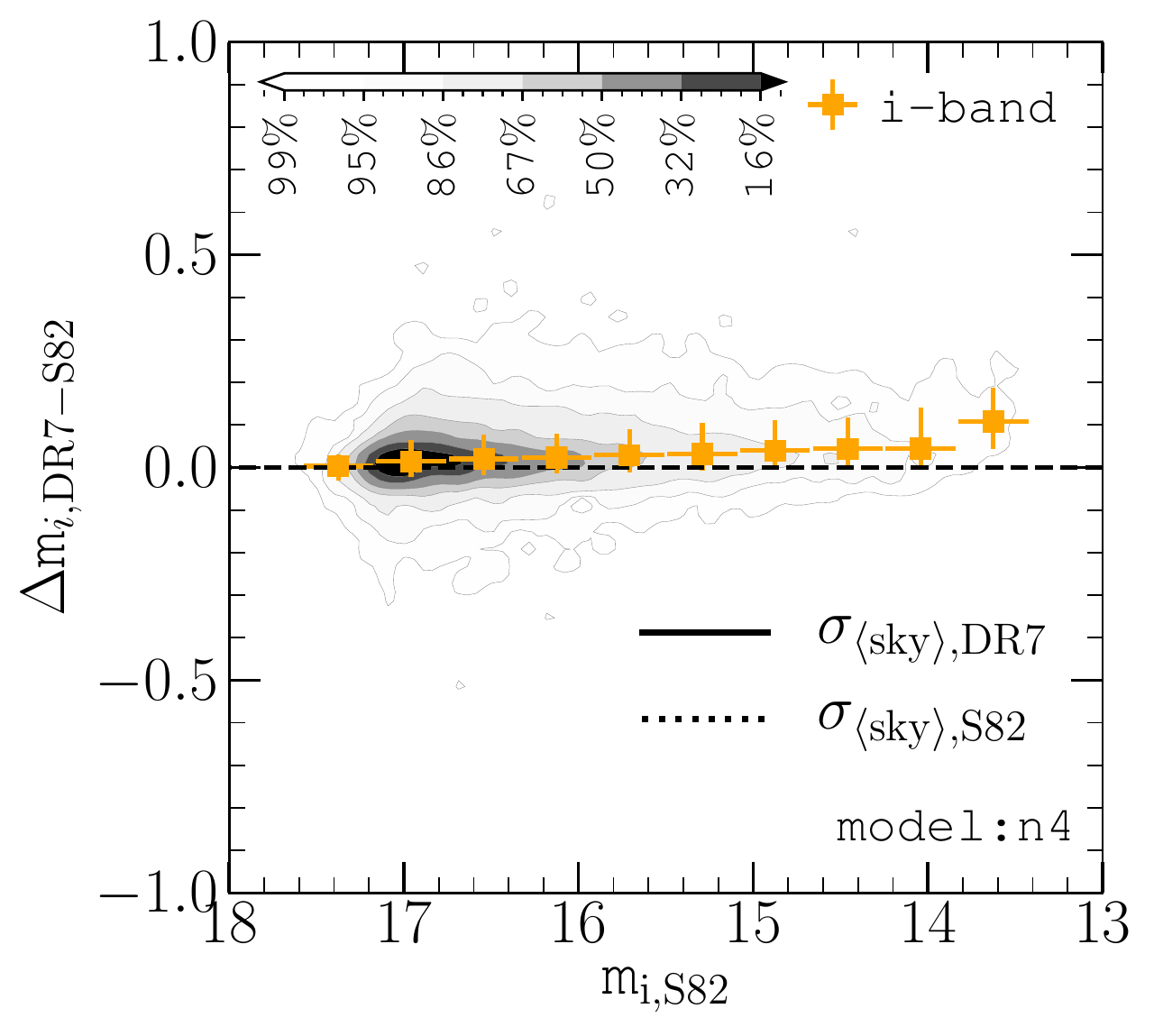}
  \includegraphics[width=0.33\linewidth]{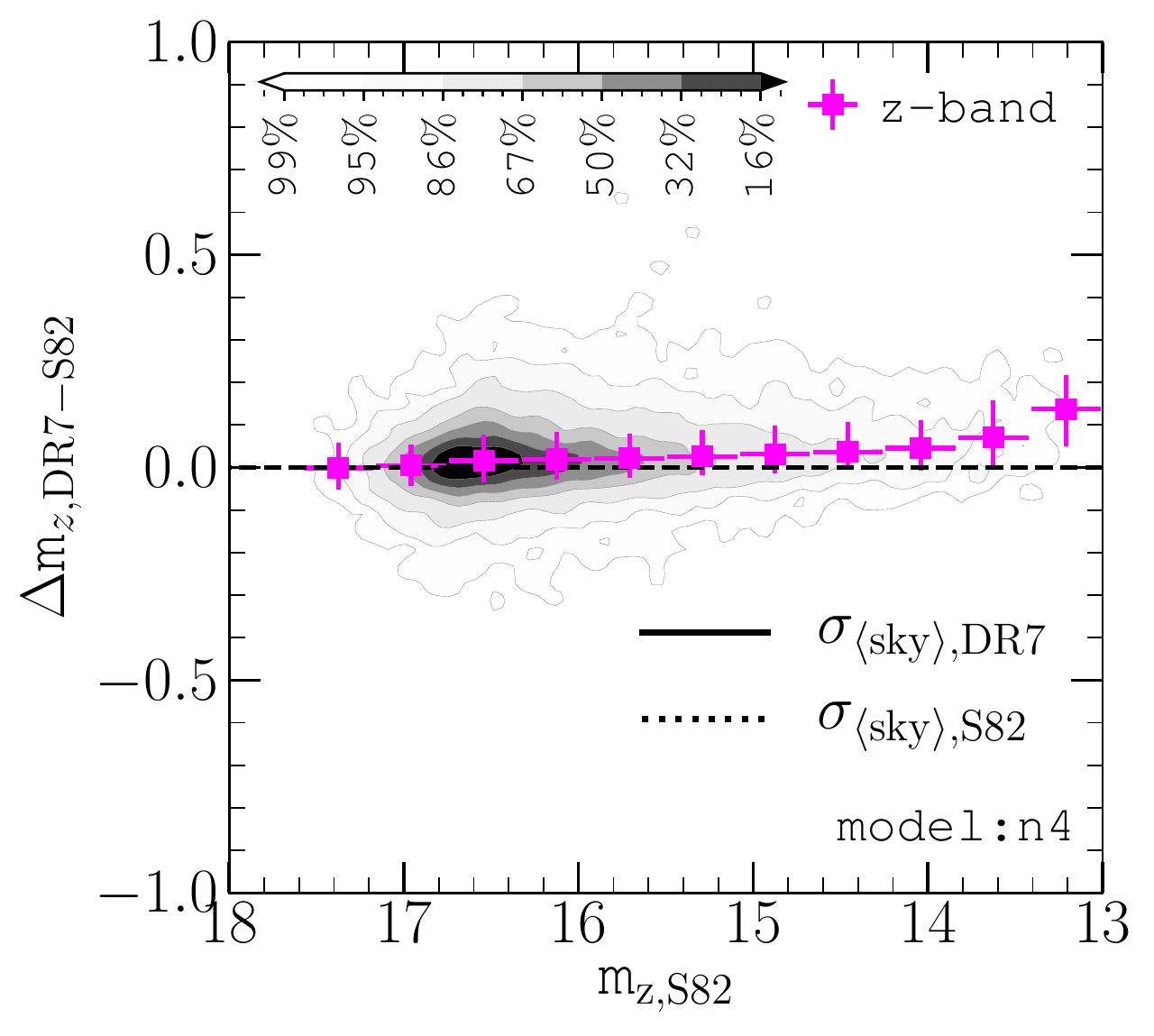}
\caption[Total apparent magnitudes]{Difference in total apparent magnitudes of galaxies obtained from Stripe 82 co-adds and DR7 Legacy images with the \texttt{n4} decomposition model. The $r-$band magnitudes are taken from the $gr$ decomposition catalogs. Contour lines show percentiles of the magnitude difference distributions, $\Delta m_{x,\mathrm{DR7-S82}}$, in each panel. Dashed lines denote $\Delta m_{x,\mathrm{DR7-S82}}=0$. Coloured markers show the median in each bin with error bars corresponding to the $16-84$ percentile range in bins of 0.4 mag along $m_{x,\mathrm{S82}}$. Lines extending from the magnitude axis denote (from left to right) the \{95\%, 84\%, 68\%, 50\%, 32\%, 16\%, 5\%\} percentiles of the local \emph{mean} sky measurement uncertainties, $\sigma_{\langle \mathrm{sky} \rangle}$ (mag) in the Stripe 82 fits (dotted, not visible in any panel) and Legacy fits (solid, only visible in the $u-$band). As a pedagogical example of how to interpret these lines, 5\% of DR7 $u$-band mean sky surface brightness uncertainties correspond to magnitudes brighter than $m_u\approx20.2$ mag. Note that the Legacy 50\% $\sigma_{\langle \mathrm{sky} \rangle,u}$ peak coincides with the sudden decrease in Stripe 82 $u-$band brightnesses relative to Legacy. Legacy brightness measurements have difficulty penetrating the sky noise at these limits.}
\label{fig:mags}
\end{figure*}

\begin{figure*}
  \includegraphics[width=0.19\linewidth]{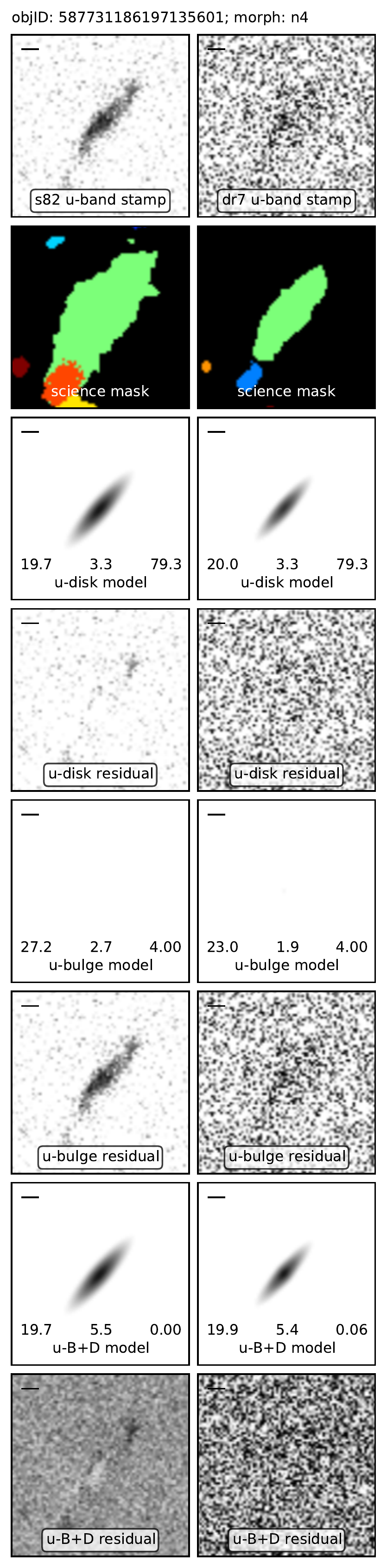}
  \includegraphics[width=0.19\linewidth]{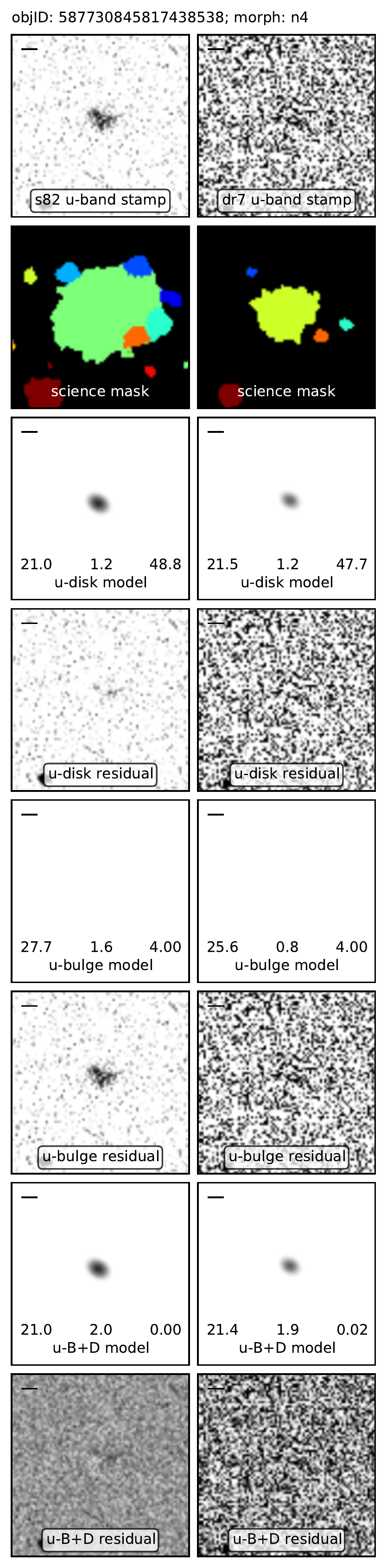}
  \includegraphics[width=0.19\linewidth]{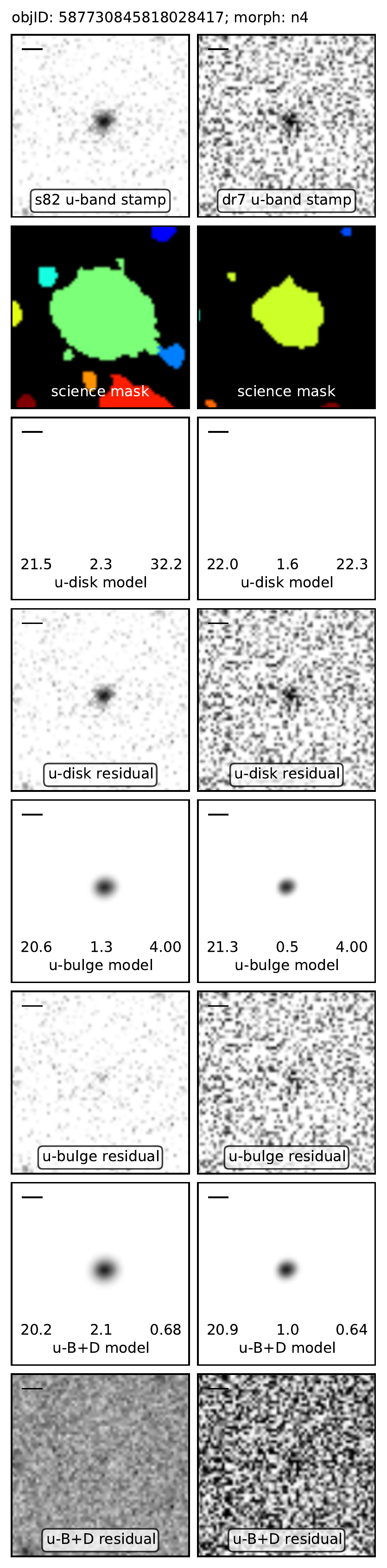}
  \includegraphics[width=0.19\linewidth]{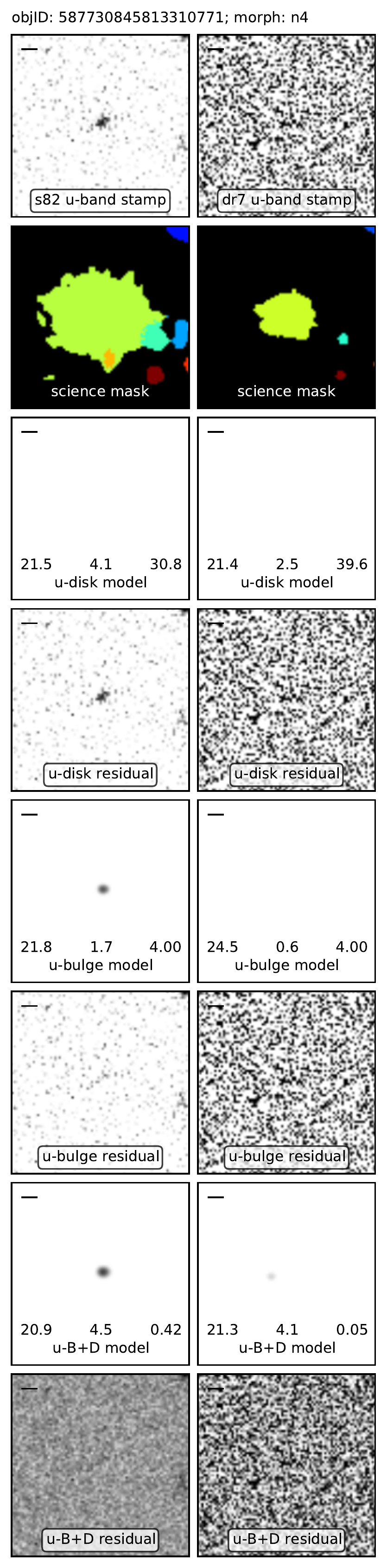}
  \includegraphics[width=0.19\linewidth]{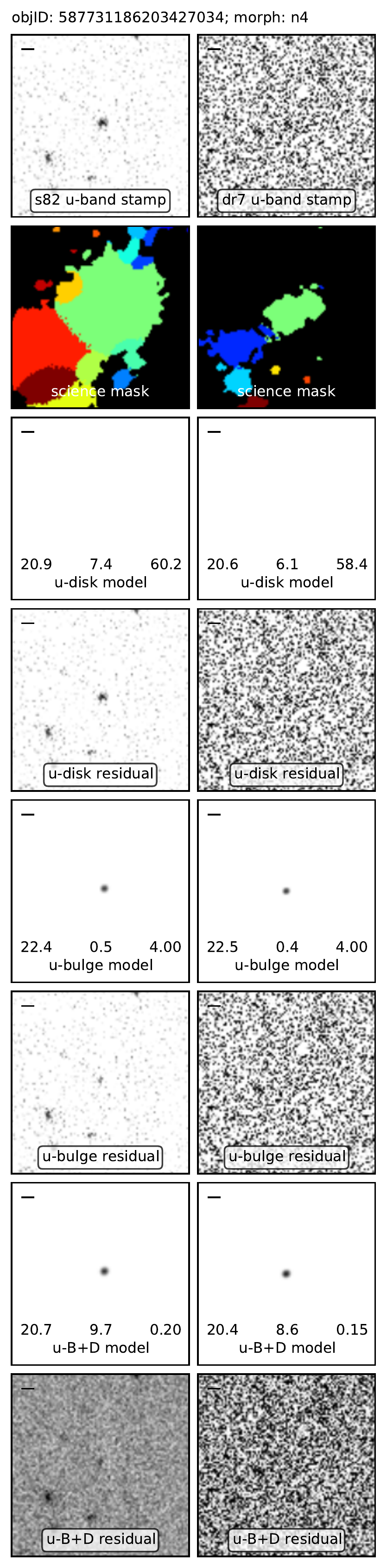}
\caption[$u$-band mosaic]{Similar to Figure \ref{fig:bands} but now each pair of columns pertains to an individual galaxy's $u-$band Stripe 82 and Legacy decompositions. The first pair of columns shows a galaxy that is generally representative of the positive $\Delta m_{u,\mathrm{DR7-S82}}$ scatter at faint $u-$band magnitudes in Figure \ref{fig:mags} owing to substantially improved visibility in the Stripe 82 co-adds. Despite this improvement, the change in total $u-$band apparent magnitude for this galaxy is relatively small $\Delta m_{u,\mathrm{S82-DR7}}\approx0.1$ mag. The second, third, and fourth pairs of columns show galaxies with $m_{u,\mathrm{DR7}}>20.5$ mag and larger positive offsets in $\Delta m_{u,\mathrm{DR7-S82}}$. In the second and fourth pairs of columns, the galaxies are virtually invisible in the Legacy images. In the third case, the galaxy is visible but is still noise-dominated -- resulting in similar under-estimated total apparent magnitude. The fifth pair of columns shows an example characteristic of the downward systematic in the $u-$band panel of Figure \ref{fig:mags}. It shows a galaxy for which the Stripe 82 total apparent magnitude is significantly fainter than in Legacy and practically invisible in the shallower image.}
\label{fig:uband}
\end{figure*}

Our galaxy sample selection and consistency in methodology enable direct comparison of decomposition results from the co-adds and shallower Legacy images. Changes to \emph{integrated} galaxy properties (accounting total flux contributions from all model components -- whether single-component or multi-component) such as total galaxy magnitudes and galaxy half-light radii are not expected to change drastically among galaxies with surface brightnesses that are well within the Legacy detection limits of each bandpass. For example, Figure \ref{fig:bands} shows a mosaic of \texttt{n4} decomposition results for a single randomly selected galaxy in the range $15<m_r<16$ covering each band for both the Stripe 82 co-adds and corresponding Legacy images. Despite a visibly significant improvement in image quality in the co-adds, total magnitude and half-light radius measurements in each band are largely unchanged -- even in the $u-$ and $z-$ bands where the starkest contrast in image quality is seen.\footnote{Although, note that if the $gr$ structural parameters are not changed, then the reason that $uiz$ half-light radii do not change \emph{can} be partly by construction -- the bulge and disc structures in these bands are tied to the $gr$ bulge and disc structures. However, the magnitude of each component is fully independent in each band. Therefore, one can still have band-specific changes in half-light radius even where the structures of bulge and disc components are fixed.} In the next section, we will show that this consistency holds for most galaxies in our comparison. However, galaxies at the thresholds of the Legacy detection limits stand to see improved constraints on their integrated properties. As for the components, it is of particular interest whether the added depth can better discriminate the light emerging from the bulge and disc. In this section, we compare the \texttt{n4} decompositions for galaxies in Stripe 82 co-adds and Legacy images.

\subsection{Total magnitudes}

The total flux from all model components should always be conserved in any decomposition that uses the same degrees of freedom and model parametrization. In other words, similar models fitting the same set of pixels should recover the same total flux -- even when other best-fitting model parameters may differ (e.g., degeneracy). The number of degrees of freedom in a 2D surface brightness decomposition is equal to the number of pixels used in the fit, $N_{\mathrm{pixfit}}$, subtracted by the number of free parameters in the model. This quantity is not conserved in our comparisons, as the increased $S/N$ in the co-adds extends the source masks when deblending objects from sky -- increasing $N_{\mathrm{pixfit}}$ often substantially (for example, see the second row of panels in Figure \ref{fig:bands}). However, since the total magnitudes are computed by integrating the model fluxes to infinity, increases to the sizes of the object footprints should not change their measured fluxes unless the signal at extended radii necessitate changes to the structural parameters. It is one of the goals of the current work to examine whether the added depth imparts such changes.

Figure \ref{fig:mags} shows that the total galaxy light is largely conserved. The median offsets in total apparent magnitude for all galaxies in each band are $\Delta m_{u,g,r,i,z}^{50\%} = (+0.037,+0.022,+0.021,+0.020,+0.019)$ mag (DR7-S82). The most notable exception is the $u-$band which has the poorest response and in which many galaxies are also intrinsically fainter. Mild median systematic trends in $\Delta m_{x,\mathrm{DR7-S82}}$ are clear in each bandpass. The brightest galaxies get boosts to their total fluxes in the co-add measurements with respect to Legacy. At the faint end, a systematic trend is notable in the $u-$band -- where the median systematic magnitude offsets first increase slightly (brighter in the co-adds) at $m_u\lesssim19.5$ mag then drop rapidly at $m_u\lesssim21.2$ mag (becoming fainter in the co-adds). To understand this systematic, Figure \ref{fig:mags} also shows percentiles in the DR7 (solid) and Stripe 82 (dotted) mean local sky uncertainties, $\sigma_{\langle\mathrm{sky}\rangle}$, expressed as magnitudes (rather than relative magnitude \emph{offsets} as used in, for example, Appendix \ref{sec:appB}). These are computed:
\begin{align}
\nonumber \sigma_{\langle \mathrm{sky} \rangle} \mathrm{(mag)} = -2.5\log_{10}\Bigl( & \sigma_{\mathrm{sky}}(\mathrm{maggies/arcsec}^2) \\
&\times \sqrt{N}_{\mathrm{pixfit}} \times \mathrm{\texttt{pixel\_scale}}^2\Bigr)
\end{align}
where $\sigma_{\mathrm{sky}}$ is the local \emph{sample} sky surface brightness uncertainty we measure (see Appendix \ref{sec:appB}, Eq. \ref{eq:sky}), $N_{\mathrm{pixfit}}$ is the number of pixels used in the fit, and \texttt{pixel\_scale} is the CCD plate scale of 0.396 arcsec/pixel. While these uncertainties are too small to be seen in the panels for other bandpasses, the sudden systematic drop in median Stripe 82 $u$-band brightness at $m_u \approx21$ mag coincides with the median in $\sigma_{\langle\mathrm{sky}\rangle,\mathrm{DR7}}$ in that band. So, while the Legacy flux measurements for the faintest targets in the $u-$band can be limited in their capacity to penetrate the sky noise, Stripe 82 measurements go deeper and are able to properly characterize these faint sources. In other words, given that the Legacy $u-$band mean sky uncertainty measurements lay closer to the distribution of galaxy $u$-band brightnesses, it is the $u$-band magnitudes that stand to be most improved by the added depth of the Stripe 82 co-adds.\footnote{One should recall, however, that \emph{structural} parameters in $ur$, $ir$, and $zr$ in the \texttt{n4} fits are fixed to the $gr$ results in our experimental design. So these parameters are not affected by any improvements in the photometry of $uiz$ bandpasses.} Visual inspection of the science images and decomposition results for faint $u-$band galaxies with $m_{u,\mathrm{DR7}}>20.5$ mag in the co-adds and Legacy revealed that the vast majority of these galaxies are indeed barely (if at all) discernible from the sky background in the Legacy $u-$band images. Figure \ref{fig:uband} highlights a few such cases in a mosaic of decompositions for five individual galaxies in the $u-$band. Cases in which a galaxy is barely detectable can generate both negative \emph{and} positive scatter in $\Delta m_{x,\mathrm{DR7-S82}}$, but have a stronger tendency towards negative offsets as the model likely attempts to fit features of the noise.

\begin{figure}
  \includegraphics[width=0.49\linewidth]{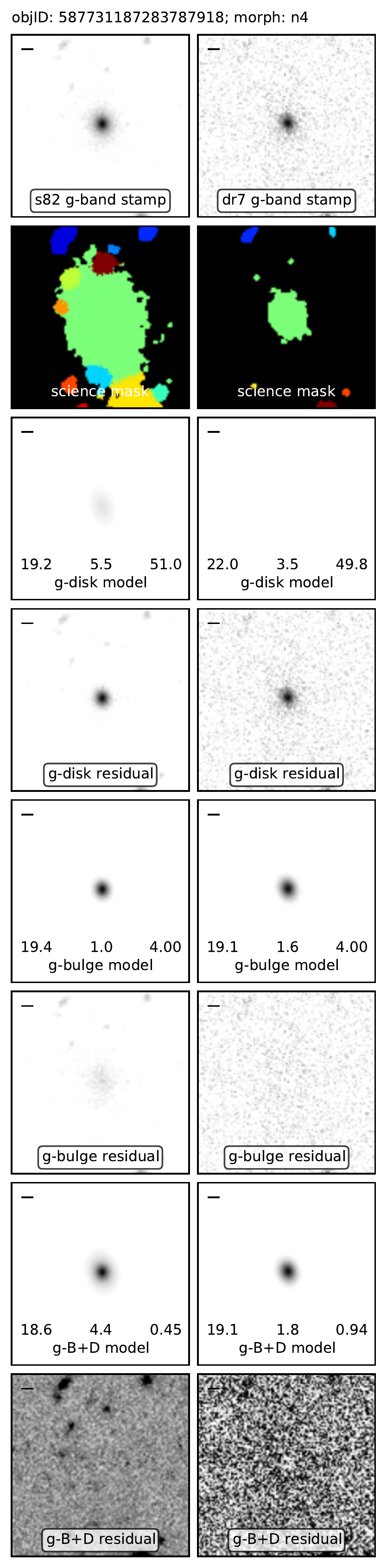}
  \includegraphics[width=0.49\linewidth]{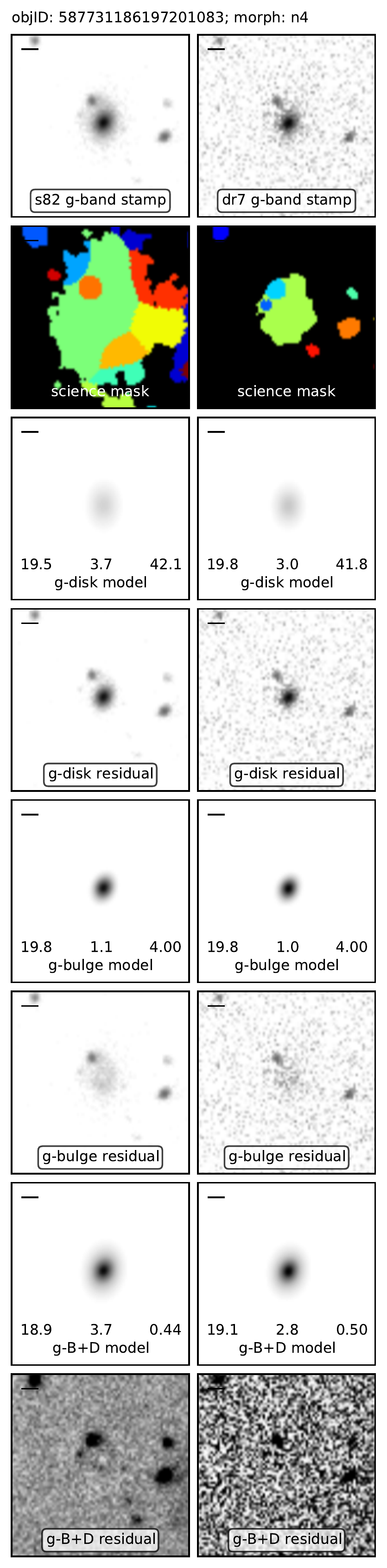}
\caption[Systematics on Apparent Magnitudes]{Decomposition mosaics for two galaxies whose faint $g-$band total magnitudes were brightened by the co-adds. Improved characterization of the faint structure surrounding the central galaxy is afforded by the co-adds in each case. The galaxy in the left two columns switches from bulge-dominated in Legacy to a composite bulge+disc system in the co-adds. An extended disc is assigned to characterize the faint structure in the co-add images. This new extended structure accounts for $\Delta m_{g,\mathrm{S82-DR7}}=-0.5$ mag difference between the best-fitting bulge+disc models. The panels on the right also shows a galaxy for which the disc becomes brighter and larger in order to characterize the fainter structure. The bulge model remains largely consistent. Additionally, a possible tidal stream is revealed between the central galaxy and the interloper to its upper left. It is most visible in the disc-subtracted co-add image.
}
\label{fig:mag_sys}
\end{figure}

Figure \ref{fig:mag_sys} shows two examples that are representative of many faint galaxies whose fluxes were increased in the co-add measurements. Despite the differences to the model parameters, the residuals are identical in quality. This result suggests that we are not incorrectly over-estimating total brightnesses in these faint systems in Stripe 82. Rather, the enhanced fluxes in the models reflect true changes to the faint end of the galaxy surface-brightness distributions with respect to what is accessible in Legacy. It is possible that changes to the model parameters are partially due to degeneracies between structural parameters at these limits of resolution and $S/N$. But if degeneracy was the dominant factor, the total magnitudes would be unchanged as long as the residuals are consistent. A statistical approach to characterize the role of degeneracy in the decompositions follows in Section \ref{sec:ftest}. For now, we assert that the total fluxes in the co-adds exhibit reduced systematics on model parameters due to surface-brightness limits relative to the Legacy images for our galaxy sample. We further investigate this assertion by directly examining the total photometric uncertainties for measurements of total and component magnitudes and bulge-to-total fractions in Appendix \ref{sec:appA}. 

The Stripe 82 magnitudes of the brightest galaxies in each band can be up to 0.1 mag ($\sim10\%$) brighter than the DR7 magnitudes. This systematic difference in total flux is likely due to changes to the measured light profiles at extended radii -- afforded by the increased $S/N$ and consequent galaxy footprint size in the mask image. Given that our galaxy magnitudes are based on integrating the best-fit models out to infinity, an improved characterization of low-surface brightness wings in the light profiles of bright galaxies may easily add up to a 10\% total flux difference for bright galaxies. Another likely culprit is that the local sky level around each galaxy is more accurately estimated in the Stripe 82 images than in Legacy. The increased footprint sizes of sources in the mask images (by an average of 246\% relative to Legacy) mean that the local sky statistics are computed from an ensemble of pixels that are farther from \emph{all} sources in an image. Given the environments that the brightest and largest galaxies tend to live in and their tendency to have highly extended stellar halos (e.g., \citealt{2011ApJ...731...89T}), it is important to find balance between a obtaining a truly local estimate of the sky and ensuring that the sky estimate is not overly contaminated by the extended flux from bright targets or their neighbours.\footnote{To justify this claim, we have separately confirmed that $\Delta m_{x,\mathrm{DR7-S82}}$ and the Legacy residual sky offset between the local \textsc{gim2d} sky and the full-frame \textsc{SExtractor} sky, $db$, are positively correlated for galaxies with $m_r<14.5$ mag. However, the systematic suppression of total flux is limited primarily to targets in this magnitude range and that \emph{also} have high bulge-to-total fractions (high S\'ersic indices) as predicted by \cite{2014ApJS..210....3M} using artificial galaxy simulations.} From Figure \ref{fig:mags}, we highlight while this contamination is undoubtedly the source of suppressed fluxes for the brightest galaxies in \citetalias{2011ApJS..196...11S}'s Legacy decompositions relative to Stripe 82, its effect is limited to $\sim0.02$ mag with the exception of the brightest bins in each band.

\begin{figure*}
  \includegraphics[width=\linewidth]{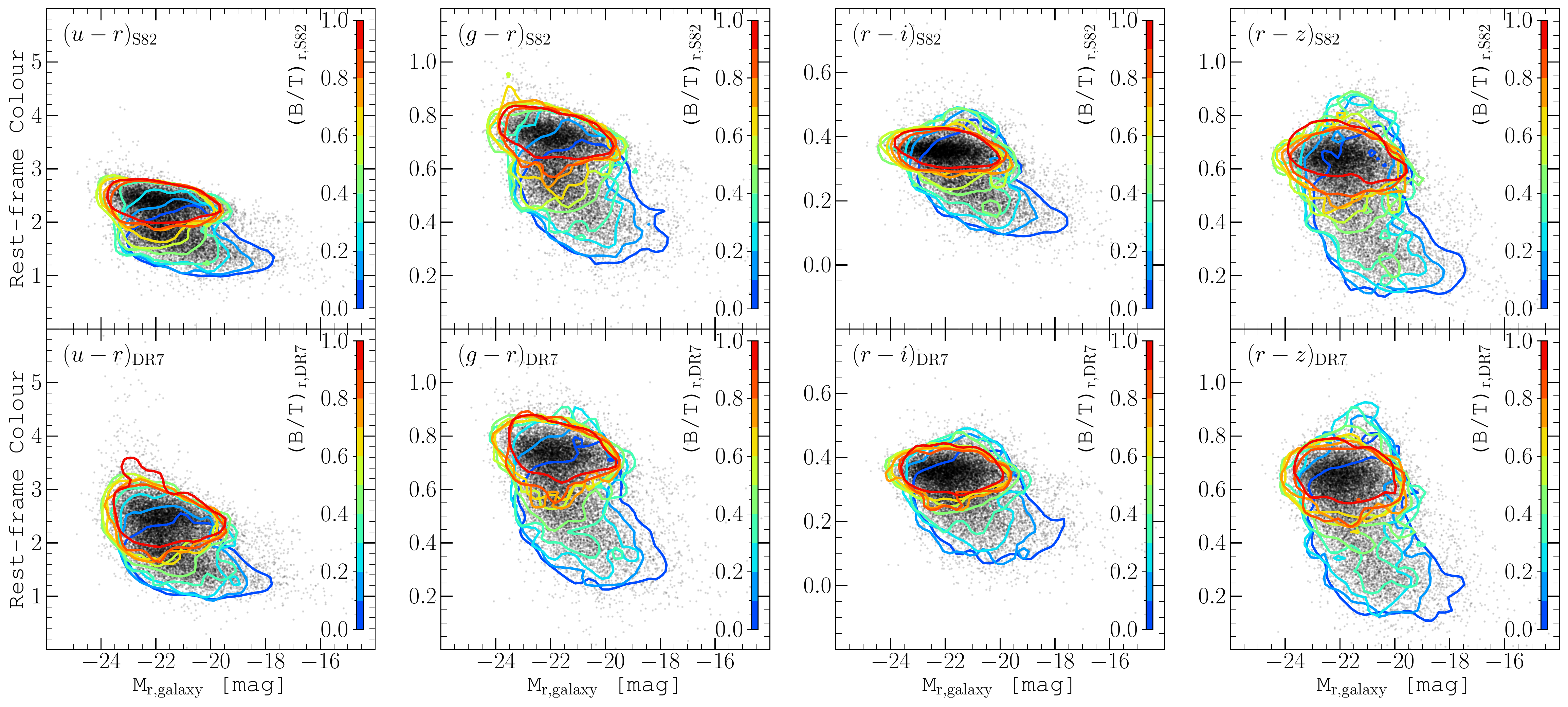}
\caption[Colour-Magnitude Diagrams]{Rest-frame Colour-Magnitude Diagrams from \texttt{n4} decompositions of galaxies in the co-adds (upper row of panels) and DR7 Legacy images (lower row of panels). Colours from each simultaneous $xr$ decomposition are only correlated insofar as forcing the structural parameters to be the same as those derived from the $gr$ fits. Brightnesses of each component in each bandpass are free to vary independently. Black markers show the full distribution of galaxy colours each panel. Coloured lines show the 86\% density contours of these distributions in 10 discrete bins of $(B/T)_r$ following the colourmaps displayed in each panel. For example, the dark red line shows the 86\% density contour for galaxies in the last $B/T$ bin the range $0.9\leq(B/T)_r\leq1$.
 }
\label{fig:colours}
\end{figure*}

\subsection{Galaxy colours} 

The scatter in $\Delta m_{x,\mathrm{DR7-S82}}$ is related to the photometric depth in each bandpass and to the intrinsic brightnesses of galaxies in each bandpass. The scatter is greatest in the $u-$band, where the Legacy photometry is the poorest. Brighter than $m_{u,\mathrm{S82}}\approx20$ mag, this scatter is symmetric and the median systematic offset is $\Delta m_{u,\mathrm{DR7-S82}}\lesssim0.02$ mag. Where it is symmetric, the scatter in $\Delta m_{x,\mathrm{S82-DR7}}$ should be the sum in quadrature of the total random and systematic measurement uncertainties from the respective co-add and Legacy decompositions with respect to the galaxies' intrinsic magnitudes -- which are not known. 

To illustrate which of the Stripe 82 and Legacy decompositions contribute most to the scatter shown in Figure \ref{fig:mags}, we compare their respective rest-frame colour-magnitude diagrams (CMDs) in Figure \ref{fig:colours}. The tightness of the red-sequence is a common metric for the constraints on global galaxy colours and the precision of total flux measurements (e.g., \citealt{2011ApJS..196...11S}). Figure \ref{fig:colours} shows the distributions of Stripe 82 and Legacy galaxies for four colours. Coloured lines show contours of these distributions in bins of $r-$band $B/T$. Most bulge-dominated galaxies live on the red sequence. But a large number of red discs and red two-component systems are also found there. The tightness in the red sequence is most improved in the $u-r$ colours in the co-adds compared to the Legacy colours. Note the large number of outliers with high $u-r$ in Legacy that are no longer outliers using the co-add measurements. In the $g-r$, $r-i$, and $r-z$ colours, the visible tightness of the red sequence is mildly improved in the co-add decompositions. In general, our results are consistent with a scenario in which the scatter in the magnitude offsets, $\Delta m_{x,\mathrm{DR7-S82}}$, is dominated by larger measurement uncertainties in the Legacy decompositions. Both sets of decompositions, however, identify the well-established trend that galaxies are typically redder as they become more bulge-dominated \citep{2001AJ....122.1861S,2003MNRAS.341...33K,2003ApJ...594..186B,2004ApJ...600..681B,2004ApJ...615L.101B,2006MNRAS.368..414D,2007ApJ...665..265F} -- consistent with quenching scenarios which revolve around growth of a compact stellar component \citep{2003MNRAS.346.1055K,2006MNRAS.367.1394K,2007ApJS..173..315S,2008ApJ...682..355B,2012ApJ...760..131C,2013ApJ...776...63F,2014ApJ...788...11L,2014MNRAS.440..843O,2014MNRAS.441..599B,2015MNRAS.448..237W,2016MNRAS.462.2559B,2016MNRAS.457.2086T}. 

The tightness of the red sequence is also a general indicator of the effectiveness of deblending algorithms in masking the light emanating from nearby sources. If the deblending is poor and the colours of nearby objects differ from those of the target galaxies, then the scatter in the red sequence will be inflated. The visibly improved tightness of the red sequence in the co-add colours indicates that the increased $S/N$ in the co-adds may improve deblending of interloping light for crowded targets. This result may be particularly useful for analysis of galaxy pair colours and the masses that can be inferred from these colours.

\subsection{Galaxy sizes}
\begin{figure}
  \includegraphics[width=1.0\linewidth]{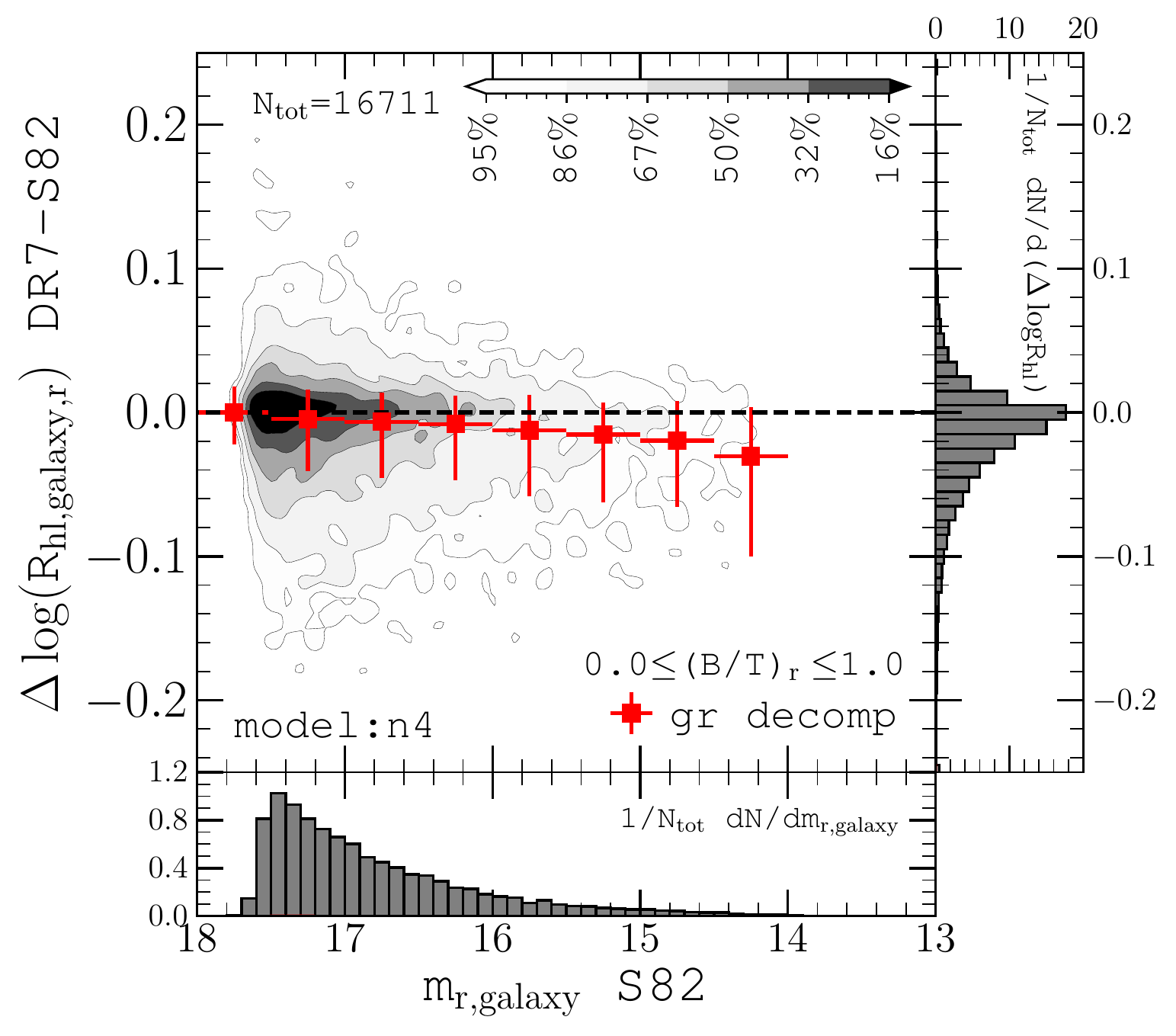}
\caption[Galaxy half-light radii]{Galaxy half-light radii measured the $r-$band from the \texttt{n4} decompositions. Median systematic offsets for all galaxies taken together are typically small, but can be larger for particular $B/T$ ranges. The largest median systematics were found to be -0.05 dex at the bright end of $0.4<$B/T$<0.6$. 
}
\label{fig:sizes}
\end{figure}

As with the total magnitudes, total galaxy sizes (e.g., Petrosian and galaxy half-light radii) should be mostly conserved barring galaxies at the detection limits. However, even for galaxies with good photometry in Legacy, half-light radii in the co-adds could benefit from improved constraints on the extended light from the bulge or disc. In the previous section, we asserted that the systematic increase in galaxy fluxes measured in the co-adds was due to the increased $S/N$ at faint surface-brightnesses. We should then expect to see a corresponding systematic increase in galaxy half-light radii. Figure \ref{fig:sizes} shows this expected increase in galaxy sizes using $r-$band galaxy half-light radii, $r_{\mathrm{hl,galaxy,r}}$, from the $gr$ \texttt{n4} decompositions. Half-light radii are computed as the semi-major axis radius of the ellipse in which half of the total model flux is contained (circular aperture half-light radii are also available in our catalogs). Similar to Figure \ref{fig:mags}, we plot the offset of the co-add sizes from the Legacy sizes. The median systematic peaks at the bright end with $\Delta r_{\mathrm{hl,galaxy,r}}\approx-0.03$ dex or $\sim 7\%$ and shallows at fainter magnitudes. Figure \ref{fig:sizes} is focused on the $r$-band, but the same trend was found in each other band (though, as for the magnitude differences, with larger scatter). On some level the consistency of this trend across each band is by construction in the \texttt{n4} fits -- as the structural parameters of the bulge and disc components are fixed to the $gr$ results. However, because the magnitudes of the components (and consequently $B/T$) are free to vary independently in each band, half-light radii (which measure half-light radius of the full model) can still change in each band due to varying bulge and disc fractions.

The scatter in $\Delta r_{\mathrm{hl,galaxy,r}}$ will be partially driven by differences in the segmentation maps. As discussed in previous sections, the sizes of the \textsc{SExtractor} footprints for galaxies in the co-adds are larger on average than in the Legacy images. There is an important consequence of having larger maps and higher $S/N$ in the deep stacks on galaxy size. The increased $S/N$ enables detection and deblending of sources previously unidentified in the Legacy segmentation maps. By deblending faint sources from the galaxy flux, their typically positive systematic on galaxy sizes can be reduced.

\subsection{Bulge-to-total light fractions}\label{sec:btr}

\begin{figure}
  \includegraphics[width=0.9\linewidth]{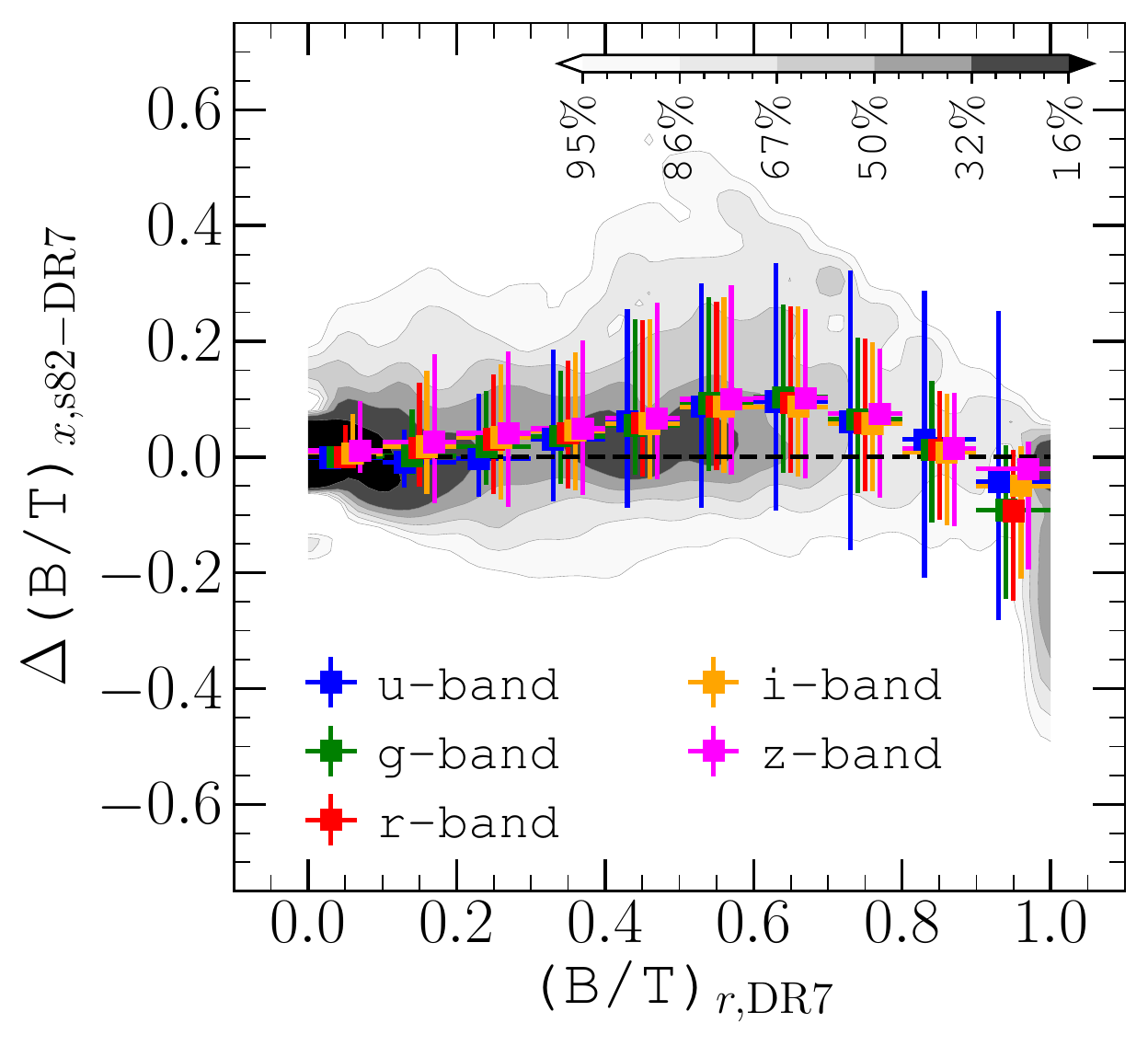}
  \includegraphics[width=\linewidth]{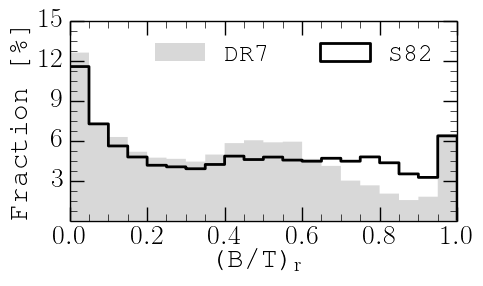}
\caption[Galaxy bulge-to-total light ratios]{Galaxy bulge-to-total light ratios. The upper panel shows the $B/T$ offsets, $\Delta(B/T)$, as a function of the $B/T$ measured using Legacy photometry. The background contour map shows the two-dimensional distribution in the $r-$band. Coloured markers show the median and 16-84 percentile ranges in $\Delta(B/T)$ measured in each bandpass. The lower panel shows the one-dimensional $B/T$ distributions from the Legacy (gray, shaded) and Stripe 82 (black, empty) decompositions in the $r-$band.
}
\label{fig:btr}
\end{figure}

So far we have focused on characterizing the integrated properties of galaxies. We now shift our attention to the the bulge and disc components starting with bulge-to-total fraction, $B/T$ -- a useful quantitative indicator of galaxy morphology. Figure \ref{fig:btr} shows the changes in $B/T$ in the co-add and Legacy images. The lower panel shows the $B/T$ distribution for each set of decompositions taken from the respective \texttt{n4} $gr$ tables. Interestingly, the peak at $B/T=0.5$ in the Legacy images is largely removed in the co-add results. The upper panel shows the difference between the co-add and Legacy bulge-to-total fractions in each band, $\Delta(B/T)$, plotted against the Legacy $B/T$ estimates in the $r$-band. Here we see that the $B/T$ values which previously resided at the peak are enhanced in the deeper images and that they fill up the deficit between $0.6\lesssim B/T \lesssim0.95$. These results support the idea that the peak at $B/T=0.5$ in the Legacy decompositions is an artifact that arises wherever there are very poor constraints on a galaxy's structure in the Legacy photometry. Such cases will naturally tend to $B/T\sim0.5$ because it is the median value of a flat $B/T$ posterior probability distribution (i.e. unconstrained $B/T$).

Note also the large number of galaxies with $(B/T)_{\mathrm{DR7}}=1$ in the upper panel of Figure \ref{fig:btr} which now have negative offsets as large as $\Delta(B/T)=-0.4$. While it should be noted that values of $\Delta(B/T)>0$ are by definition impossible for $(B/T)_{\mathrm{DR7}}=1$, the same is true for $\Delta(B/T)<0$ where $(B/T)_{\mathrm{DR7}}=0$ and no such strong clustering of positive $\Delta(B/T)$ is seen there. Are these discs being revealed by deeper photometry of predominantly bulge-dominated galaxies? To answer this question, we examine the underlying hypothesis that the deeper images enable better discrimination between single-component and two-component systems. 

\subsection{$F$-test statistics}\label{sec:ftest}

Until now we have not considered the decomposition results in the other tables. The integrated quantities (total magnitudes and galaxy half-light radii) for the \texttt{ps} and \texttt{fn} model decompositions broadly mirror the results from the \texttt{n4} decompositions highlighted in the above subsections and compare similarly with the matching values in Legacy. The quality of the decomposition of a galaxy is characterized quantitatively by the reduced $\chi^2_{\nu}$ statistic of the model with respect to the data and can vary significantly between models. However, a direct comparison between the $\chi^2_{\nu}$ for each model in our hierarchy will always yield better $\chi^2_{\nu}$ for the model with the greatest number of degrees of freedom (up to Monte Carlo error). 

In order to discern whether the data demand a more complex model over a simpler model, we compare the results from different models using the $F$-statistic \citep{2011ApJS..196...11S,2013MNRAS.433.1344M,2014ApJS..210....3M}. For each decomposition model, we compute $\chi^2_{\nu}$, where $\nu$ is taken as the number of resolution elements, $N_\mathrm{pix}/(\pi\; \mathrm{HWHM^2_{psf}})$, subtracted by the number of free parameters in the model.\footnote{The number of degrees of freedom takes a different definition here than it does in the decompositions themselves, where $\nu$ is the number of target pixels (those flagged as the target galaxy) and sky pixels in the science image cut-out minus the number of free parameters in the model.} The HWHM is the half-width at half-maximum of the PSF. The $F$ value is then the ratio of the $\chi^2_{\nu}$ for each pair of models. These $F$ values are converted to corresponding probabilities via the $F$-distribution. In our implementation, the $F$ values convert to probabilities that the more complex model is \emph{not} required to properly model the galaxy structure. In this way, a more complex model is only favoured under the test if the additional degrees of freedom offer \emph{substantial} improvement to $\chi^2_{\nu}$ over a simpler model. The \texttt{n4} table therefore includes the $P_{pS}$ probability that an \texttt{n4} decomposition is \emph{not} required to properly characterize the surface brightness distribution of a galaxy relative to the \texttt{ps} decomposition. Similarly, the \texttt{fn} table includes both $P_{pS}$ and $P_{n4}$ probabilities that the \texttt{fn} decomposition is \emph{not} required relative to the \texttt{ps} and \texttt{n4} models, respectively.

\begin{figure*}
  \includegraphics[width=\linewidth]{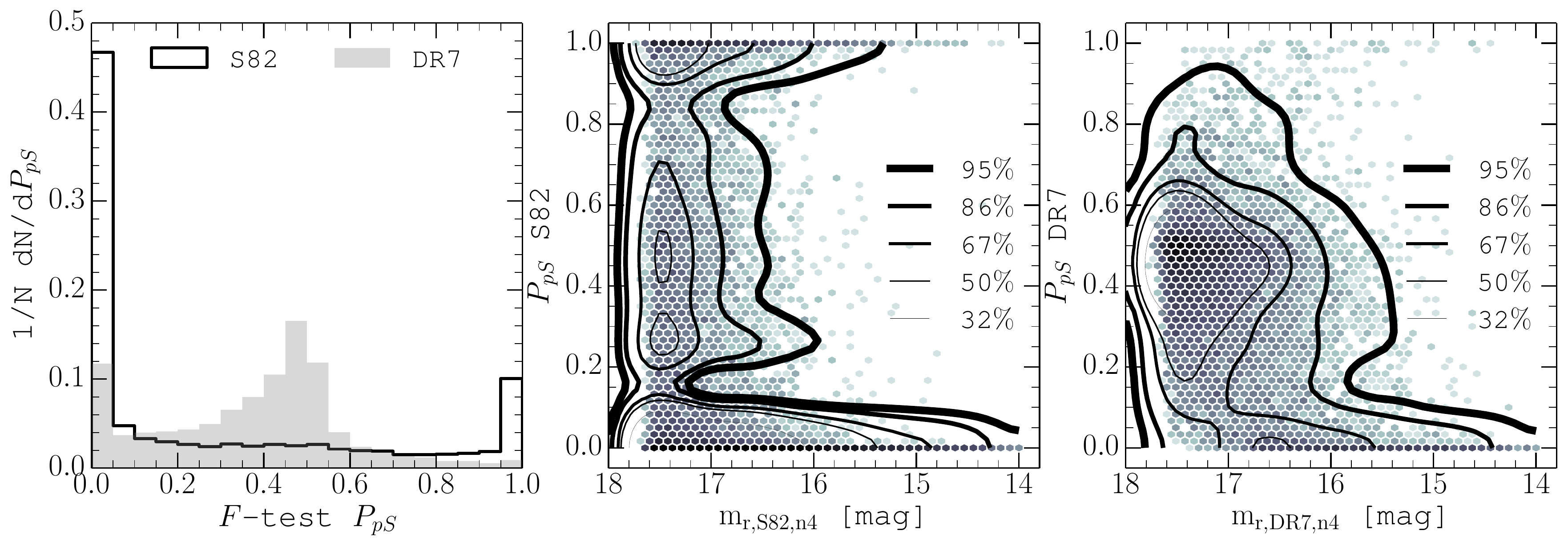}
  \includegraphics[width=\linewidth]{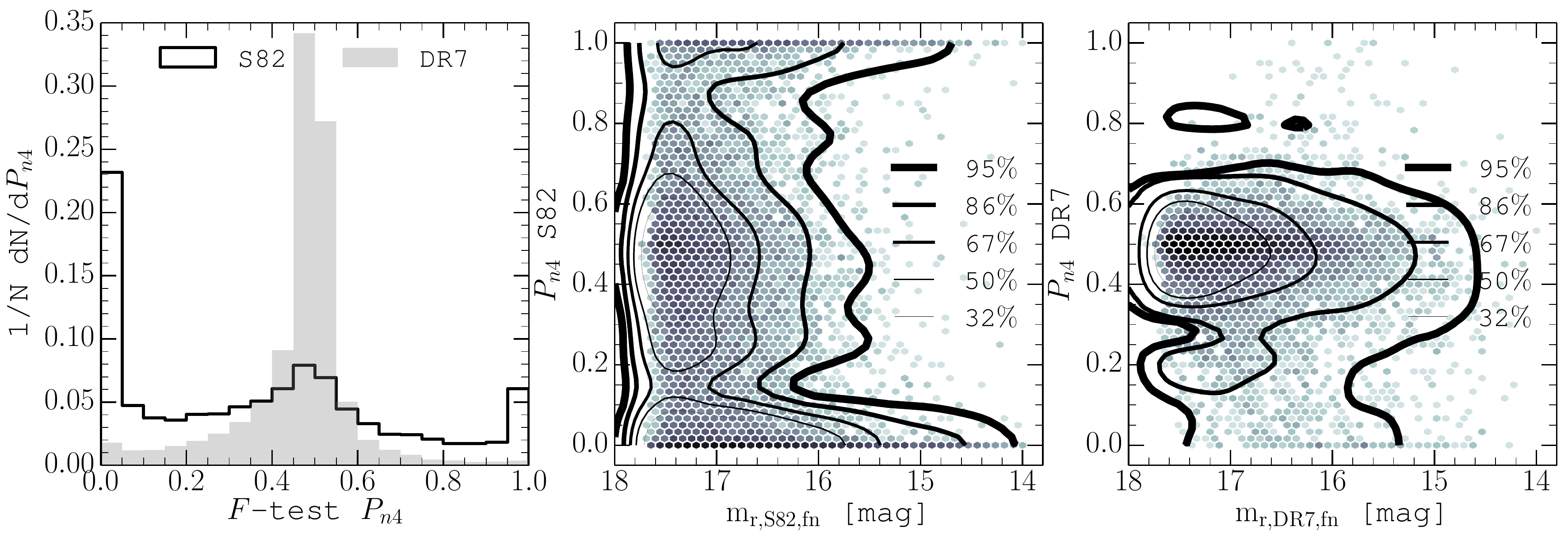}
\caption[F-test results]{F-test statistics comparing between models of increasing number of degrees of freedom. \emph{Upper left}: one-dimensional histograms of $P_{pS}$ for Stripe 82 (solid, empty) and DR7 Legacy (gray, shaded). \emph{Upper center and right}: histograms of $P_{pS}$ in Stripe 82 (left) and DR7 Legacy (right) and their respective galaxy total magnitudes measured in the \texttt{n4} decompositions. Percentile legends for the black contour lines are inset on the middle right of each panel. \emph{Lower row}: same as upper row of panels but comparing the $P_{n4}$ statistics derived Stripe 82 and DR7 Legacy photometric decompositions. 
}
\label{fig:ftest}
\end{figure*}

Figure \ref{fig:ftest} shows the $F$-test results for our Stripe 82 galaxy sample in a comparison with Legacy values from \citetalias{2011ApJS..196...11S}. Upper panels show the $P_{pS}$ probabilities computed from the $\chi^2_{\nu}$ of the best-fitting \texttt{n4} and \texttt{ps} from the $gr$ decompositions. An ideal indicator of whether a galaxy is a single- or two-component system would yield a binary classification. In practice, this idealized binary classification scenario reduces to assignment of probabilities due to the presence of additional structures in galaxies (bars, stellar nuclei, etc.) and the limitations of the photometry to reveal faint underlying bulge or disc components. The upper left panel of Figure \ref{fig:ftest} shows that $P_{pS}$ from Legacy are centrally peaked and largely concentrated around $P_{pS}=0.5$. This result demonstrates that Legacy photometry offers little discriminating power between the bulge+disc and single-component models. The ambiguity in Legacy is largely eliminated in the co-add decomposition results -- which offer substantially improved classification of galaxies as a single- or two-component systems. Panels to the right show the sensitivity of $P_{pS}$ discriminating power to total apparent magnitude in the co-adds (middle) and Legacy images (right). Both maps become denser around $P_{pS}=0.5$ for fainter galaxies. But the improved $S/N$ in the co-adds greatly extends the magnitude range over which the $F$-test probabilities yield a binary-like classification.

Lower panels of Figure \ref{fig:ftest} show the $P_{n4}$ values computed from the $\chi^2_{\nu}$ of the best-fitting bulge+disc models using free and fixed bulge S\'ersic indices. Analogous to $P_{pS}$, the Legacy images offer little discriminating power between bulge+disc models of free and fixed $n_b$. Legacy decomposition $P_{n4}$ values are concentrated around $P_{n4}=0.5$ with almost no power at the $P_{n4}\approx0$ or $P_{n4}\approx1$. While many of these ambiguous $P_{n4}$ are preserved in the co-add decompositions, a large fraction are not. As with the $P_{pS}$, the ability to discriminate between bulge models is sensitive to a galaxy's intrinsic brightness. The concentration in the co-add $P_{n4}-m_{\mathrm{r,s82,fn}}$ map is still broad, however. Nearly 50\% of galaxies still have $0.25<P_{n4}<0.75$ in the co-adds. This number is less than $25\%$ for $P_{pS}$. 

\subsection{Galaxy and bulge S\'ersic indices}

We have examined the differences in the distributions of galaxy (\texttt{ps} fits) and bulge (\texttt{fn} fits) S\'ersic indices in the Legacy and the co-adds. First, we note the criterion of our methods that although S\'ersic index is allowed to vary in each bandpass pair of the \texttt{ps} fits, the effective radius and axis ratio are held fixed to the $gr$ results. Furthermore, since the $r-$band tends to have the best $S/N$, the optimization of S\'ersic index will rely most strongly on the $r-$band photometry in cases when a target is faint in neighbouring bands. \citetalias{2011ApJS..196...11S} showed that the $S/N$ and spatial resolution in Legacy is often limited in its capacity to characterize galaxy and bulge S\'ersic indices -- particularly for faint targets (hence a fiducial fixed $n_b=4$ bulge in their two-component model). \citetalias{2011ApJS..196...11S} also argued and that the \emph{particularly} large bump in the $n_b$ distribution that they saw at $n_b=4-5$ (compared to $n_g$ which showed a more minor bump) is not some general property of physical bulges but is more likely a statistical artifact of computing the median of bulge S\'ersic index, marginalized over all other parameters, in a likelihood space that does not respond sensitively to changes in $n_b$ (i.e. a uniform posterior probability distribution with hard limits $0.5\leq n_b\leq 8$). Here, we determine whether the added depth from the co-adds offers an improved characterization of galaxy and bulge S\'ersic indices given our foreknowledge of these existing systematics.

\begin{figure}
  \includegraphics[width=\linewidth]{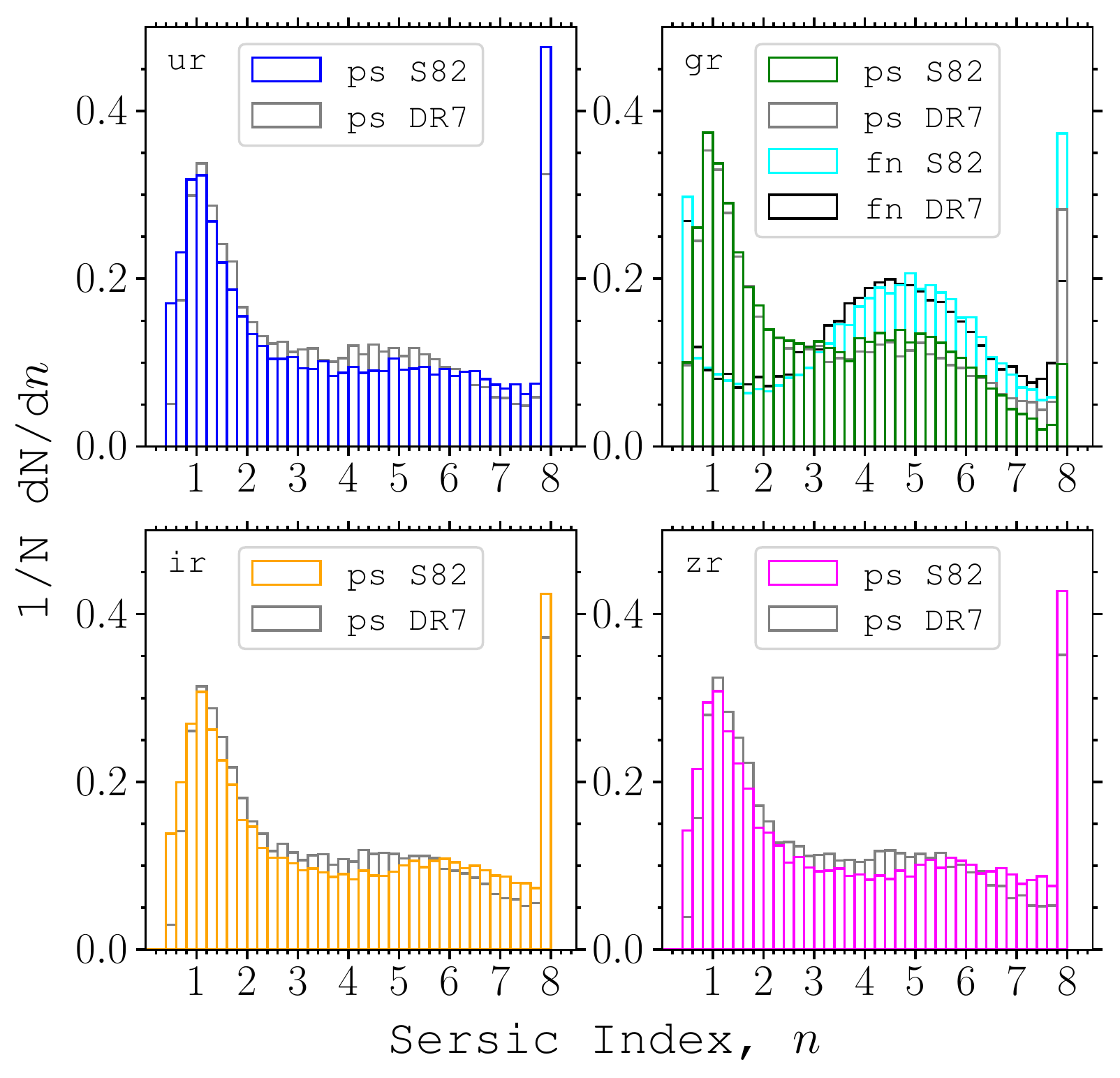}
  \includegraphics[width=\linewidth]{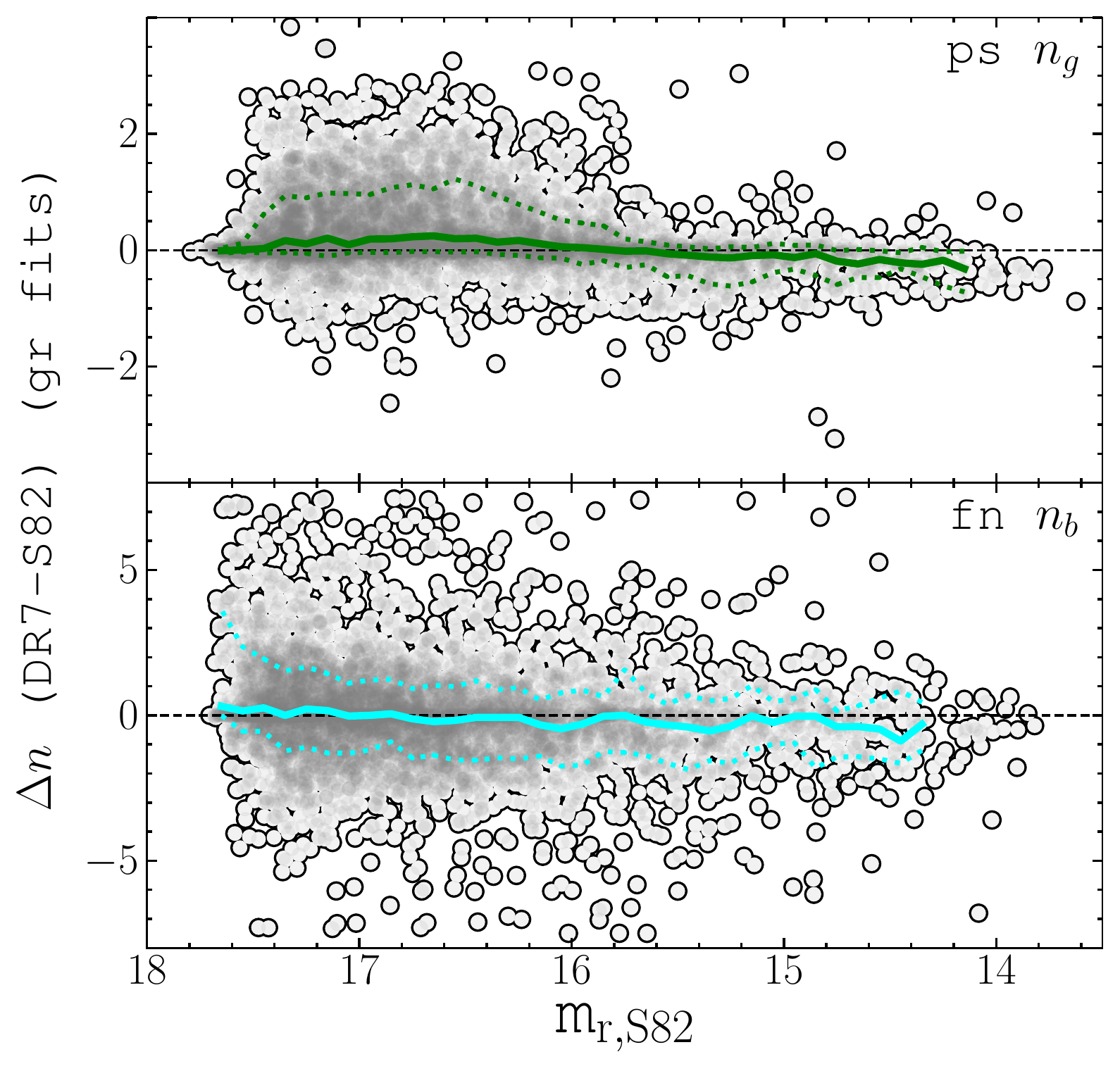}
\caption[Sersic Indices]{Comparison of S\'ersic indices, $n$, in Stripe 82 and DR7 Legacy. The top four panels compare distribution functions of \texttt{ps} galaxy S\'ersic indices, $n_g$, in the $ur$, $gr$, $ir$, and $zr$ fits. Stripe 82 histograms are coloured according to each band pairing and the corresponding DR7 Legacy histograms are behind in grey. The $gr$ panel to the upper right also compares the distributions of \texttt{fn} bulge S\'ersic indices, $n_b$, from the corresponding $gr$ fits (Stripe 82 -- cyan; DR7 Legacy -- black). The lower panel more directly compares the Stripe 82 and Legacy $gr$ galaxy S\'ersic indices from the \texttt{ps} fits (upper, green) and bulge S\'ersic indices from the \texttt{fn} fits (lower, cyan) as functions of \emph{total} $r-$band magnitude from the corresponding model. Solid lines show the median $\Delta n$ as a function of apparent magnitude. Dashed lines show the 16th and 84th percentiles. Density maps in the background show the full sample distributions.}
\label{fig:sersic}
\end{figure}

The upper four panels of Figure \ref{fig:sersic} compare the distributions of galaxy S\'ersic indices, $n_g$, from the \texttt{ps} model fits to each pair of bandpasses ($ur$, $gr$, $ir$, $zr$). The \texttt{ps} S\'ersic index distributions for the $ur$, $ir$, and $zr$ fits each show some modest differences that are qualitatively similar to the trend in $B/T$ reported in Section \ref{sec:btr}. Fewer galaxies reside in the bump at $n_g=4-5$ and the distribution is slightly more uniform at $n_g\gtrsim3$ -- with some galaxies moving from the bump to higher and lower $n_g$ in the co-add fits. Interestingly, this qualitative trend is not shared in the $gr$ \texttt{ps} fits (upper right panel, green and grey) -- despite the fact that the $r-$band is used in every other fit \emph{and} the other structural are fixed to the $gr$ result. Indeed, the $gr$ \texttt{ps} distributions for Stripe 82 appear very similar apart from the right-most bins at $n_g\gtrsim7$ where there are $\emph{fewer}$ targets in the Stripe 82 relative to Legacy -- contrary to what is seen in every other panel. 

The upper half of the lower panel in Figure \ref{fig:sersic} explores the apparent discrepancy in $gr$ more closely by plotting the difference in Legacy and Stripe 82 $n_g$ as a function of Stripe 82 \texttt{ps} apparent magnitude. At bright magnitudes, the median Stripe 82 \texttt{ps} $n_g$ are larger. In particular, this result, combined with the increase in total brightness and sizes seen in Figures \ref{fig:mags} and \ref{fig:sizes} in the deeper images, supports findings by other comparative works that have suggested that the \textsc{gim2d} sky estimation method used in \citetalias{2011ApJS..196...11S} can be sub-optimal for bright targets with highly extended surface brightness profiles \citep{2014MNRAS.443..874B,2014ApJS..210....3M,2015MNRAS.446.3943M}. This systematic is less important in Stripe 82 (even though the same background estimation method and segmentation parameters are used) because the increase in $S/N$ expands the sizes of the source masks in the segmentation images -- forcing the background estimates to be made from pixels significantly farther from the target and every other source in Stripe 82 images. As shown by \cite{2014ApJS..210....3M} (in their Appendix B) using artificial galaxy simulations, the systematic is most likely to arise in the \citetalias{2011ApJS..196...11S} fits to galaxies whose profiles are dominated by a component with intrinsically high S\'ersic indices, $n\gtrsim5$. Combining information from Figures \ref{fig:mags}, \ref{fig:sizes}, and \ref{fig:sersic} we caution that \citetalias{2011ApJS..196...11S} measurements for Legacy galaxies with high $n_g$ and $m_r\lesssim14.5$ mag \emph{can} be compromised by these systematics which we have now quantified using deeper imaging. Our results highlight the importance of artificial galaxy simulation recovery analysis and/or deep imaging compliments to morphological analyses.

The upper half of the lower panel in Figure \ref{fig:sersic} also shows the median trend that fainter targets tend to have mildly lower $n_g$ in the \texttt{ps} fits for Stripe 82 relative to Legacy. Judging from the one-dimensional $n_g$ distributions in the upper right panel for \texttt{ps}, these must be targets that had $n_g\gtrsim7$ in Legacy but now have lower $n_g$ in Stripe 82. The asymmetry of the scatter at the faint end supports this assertion. Given that this trend is reversed in the $ur$, $ir$, and $zr$ fits, it is unclear what is driving this mild decrease in $gr$ $n_g$ for faint targets -- particularly because the other structural parameters of the \texttt{ps} fits in those other bandpass pairs (effective radius, axis ratio, and position angle) are fixed to the $gr$ results. The Legacy distributions, for example, do not differ in this way -- they are practically identical in all bandpass pairs. The fact that Legacy $n_g$ distributions do not change in these other bandpass pairs is mostly likely a consequence of fixing other structural parameters to the $gr$ results and dominance of the higher $S/N$ $r-$band images to the optimization of $n_g$ in cases where there is low $S/N$ in the $u$, $i$, or $z$ images. As such, it could be argued that the improved $S/N$ in the $uiz$ images increases the covariance between $uiz$ images \emph{and} their $r-$band counterparts in optimizing $n_g$ to both bands -- resulting in unique distributions in each bandpass pair. But this argument does not also explain why the trend in the $gr$ fits at high $n_g$ is so suddenly reversed relative to fits in the other bandpass pairs. At this time, it is not immediately clear why the fainter sources have suppressed $n_g$ in Stripe 82 relative to Legacy. However, we point out that this suppression is very modest -- with the median offset not exceeding $\Delta n_g = 0.15$.

The upper right panel of Figure \ref{fig:sersic} also compares Stripe 82 and Legacy bulge S\'ersic indices, $n_b$, from the \texttt{fn} fits to the $gr$ images (cyan and black). The differences are minor with the exception of an increased number of objects in the final $n_b\approx8$ bin. As with the \texttt{ps} fits to the $gr$ bandpasses, we examine the difference more closely in the lower half of the lower panel of Figure \ref{fig:sersic}. The $\Delta n$ axis has been broadened to better encompass the scatter (that predominantly arises due to the fact that we have not made any cuts on bulge brightness). We find similar systematics in $n_b$ for \texttt{fn} fits to bright sources that were reported and discussed for the \texttt{ps} fits. As total galaxy brightness decreases, the scatter increases rapidly but with reasonable symmetry and no other particular systematics between deep and shallow images. Ultimately, we caution that the similarity in one-dimensional distributions of $n_b$ in the \texttt{fn} decompositions implies that the majority of $n_b$ measurements from the deep images are likely to be affected by the same systematics that affected $n_b$ measurements in the shallow images (also supported by the $P_{n4,S82}$ results in the previous section). Cuts in $B/T$ (to focus on bulge brightness) or the more discriminate $P_{n4,S82}$ (to focus on objects for which the free $n_b$ made a significant difference to the fitting result) can be used to suppress this scatter or to glean galaxies that are less affected (e.g., as in Figure 15 of \citetalias{2011ApJS..196...11S}) but may introduce biases depending on the science case.

\subsection{Bulge and disc sizes}

\begin{figure*}
  \includegraphics[width=0.49\linewidth]{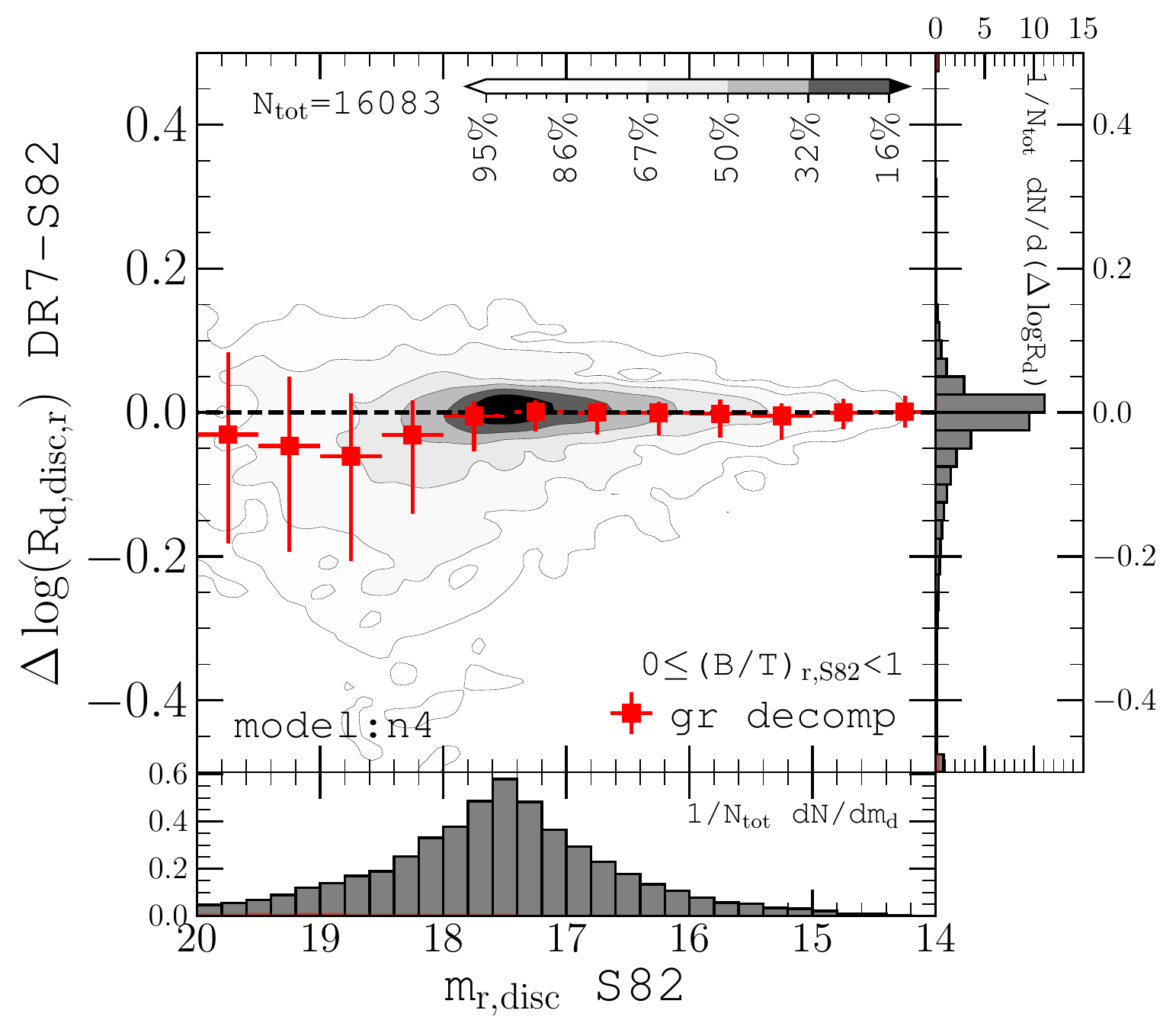}
  \includegraphics[width=0.49\linewidth]{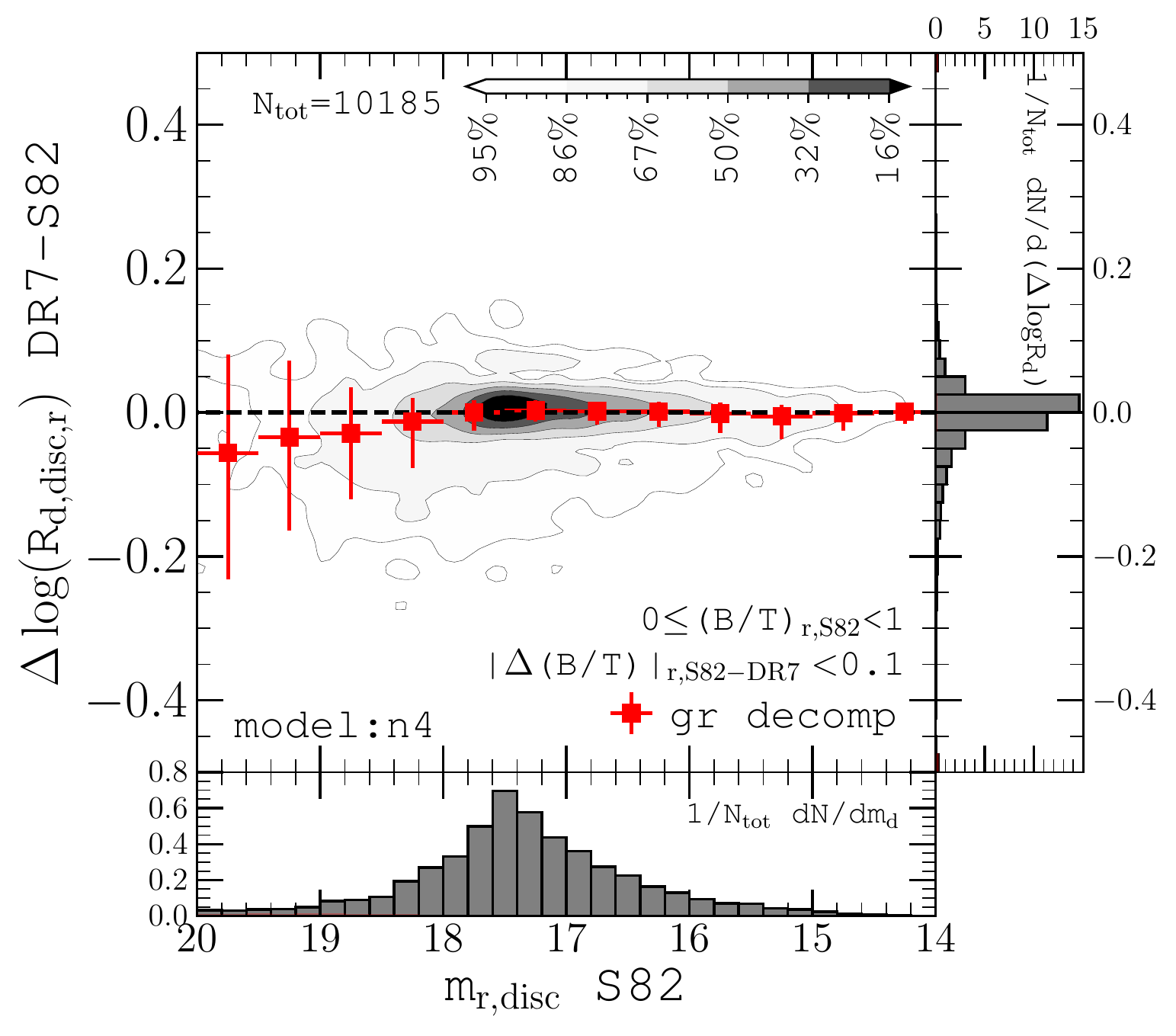}
  \includegraphics[width=0.49\linewidth]{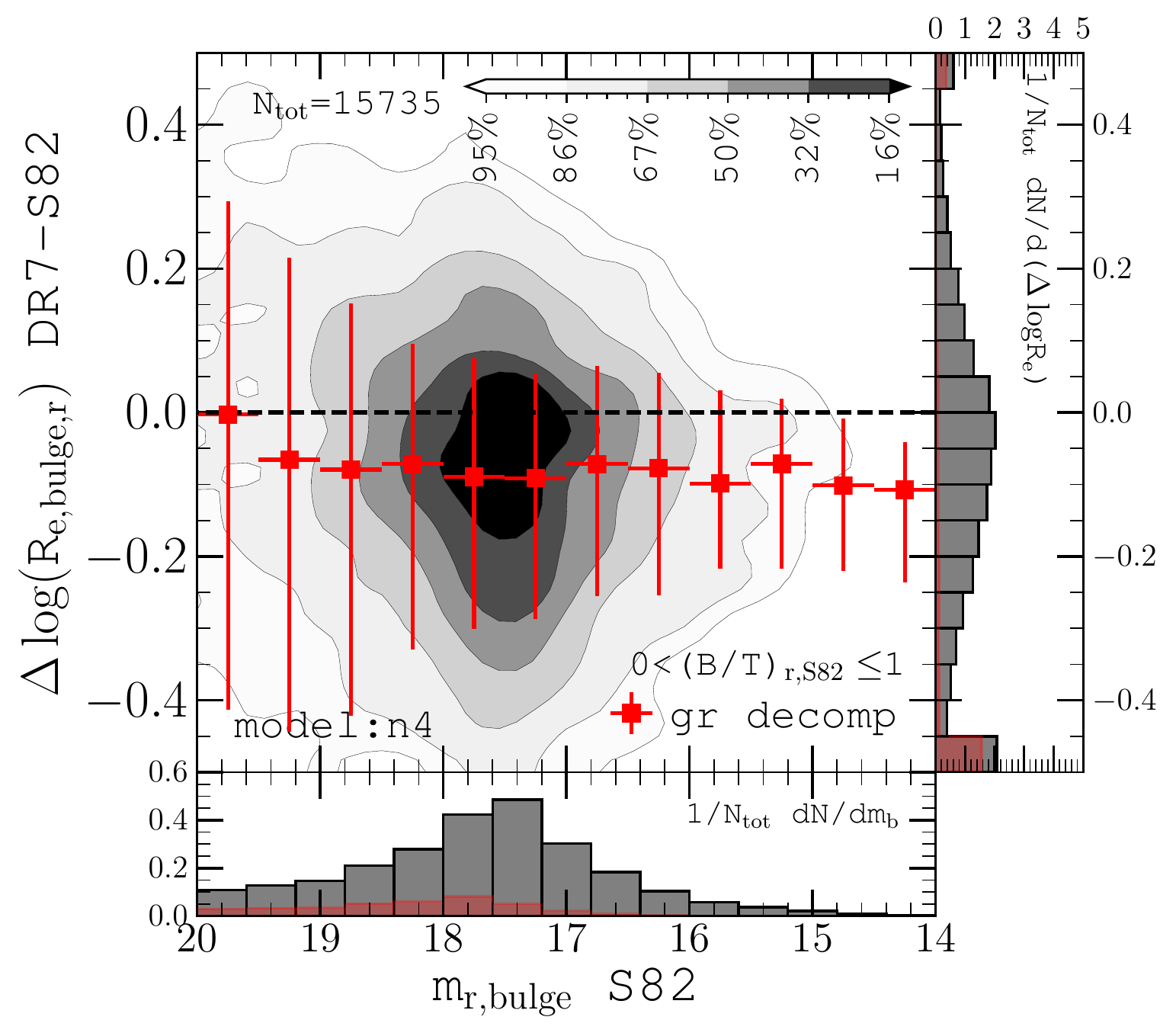}
  \includegraphics[width=0.49\linewidth]{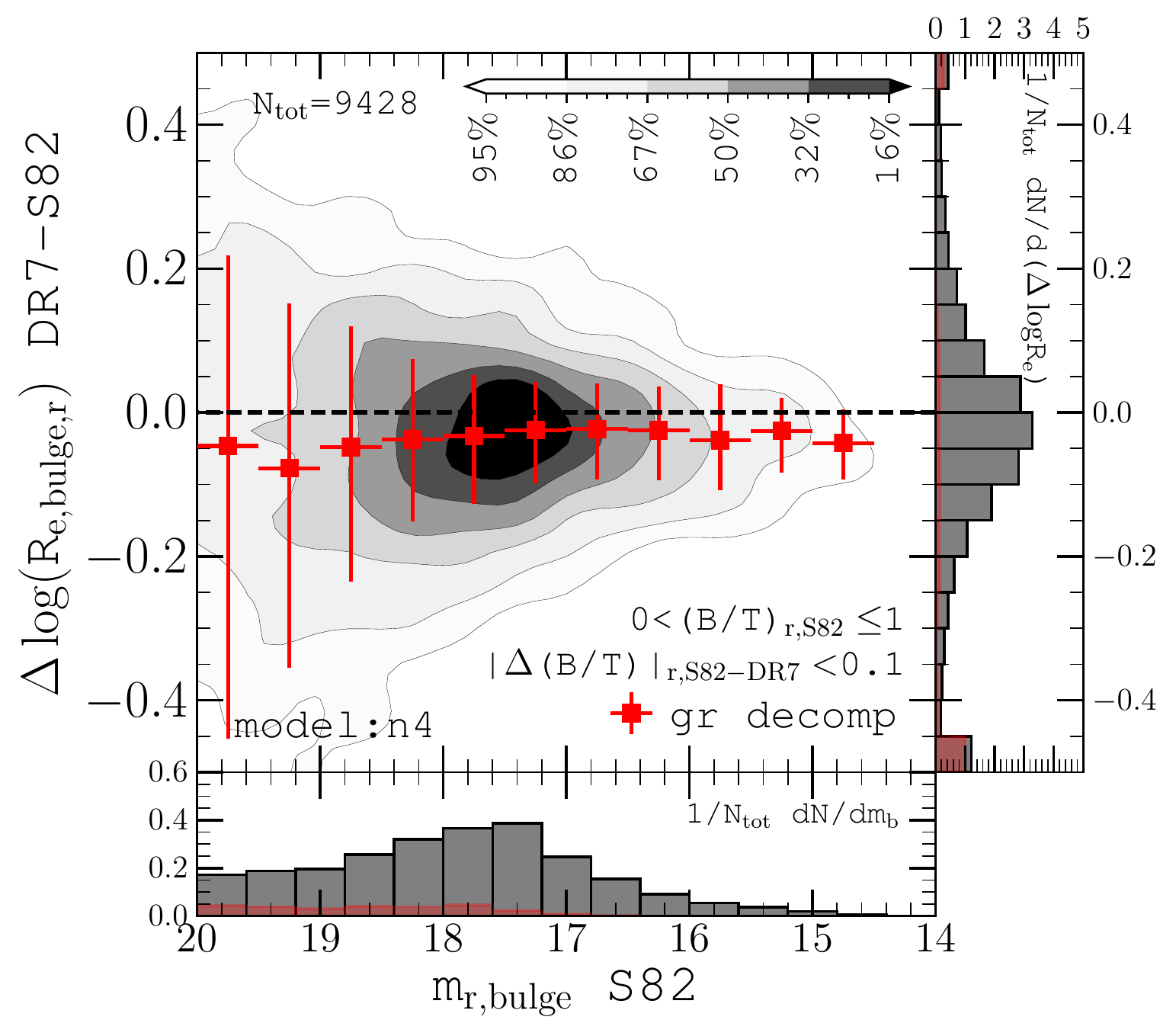}

\caption[Bulge and disc sizes]{Comparison of disc scale lengths and bulge effective radii in Stripe 82 and Legacy. Markers show the median and 16-84 percentile range for $\Delta \log r_{\mathrm{comp}}$ in bins of component magnitude. Two-dimensional distributions are flanked by marginal histograms for each axis -- where all values extending beyond the binning range have been clipped and added to the corresponding edge of the distribution. Red shading in marginal histograms indicate the relative fraction of galaxies per bin for which the Stripe 82 or DR7 size is smaller than the HWHM of the PSF following Equation \ref{eq:fracred}. All measurements are taken from the \texttt{n4} $gr$ decomposition table and are by construction identical in each other bandpass pair (excepting their magnitude distributions). Markers show the median and 16th-84th percentiles in each 0.5 magnitude bin. Left panels: $\Delta \log r_{d,\mathrm{disc}}$ and $\Delta \log r_{e,\mathrm{bulge}}$ plotted against respective Legacy disc and bulge apparent magnitudes. $B/T$ ranges are indicated at the lower left of each panel. Right panels: selection of galaxies from left panels with $|\Delta(B/T)|_{\mathrm{r,DR7-S82}}<0.1$. This selection reduces systematics in bulge sizes and reduces the scatter in both bulge and disc comparisons.
}
\label{fig:rd}
\end{figure*}

The $F$-test statistics show that the co-add images offer improved discrimination between single- and two-component systems compared to Legacy. We now compare the bulge and disc sizes to see how the component properties are affected. The upper left panel of Figure \ref{fig:rd} shows the change in disc scale-length, $\Delta \log r_d$, plotted against disc apparent magnitude for the full range in $B/T$. There is excellent agreement between the median co-add and Legacy disc sizes all the way to $m_r\approx18$ mag. The 16-84 percentile range in $\Delta \log r_d$ for discs brighter than $m_r\approx18$ mag is also remarkably tight between the co-add and Legacy decompositions. However, at fainter disc magnitudes, the scatter is increased and Stripe 82 disc components are systematically larger than their Legacy counterparts. The median offset at faint disc magnitudes can be as large as $\Delta \log r_d \approx -0.07$ dex. 

The lower left panel of Figure \ref{fig:rd} shows the change in bulge effective radius $\Delta \log r_e$. The red shading in the marginals shows the relative fraction of bulges in each bin whose measured effective radii in the Stripe 82 or Legacy decomposition are less than associated the Half-Width at Half Maximum (HWHM) of the PSF:
\begin{align}
\label{eq:fracred}
F_{\mathrm{bin,red}} = \frac{N_{\mathrm{bin}}(r_{\mathrm{e,S82}}<\mathrm{HWHM}_{\mathrm{S82}}\; \mathrm{OR}\; r_{\mathrm{e,DR7}}<\mathrm{HWHM}_{\mathrm{DR7}})}{N_{\mathrm{bin,tot}}}
\end{align}
The same is shown for the discs, but very few disc sizes from either the co-adds or Legacy are below the HWHM of the PSF. Galaxies with the largest offsets in bulge size (those clipped to the edges of the right marginals) are those that are unresolved and faint (see lower marginals). 

Stripe 82 bulge sizes are systematically larger than in Legacy for all bulge magnitudes (with median offsets as large as $\Delta \log r_e\approx-0.1$ dex). The scatter in the bulge size comparison is also far larger than for disc sizes. In general, bulges are intrinsically more difficult to fit than discs due to their high sensitivity to spatial resolution, sky subtraction, and the presence of additional substructures and components (such as bars). But there are several factors which can drive increased scatter and potentially introduce systematics. 

First, changes in the sizes of intrinsically small and faint discs or bulges are exacerbated on a logarithmic scale. Since bulges are generally more compact than discs for the same luminosities (e.g., \citealt{2003AJ....125.2936G}), the scatter in $\Delta \log r_e$ will generally be larger than in $\Delta \log r_d$.

Second, if a physical component is so small in angular size that it is spatially unresolved then scatter will be introduced by mismatches between the reconstructed PSF with which our models are convolved and the true PSF. Since the spatial resolution is $8\%$ poorer on average in the co-adds than in Legacy, there is a slightly greater likelihood of having such mismatches in the co-adds. 

Third, since galaxies in our sample all have $m_{\mathrm{r,galaxy}}\leq17.77$ mag, a bulge (or disc) with brightness $m_{\mathrm{r,comp}}>17.77$ mag often belongs to a system whose surface brightness distribution is dominated by the disc (or bulge). Applying a cut of $0.5\leq B/T <0.9$ in the disc comparison or $0.1<B/T<0.5$ to the bulge comparison reveals that there is greater scatter in $\Delta \log r_d$ for discs and $\Delta \log r_e$ for bulges that are embedded in two-component systems in which the secondary component dominates. Applying a $0\leq B/T<0.3$ cut to select only disc dominated systems reduces the scatter to $\Delta \log r_d$ to 0.05 dex. However, the same cannot be said of the bulge size comparison using a cut of $0.7<B/T\leq1$. We found that the scatter and systematics in the lower left panel of Figure \ref{fig:rd} largely persist despite the cut.

The persistent systematics and biases among bulges raises a fourth driving source of contrast between the co-adds and Legacy component sizes -- the increased bulge fractions in the co-adds for many galaxies with intermediate $B/T$. In the right panels of Figure \ref{fig:rd}, we control for this bias by strictly selecting galaxies which have a change in $|\Delta(B/T)|_{\mathrm{DR7-S82}}<0.1$. Since total galaxy magnitudes are largely the same in the co-adds and Legacy, such a cut is the equivalent of saying that the component brightnesses must be the same as well -- and so the result is unsurprising but demonstrates the contribution of this bias to the scatter. Most notably, the systematic in the bulge size comparison at faint bulge magnitudes shown in the lower left panel of Figure \ref{fig:rd} is almost entirely eliminated in the lower right panel where $|\Delta(B/T)|_{\mathrm{DR7-S82}}$ is controlled. Thus the systematic at faint magnitudes and inflated scatter is only the consequence of the different (but typically larger) bulge-fractions measured in the co-adds. 

The remaining scatter after our cut in $|\Delta(B/T)|$ is qualitatively consistent with sources we have already discussed in the previous paragraphs of this section. Now, whether the tendency towards higher bulge-fractions are indeed a \emph{correction} to systematically under-estimated bulge fractions for the fainter Legacy galaxies with intermediate $B/T$ is not certain. But consider again the arguments we made in Section \ref{sec:btr} regarding the artificiality of the bump at $B/T=0.5$ in Legacy relative to the more uniform $B/T$ distribution found in the co-adds. Our results suggest that the bump at $B/T=0.5$ (the median over the allowed range in bulge fractions) arises from poor constraints on the light profiles of faint galaxies. The characterization of these galaxies and their components is substantially improved in the deep co-adds as shown in our comparison of the $F$-test statistics. These pieces of evidence suggest correction to a systematic arising from poorer photometric constraints on faint bulges and discs in Legacy rather than a new and unexpected systematic in the co-adds. 

As with the sizes, we have compared the brightnesses of the components. Our findings are consistent with our analysis of the sizes. Many previously faint bulges get a brightness boost in the co-adds. The disc fractions are suppressed to compensate. It is worth noting that our comparison of component properties essentially functions as a convergence test. We have shown that it is possible to use our tables to estimate the brightness and component contrast (set by a galaxy's intrinsic bulge-to-disc, $(B/D)$, light ratio) for which a bulge or disc component's physical properties hold-up at fainter surface-brightness limits. 

\subsection{Residual asymmetries and non-parametric indices}
Six additional morphological indices are computed for each decomposition. The majority of these indices are aimed at quantitatively identifying visual indicators of disturbances in galaxy structures. The first two are the asymmetry, $A$, and concentration, $C$, indices from the automatic classification system (CAS) proposed by \cite{1994ApJ...432...75A,1996ApJS..107....1A}. Two more are $D_z$ and $A_z$, defined in Section 5.6 of \cite{2002ApJS..142....1S}. $A_z$ measures the flux from all pixels that are $n\sigma_{\mathrm{sky}}$ higher than their symmetric counterparts when rotated $180^{\circ}$ about the target's barycenter, normalized by the total object flux. $A_z$ is computed for $n={2,3,5}$ within circular apertures extending either one or two half-light radii from the target centroid. Similarly, $D_z$ is the sum of the fluxes of target object pixels, as determined from the \textsc{SExtractor} segmentation image, whose symmetric counterparts are not also target object pixels. $D_z$ is similar in nature to the shape asymmetry parameter first proposed by \cite{2016MNRAS.456.3032P} which is sensitive to crowding by neighbouring sources and potentially useful in quantitatively identifying close galaxy pairs.

The remaining two indices are based on the $R_{T}$ and $R_{A}$ indices used in local studies of spiral arm patterns by \cite{1992ApJS...79...37E} and first applied to distant galaxies by \cite{1995ApJ...451L...1S} as part of the Canada-France Redshift Survey \citep{1995ApJ...455...50L}. $R_{T}$ and $R_{A}$ each quantify the residual light that is not characterized by the symmetric, analytic models used in the decomposition. For both indices, the residual model-subtracted image is either added to ($R_T$) or subtracted ($R_{A}$) by the same image rotated by 180 degrees about the galactic centre.\footnote{The $R_{T}$ and $R_{A}$ indices are known to be sensitive to the pivot point as for the $C-A$ indices \cite{2000ApJ...529..886C}. We do not include a step to find pivot point in the image about which the asymmetry indices are minimized.} As such, asymmetric features which, in particular, are known to be indicators of galaxy interactions such as tails and bridges may be identified. Calculation of $R_{T}$ and $R_{A}$ in the \textsc{gim2d} pipeline differs slightly from \cite{1995ApJ...451L...1S} in that the \textsc{gim2d} indices are computed within one, two, and three multiples of a galaxy's measured half-light radius. Table values are correspondingly named \textsc{r\{a,t\}\{1,2,3\}\_1\_\{band\}}.

\begin{figure}
  \includegraphics[width=0.975\linewidth]{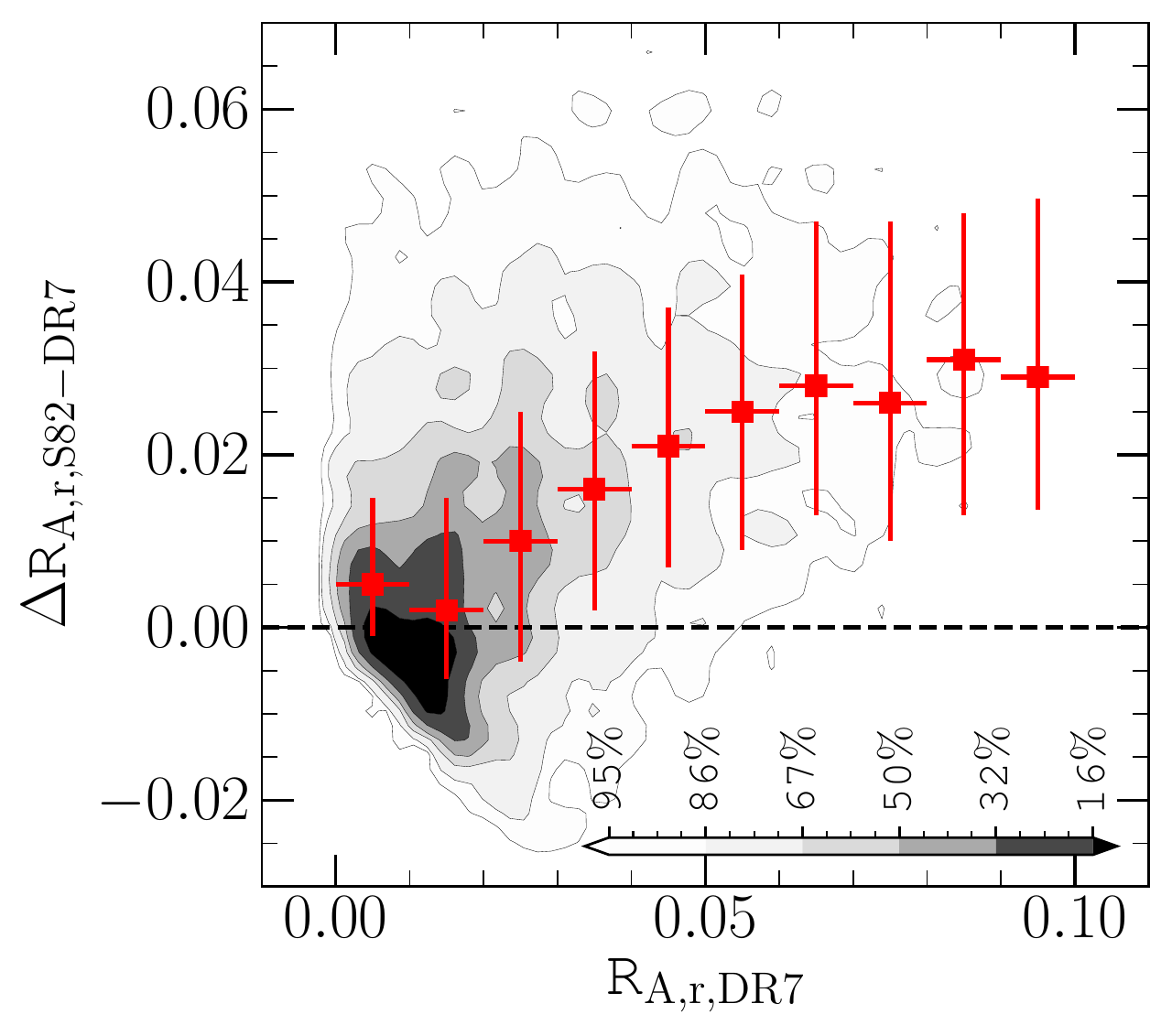}
  \includegraphics[width=\linewidth]{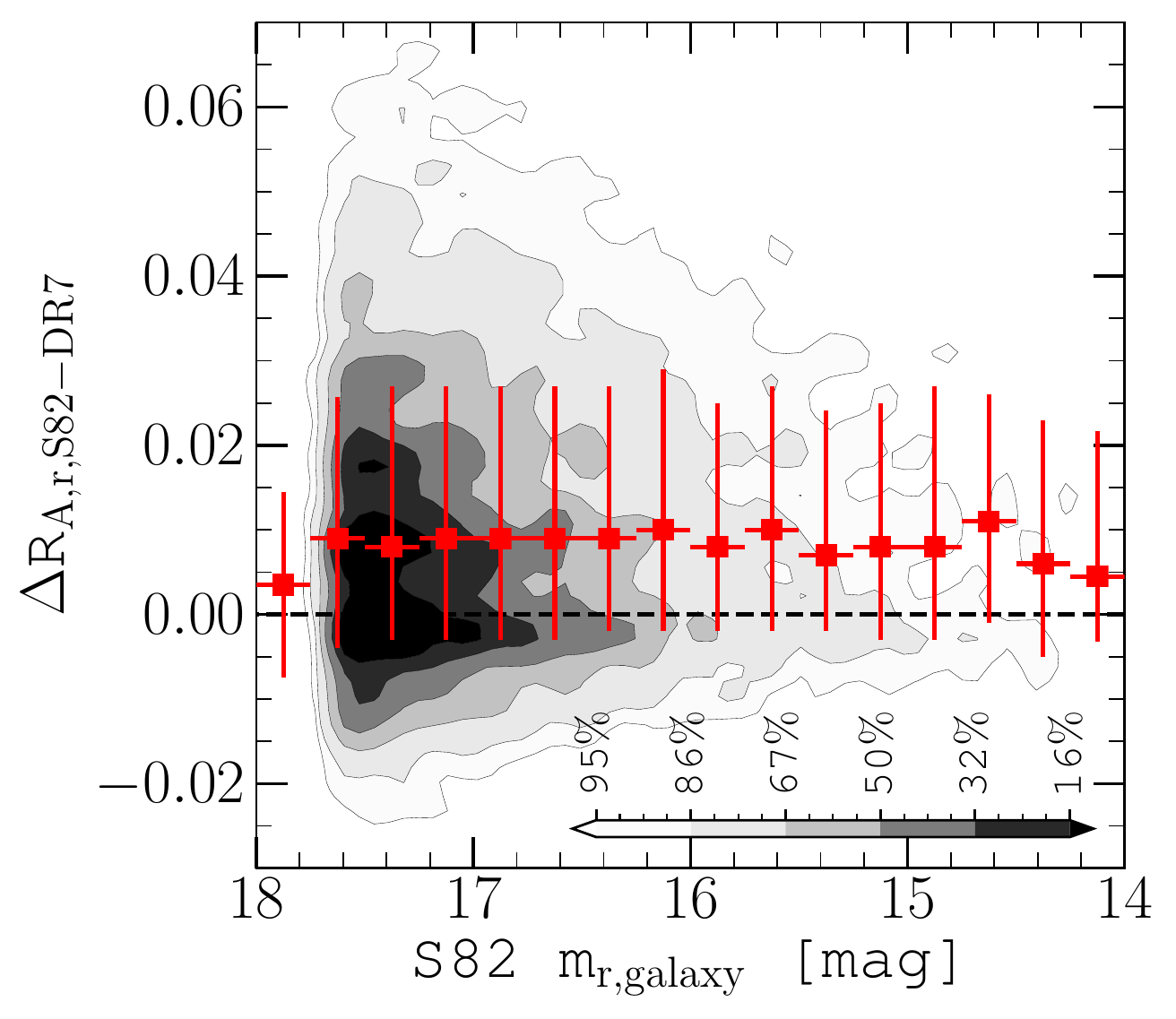}

\caption[Asymmetries]{Change in residual asymmetries, $\Delta R_A$ between Legacy and the Stripe 82 co-adds. $R_A$ is computed within three half-light radii in each case. The upper panel shows the enhancement as a function of the original $R_A$ measurement from the Legacy $r$-band decompositions. The lower panel shows the enhancement as a function of Stripe 82 total apparent magnitude. Asymmetries are uniformly enhanced as a function of magnitude, but are preferentially enhanced in targets that had large asymmetries in Legacy images.}
\label{fig:ra3}
\end{figure}

Figure \ref{fig:ra3} compares the $R_A$ indicators measured for the co-add and Legacy decompositions using the difference $\Delta R_A$. The lower panel shows the comparison as a function of magnitude. There is a uniform median enhancement in the co-add $R_A$ (by $\Delta R_A\approx0.07$) across the full magnitude range. However, in the upper panel, we see that this enhancement in asymmetry for the co-adds is largely attributed to galaxies with some degree of asymmetry already measured in Legacy. We have confirmed these findings through visual inspection of galaxies with large $\Delta R_A$ which either had small or large $R_A$ originally in Legacy. In particular, the co-adds often reveal \textsc{Hii} regions, tidal tails, shells, and streams in galaxies that were hardly detectable in the shallower Legacy images. Taken together, these results demonstrate that asymmetric features are generally enhanced in the deeper images insofar as they may be distinguished from the sky background. The standard and residual asymmetric features that are revealed through deep imaging may be exploited to construct more robust samples of galaxies that are expected to contain such asymmetric features such as peculiar early-type galaxies (e.g., \citealt{2010MNRAS.406..382K}) and recent/ongoing mergers (e.g, \citealt{Ellison_CFIS}). 

\section{Summary}\label{sec:summary}
Using images constructed from multiple exposures of 275 deg$^2$ in the SDSS Stripe 82, which permit an additional $1.6-1.8$ magnitudes of depth with respect to single-exposure SDSS Legacy images, we have performed free-$n_b$ and $n_b=4$ bulge+exp. disc (\texttt{fn} and \texttt{n4}) as well as single-component S\'ersic \texttt{ps} decompositions on 16,908 galaxies -- obtaining parametric morphologies in the $u,g,r,i,z$ bands. We make all of our catalogs publicly available. Our catalog structures are consistent with those from \citetalias{2011ApJS..196...11S}, enabling straightforward comparisons of structural measurements using deep and shallow photometry. Our main findings are summarized as follows:

\begin{itemize}

\item \textbf{Integrated galaxy properties in each band such as size and luminosity are largely unchanged (Fig. \ref{fig:mags}).} For example, the median offsets in total apparent magnitude for all galaxies in each band are $\Delta m_{u,g,r,i,z}^{50\%} = (0.037,0.022,0.021,0.020,0.019)$ mag (DR7-S82) for the \texttt{n4} decompositions. Exceptions arise for the brightest galaxies in each bandpass, where broader target masks and correspondingly improved local sky estimation yields brighter results in the co-adds. Also, many galaxies with $u$-band intrinsic brightnesses that are poorly constrained or resulted in non-detections in Legacy are actually measured and characterized in the co-adds. This systematic manifests as large negative offsets between the Legacy and Stripe 82 decomposition $u$-band magnitudes. 

\item \textbf{The colour-magnitude diagrams for the co-adds have a tighter red-sequence in all bands with respect to Legacy decompositions (Fig. \ref{fig:colours}).} The removal of outliers is particularly noticeable in the $u$-band relative to the Legacy colours. The majority of the outliers in Legecy have large, positive $(u-r)$ colours. Given that their $r$-band magnitudes are unchanged in the co-add decompositions, the high $(u-r)$ colours indicate that many galaxies' $u$-band magnitudes were systematically underestimated in Legacy. The reduced scatter in the red-sequences is also an indicator that deblending of light for crowded targets is improved -- as satellites and other potential interlopers are now more efficiently masked due to the high $S/N$ in the co-adds.

\item \textbf{Bulge-to-total fractions for many galaxies with intermediate $B/T$ are enhanced in the co-add decompositions, but are largely unchanged at low $B/T$ (Fig. \ref{fig:btr}).} Importantly, the changes to $B/T$ appear to remove a systematic aggregation of galaxies around $B/T=0.5$. We argue that the grouping at $B/T=0.5$ is not physical, but is a systematic in the \citetalias{2011ApJS..196...11S} decompositions owing to poor constraints on the structures of galaxies at the limits of Legacy photometry. 

\item \textbf{A number of galaxies with $B/T=1$ in the shallow images get a boost to their disc fractions in the deeper images (Fig. \ref{fig:btr}).} The median $\Delta(B/T)_{\mathrm{S82-DR7,r}} \approx -0.09$ for galaxies with $B/T\approx1$ in Legacy. However, this reduction in bulge fraction can be up to 40\% for such galaxies.

\item \textbf{We find that the deep imaging improves the discriminating ability of the $F$-test in determining whether a more complex decomposition model is favoured with respect to a simpler model (Fig. \ref{fig:ftest}).} Where the majority of galaxies in the Legacy decompositions have $P_{pS,DR7}\approx0.5$ and $P_{n4,DR7}\approx0.5$ (indicating no statistical preference between decomposition models), the co-add $P_{pS,S82}$ and $P_{n4,S82}$ are bimodal over a large range in galaxy apparent magnitudes. We argue the improved photometry in the co-adds enables better characterization of the components and, in particular, an improved capacity to determine whether a two-component model is favoured over a single-component model.

\item \textbf{We show that systematics on galaxy and bulge S\'ersic index in the \texttt{ps} and \texttt{fn} fits are suppressed in the deeper images (Fig. \ref{fig:sersic}).} More accurate sky measurements and reduced sky uncertainties yield improved characterization of the extended profiles of galaxies with both $m_r<14.5$ \emph{and} high S\'ersic indices whose fluxes, sizes, and S\'ersic indices can be suppressed by over-estimated skies in the \citetalias{2011ApJS..196...11S} Legacy measurements. These findings using deep imaging agree with systematic uncertainty predictions by \cite{2014ApJS..210....3M} from artificial galaxy recovery analysis. They also agree with results of comparisons between the \citetalias{2011ApJS..196...11S} Legacy catalogs to other decomposition analyses with alternative sky-estimation procedures (e.g., \citealt{2014MNRAS.443..874B,2015MNRAS.446.3943M}). 

\item \textbf{Bulge and disc sizes are consistent out to $m_{r,\mathrm{comp}}\approx17$ (Fig. \ref{fig:rd}).} We assert that this is the magnitude limit at which Legacy measurements of components can be trusted before high measurement uncertainties and systematics dominate. The scatter in the component sizes increases rapidly with decreasing brightness. We showed that this scatter arises from systems in which the one component dominates over the other and where the photometry of a component is limited in the shallower Legacy images. Using cuts in $B/T$, the scatter in the disc sizes can be reduced to $\Delta r_{d,disc,r}< 0.05$ dex. The higher bulge fractions in many faint galaxies drive a systematic enhancement of bulges sizes in the co-adds. We caution, however, that users of our catalogs must consider their science case before making such cuts -- which may introduces biases.  

\item \textbf{Measures of asymmetry are enhanced in the deeper Stripe 82 co-adds with respect to Legacy measurements (Fig. \ref{fig:ra3}).} We compute a range of asymmetry parameters for each galaxy and decomposition that are included in the catalogs. We find that asymmetries are generally enhanced in the Stripe 82 co-adds and that the enhancement is strongest for galaxies that were already asymmetric in Legacy. 

\end{itemize}

In general, the changes to morphological and statistical parameters in the Stripe 82 co-adds highlight the importance of deep imaging compliments to large photometric galaxy surveys. In particular, such deep imaging enables valuable and straight-forward validation of galaxy structural measurements and the magnitude limits to which they are robust.

\section*{Acknowledgements}

We thank James Annis, J\"urgen Fliri, and Ignacio Trujillo for useful discussions on their respective Stripe 82 stacks. CB acknowledges the support of the Natural Sciences and Engineering Research Council of Canada (NSERC). CB also thanks McGill University and the McGill Space Institute for being generous hosts during the completion of this project. Parts of this research were conducted by the Australian Research Council Centre of Excellence for All Sky Astrophysics in 3 Dimensions (ASTRO 3D), through project number CE170100013. We also thank the anonymous referee whose suggestions helped to improve the quality of this paper.

This project made use of public data from the Sloan Digital Sky Survey. Funding for the Sloan Digital Sky Survey IV has been provided by the Alfred P. Sloan Foundation, the U.S. Department of Energy Office of Science, and the Participating Institutions. SDSS-IV acknowledges
support and resources from the Center for High-Performance Computing at
the University of Utah. The SDSS web site is www.sdss.org.

SDSS-IV is managed by the Astrophysical Research Consortium for the 
Participating Institutions of the SDSS Collaboration including the 
Brazilian Participation Group, the Carnegie Institution for Science, 
Carnegie Mellon University, the Chilean Participation Group, the French Participation Group, Harvard-Smithsonian Center for Astrophysics, 
Instituto de Astrof\'isica de Canarias, The Johns Hopkins University, 
Kavli Institute for the Physics and Mathematics of the Universe (IPMU) / 
University of Tokyo, Lawrence Berkeley National Laboratory, 
Leibniz Institut f\"ur Astrophysik Potsdam (AIP),  
Max-Planck-Institut f\"ur Astronomie (MPIA Heidelberg), 
Max-Planck-Institut f\"ur Astrophysik (MPA Garching), 
Max-Planck-Institut f\"ur Extraterrestrische Physik (MPE), 
National Astronomical Observatories of China, New Mexico State University, 
New York University, University of Notre Dame, 
Observat\'ario Nacional / MCTI, The Ohio State University, 
Pennsylvania State University, Shanghai Astronomical Observatory, 
United Kingdom Participation Group,
Universidad Nacional Aut\'onoma de M\'exico, University of Arizona, 
University of Colorado Boulder, University of Oxford, University of Portsmouth, 
University of Utah, University of Virginia, University of Washington, University of Wisconsin, 
Vanderbilt University, and Yale University.





\bibliographystyle{mnras}
\bibliography{BIB/references_full.bib}

\begin{thebibliography}{}
\makeatletter
\relax
\def\mn@urlcharsother{\let\do\@makeother \do\$\do\&\do\#\do\^\do\_\do\%\do\~}
\def\mn@doi{\begingroup\mn@urlcharsother \@ifnextchar [ {\mn@doi@}
  {\mn@doi@[]}}
\def\mn@doi@[#1]#2{\def\@tempa{#1}\ifx\@tempa\@empty \href
  {http://dx.doi.org/#2} {doi:#2}\else \href {http://dx.doi.org/#2} {#1}\fi
  \endgroup}
\def\mn@eprint#1#2{\mn@eprint@#1:#2::\@nil}
\def\mn@eprint@arXiv#1{\href {http://arxiv.org/abs/#1} {{\tt arXiv:#1}}}
\def\mn@eprint@dblp#1{\href {http://dblp.uni-trier.de/rec/bibtex/#1.xml}
  {dblp:#1}}
\def\mn@eprint@#1:#2:#3:#4\@nil{\def\@tempa {#1}\def\@tempb {#2}\def\@tempc
  {#3}\ifx \@tempc \@empty \let \@tempc \@tempb \let \@tempb \@tempa \fi \ifx
  \@tempb \@empty \def\@tempb {arXiv}\fi \@ifundefined
  {mn@eprint@\@tempb}{\@tempb:\@tempc}{\expandafter \expandafter \csname
  mn@eprint@\@tempb\endcsname \expandafter{\@tempc}}}

\bibitem[\protect\citeauthoryear{{Abazajian} et~al.,}{{Abazajian}
  et~al.}{2009}]{2009ApJS..182..543A}
{Abazajian} K.~N.,  et~al., 2009, \mn@doi [\apjs]
  {10.1088/0067-0049/182/2/543}, \href
  {http://adsabs.harvard.edu/abs/2009ApJS..182..543A} {182, 543}

\bibitem[\protect\citeauthoryear{{Abraham}, {Valdes}, {Yee}  \& {van den
  Bergh}}{{Abraham} et~al.}{1994}]{1994ApJ...432...75A}
{Abraham} R.~G.,  {Valdes} F.,  {Yee} H.~K.~C.,   {van den Bergh} S.,  1994,
  \mn@doi [\apj] {10.1086/174550}, \href
  {http://adsabs.harvard.edu/abs/1994ApJ...432...75A} {432, 75}

\bibitem[\protect\citeauthoryear{{Abraham}, {van den Bergh}, {Glazebrook},
  {Ellis}, {Santiago}, {Surma}  \& {Griffiths}}{{Abraham}
  et~al.}{1996}]{1996ApJS..107....1A}
{Abraham} R.~G.,  {van den Bergh} S.,  {Glazebrook} K.,  {Ellis} R.~S.,
  {Santiago} B.~X.,  {Surma} P.,   {Griffiths} R.~E.,  1996, \mn@doi [\apjs]
  {10.1086/192352}, \href {http://adsabs.harvard.edu/abs/1996ApJS..107....1A}
  {107, 1}

\bibitem[\protect\citeauthoryear{{Allen}, {Driver}, {Graham}, {Cameron},
  {Liske}  \& {de Propris}}{{Allen} et~al.}{2006}]{2006MNRAS.371....2A}
{Allen} P.~D.,  {Driver} S.~P.,  {Graham} A.~W.,  {Cameron} E.,  {Liske} J.,
  {de Propris} R.,  2006, \mn@doi [\mnras] {10.1111/j.1365-2966.2006.10586.x},
  \href {http://adsabs.harvard.edu/abs/2006MNRAS.371....2A} {371, 2}

\bibitem[\protect\citeauthoryear{{Andredakis} \& {Sanders}}{{Andredakis} \&
  {Sanders}}{1994}]{1994MNRAS.267..283A}
{Andredakis} Y.~C.,  {Sanders} R.~H.,  1994, \mn@doi [\mnras]
  {10.1093/mnras/267.2.283}, \href
  {http://adsabs.harvard.edu/abs/1994MNRAS.267..283A} {267, 283}

\bibitem[\protect\citeauthoryear{{Andredakis}, {Peletier}  \&
  {Balcells}}{{Andredakis} et~al.}{1995}]{1995MNRAS.275..874A}
{Andredakis} Y.~C.,  {Peletier} R.~F.,   {Balcells} M.,  1995, \mn@doi [\mnras]
  {10.1093/mnras/275.3.874}, \href
  {http://adsabs.harvard.edu/abs/1995MNRAS.275..874A} {275, 874}

\bibitem[\protect\citeauthoryear{{Annis} et~al.,}{{Annis}
  et~al.}{2014}]{2014ApJ...794..120A}
{Annis} J.,  et~al., 2014, \mn@doi [\apj] {10.1088/0004-637X/794/2/120}, \href
  {http://adsabs.harvard.edu/abs/2014ApJ...794..120A} {794, 120}

\bibitem[\protect\citeauthoryear{{Balcells}, {Graham},
  {Dom{\'{\i}}nguez-Palmero}  \& {Peletier}}{{Balcells}
  et~al.}{2003}]{2003ApJ...582L..79B}
{Balcells} M.,  {Graham} A.~W.,  {Dom{\'{\i}}nguez-Palmero} L.,   {Peletier}
  R.~F.,  2003, \mn@doi [\apjl] {10.1086/367783}, \href
  {http://adsabs.harvard.edu/abs/2003ApJ...582L..79B} {582, L79}

\bibitem[\protect\citeauthoryear{{Baldry}, {Glazebrook}, {Brinkmann},
  {Ivezi{\'c}}, {Lupton}, {Nichol}  \& {Szalay}}{{Baldry}
  et~al.}{2004}]{2004ApJ...600..681B}
{Baldry} I.~K.,  {Glazebrook} K.,  {Brinkmann} J.,  {Ivezi{\'c}} {\v Z}.,
  {Lupton} R.~H.,  {Nichol} R.~C.,   {Szalay} A.~S.,  2004, \mn@doi [\apj]
  {10.1086/380092}, \href {http://adsabs.harvard.edu/abs/2004ApJ...600..681B}
  {600, 681}

\bibitem[\protect\citeauthoryear{{Balogh}, {Baldry}, {Nichol}, {Miller},
  {Bower}  \& {Glazebrook}}{{Balogh} et~al.}{2004}]{2004ApJ...615L.101B}
{Balogh} M.~L.,  {Baldry} I.~K.,  {Nichol} R.,  {Miller} C.,  {Bower} R.,
  {Glazebrook} K.,  2004, \mn@doi [\apjl] {10.1086/426079}, \href
  {http://adsabs.harvard.edu/abs/2004ApJ...615L.101B} {615, L101}

\bibitem[\protect\citeauthoryear{{Barnes}}{{Barnes}}{1988}]{1988ApJ...331..699B}
{Barnes} J.~E.,  1988, \mn@doi [\apj] {10.1086/166593}, \href
  {http://adsabs.harvard.edu/abs/1988ApJ...331..699B} {331, 699}

\bibitem[\protect\citeauthoryear{{Bell}}{{Bell}}{2008}]{2008ApJ...682..355B}
{Bell} E.~F.,  2008, \mn@doi [\apj] {10.1086/589551}, \href
  {http://adsabs.harvard.edu/abs/2008ApJ...682..355B} {682, 355}

\bibitem[\protect\citeauthoryear{{Benson}, {D{\v z}anovi{\'c}}, {Frenk}  \&
  {Sharples}}{{Benson} et~al.}{2007}]{2007MNRAS.379..841B}
{Benson} A.~J.,  {D{\v z}anovi{\'c}} D.,  {Frenk} C.~S.,   {Sharples} R.,
  2007, \mn@doi [\mnras] {10.1111/j.1365-2966.2007.11923.x}, \href
  {http://adsabs.harvard.edu/abs/2007MNRAS.379..841B} {379, 841}

\bibitem[\protect\citeauthoryear{{Berg}, {Simard}, {Mendel Trevor}  \&
  {Ellison}}{{Berg} et~al.}{2014}]{2014MNRAS.440L..66B}
{Berg} T.~A.~M.,  {Simard} L.,  {Mendel Trevor} J.,   {Ellison} S.~L.,  2014,
  \mn@doi [\mnras] {10.1093/mnrasl/slu026}, \href
  {http://adsabs.harvard.edu/abs/2014MNRAS.440L..66B} {440, L66}

\bibitem[\protect\citeauthoryear{{Bernardi}, {Meert}, {Vikram},
  {Huertas-Company}, {Mei}, {Shankar}  \& {Sheth}}{{Bernardi}
  et~al.}{2014}]{2014MNRAS.443..874B}
{Bernardi} M.,  {Meert} A.,  {Vikram} V.,  {Huertas-Company} M.,  {Mei} S.,
  {Shankar} F.,   {Sheth} R.~K.,  2014, \mn@doi [\mnras]
  {10.1093/mnras/stu1106}, \href
  {http://adsabs.harvard.edu/abs/2014MNRAS.443..874B} {443, 874}

\bibitem[\protect\citeauthoryear{{Bertin} \& {Arnouts}}{{Bertin} \&
  {Arnouts}}{1996}]{1996A&AS..117..393B}
{Bertin} E.,  {Arnouts} S.,  1996, \mn@doi [\aaps] {10.1051/aas:1996164}, \href
  {http://adsabs.harvard.edu/abs/1996A%26AS..117..393B} {117, 393}

\bibitem[\protect\citeauthoryear{{Blanton} \& {Roweis}}{{Blanton} \&
  {Roweis}}{2007}]{2007AJ....133..734B}
{Blanton} M.~R.,  {Roweis} S.,  2007, \mn@doi [\aj] {10.1086/510127}, \href
  {http://adsabs.harvard.edu/abs/2007AJ....133..734B} {133, 734}

\bibitem[\protect\citeauthoryear{{Blanton} et~al.,}{{Blanton}
  et~al.}{2003}]{2003ApJ...594..186B}
{Blanton} M.~R.,  et~al., 2003, \mn@doi [\apj] {10.1086/375528}, \href
  {http://adsabs.harvard.edu/abs/2003ApJ...594..186B} {594, 186}

\bibitem[\protect\citeauthoryear{{Bluck}, {Mendel}, {Ellison}, {Moreno},
  {Simard}, {Patton}  \& {Starkenburg}}{{Bluck}
  et~al.}{2014}]{2014MNRAS.441..599B}
{Bluck} A.~F.~L.,  {Mendel} J.~T.,  {Ellison} S.~L.,  {Moreno} J.,  {Simard}
  L.,  {Patton} D.~R.,   {Starkenburg} E.,  2014, \mn@doi [\mnras]
  {10.1093/mnras/stu594}, \href
  {http://adsabs.harvard.edu/abs/2014MNRAS.441..599B} {441, 599}

\bibitem[\protect\citeauthoryear{{Bluck} et~al.,}{{Bluck}
  et~al.}{2016}]{2016MNRAS.462.2559B}
{Bluck} A.~F.~L.,  et~al., 2016, \mn@doi [\mnras] {10.1093/mnras/stw1665},
  \href {http://adsabs.harvard.edu/abs/2016MNRAS.462.2559B} {462, 2559}

\bibitem[\protect\citeauthoryear{{Bluck} et~al.,}{{Bluck}
  et~al.}{2019}]{2019MNRAS.tmp..363B}
{Bluck} A.~F.~L.,  et~al., 2019, \mn@doi [\mnras] {10.1093/mnras/stz363}, \href
  {http://adsabs.harvard.edu/abs/2019MNRAS.tmp..363B} {}

\bibitem[\protect\citeauthoryear{{Bottrell}, {Torrey}, {Simard}  \&
  {Ellison}}{{Bottrell} et~al.}{2017a}]{2017MNRAS.467.1033B}
{Bottrell} C.,  {Torrey} P.,  {Simard} L.,   {Ellison} S.~L.,  2017a, \mn@doi
  [\mnras] {10.1093/mnras/stx017}, \href
  {http://adsabs.harvard.edu/abs/2017MNRAS.467.1033B} {467, 1033}

\bibitem[\protect\citeauthoryear{{Bottrell}, {Torrey}, {Simard}  \&
  {Ellison}}{{Bottrell} et~al.}{2017b}]{2017MNRAS.467.2879B}
{Bottrell} C.,  {Torrey} P.,  {Simard} L.,   {Ellison} S.~L.,  2017b, \mn@doi
  [\mnras] {10.1093/mnras/stx276}, \href
  {http://adsabs.harvard.edu/abs/2017MNRAS.467.2879B} {467, 2879}

\bibitem[\protect\citeauthoryear{{Cheung} et~al.,}{{Cheung}
  et~al.}{2012}]{2012ApJ...760..131C}
{Cheung} E.,  et~al., 2012, \mn@doi [\apj] {10.1088/0004-637X/760/2/131}, \href
  {http://adsabs.harvard.edu/abs/2012ApJ...760..131C} {760, 131}

\bibitem[\protect\citeauthoryear{{Ciambur}}{{Ciambur}}{2015}]{2015ApJ...810..120C}
{Ciambur} B.~C.,  2015, \mn@doi [\apj] {10.1088/0004-637X/810/2/120}, \href
  {http://adsabs.harvard.edu/abs/2015ApJ...810..120C} {810, 120}

\bibitem[\protect\citeauthoryear{{Conselice}}{{Conselice}}{2006}]{2006MNRAS.373.1389C}
{Conselice} C.~J.,  2006, \mn@doi [\mnras] {10.1111/j.1365-2966.2006.11114.x},
  \href {http://adsabs.harvard.edu/abs/2006MNRAS.373.1389C} {373, 1389}

\bibitem[\protect\citeauthoryear{{Conselice}, {Bershady}  \&
  {Jangren}}{{Conselice} et~al.}{2000}]{2000ApJ...529..886C}
{Conselice} C.~J.,  {Bershady} M.~A.,   {Jangren} A.,  2000, \mn@doi [\apj]
  {10.1086/308300}, \href {http://adsabs.harvard.edu/abs/2000ApJ...529..886C}
  {529, 886}

\bibitem[\protect\citeauthoryear{{Danovich}, {Dekel}, {Hahn}, {Ceverino}  \&
  {Primack}}{{Danovich} et~al.}{2015}]{2015MNRAS.449.2087D}
{Danovich} M.,  {Dekel} A.,  {Hahn} O.,  {Ceverino} D.,   {Primack} J.,  2015,
  \mn@doi [\mnras] {10.1093/mnras/stv270}, \href
  {http://adsabs.harvard.edu/abs/2015MNRAS.449.2087D} {449, 2087}

\bibitem[\protect\citeauthoryear{{Dickinson} et~al.,}{{Dickinson}
  et~al.}{2018}]{2018ApJ...853..194D}
{Dickinson} H.,  et~al., 2018, \mn@doi [\apj] {10.3847/1538-4357/aaa250}, \href
  {http://adsabs.harvard.edu/abs/2018ApJ...853..194D} {853, 194}

\bibitem[\protect\citeauthoryear{{Djorgovski} \& {Davis}}{{Djorgovski} \&
  {Davis}}{1987}]{1987ApJ...313...59D}
{Djorgovski} S.,  {Davis} M.,  1987, \mn@doi [\apj] {10.1086/164948}, \href
  {http://adsabs.harvard.edu/abs/1987ApJ...313...59D} {313, 59}

\bibitem[\protect\citeauthoryear{{Dressler}, {Lynden-Bell}, {Burstein},
  {Davies}, {Faber}, {Terlevich}  \& {Wegner}}{{Dressler}
  et~al.}{1987}]{1987ApJ...313...42D}
{Dressler} A.,  {Lynden-Bell} D.,  {Burstein} D.,  {Davies} R.~L.,  {Faber}
  S.~M.,  {Terlevich} R.,   {Wegner} G.,  1987, \mn@doi [\apj]
  {10.1086/164947}, \href {http://adsabs.harvard.edu/abs/1987ApJ...313...42D}
  {313, 42}

\bibitem[\protect\citeauthoryear{{Driver} et~al.,}{{Driver}
  et~al.}{2006}]{2006MNRAS.368..414D}
{Driver} S.~P.,  et~al., 2006, \mn@doi [\mnras]
  {10.1111/j.1365-2966.2006.10126.x}, \href
  {http://adsabs.harvard.edu/abs/2006MNRAS.368..414D} {368, 414}

\bibitem[\protect\citeauthoryear{{Ellison}, {Viswanathan}, {Patton},
  {Bottrell}, {McConnachie}, {Gwyn}  \& {Cuillandre}}{{Ellison}
  et~al.}{2019}]{Ellison_CFIS}
{Ellison} S.~L.,  {Viswanathan} A.,  {Patton} D.~R.,  {Bottrell} C.,
  {McConnachie} A.~W.,  {Gwyn} S.~D.~J.,   {Cuillandre} J.-C.,  2019, \mnras\
  submitted

\bibitem[\protect\citeauthoryear{{Elmegreen}, {Elmegreen}  \&
  {Montenegro}}{{Elmegreen} et~al.}{1992}]{1992ApJS...79...37E}
{Elmegreen} B.~G.,  {Elmegreen} D.~M.,   {Montenegro} L.,  1992, \mn@doi
  [\apjs] {10.1086/191643}, \href
  {http://adsabs.harvard.edu/abs/1992ApJS...79...37E} {79, 37}

\bibitem[\protect\citeauthoryear{{Emsellem}, {Greusard}, {Combes}, {Friedli},
  {Leon}, {P{\'e}contal}  \& {Wozniak}}{{Emsellem}
  et~al.}{2001}]{2001A&A...368...52E}
{Emsellem} E.,  {Greusard} D.,  {Combes} F.,  {Friedli} D.,  {Leon} S.,
  {P{\'e}contal} E.,   {Wozniak} H.,  2001, \mn@doi [\aap]
  {10.1051/0004-6361:20000523}, \href
  {http://adsabs.harvard.edu/abs/2001A%26A...368...52E} {368, 52}

\bibitem[\protect\citeauthoryear{{Faber} \& {Jackson}}{{Faber} \&
  {Jackson}}{1976}]{1976ApJ...204..668F}
{Faber} S.~M.,  {Jackson} R.~E.,  1976, \mn@doi [\apj] {10.1086/154215}, \href
  {http://adsabs.harvard.edu/abs/1976ApJ...204..668F} {204, 668}

\bibitem[\protect\citeauthoryear{{Faber} et~al.,}{{Faber}
  et~al.}{2007}]{2007ApJ...665..265F}
{Faber} S.~M.,  et~al., 2007, \mn@doi [\apj] {10.1086/519294}, \href
  {http://adsabs.harvard.edu/abs/2007ApJ...665..265F} {665, 265}

\bibitem[\protect\citeauthoryear{{Falc{\'o}n-Barroso}
  et~al.,}{{Falc{\'o}n-Barroso} et~al.}{2006}]{2006MNRAS.369..529F}
{Falc{\'o}n-Barroso} J.,  et~al., 2006, \mn@doi [\mnras]
  {10.1111/j.1365-2966.2006.10261.x}, \href
  {http://adsabs.harvard.edu/abs/2006MNRAS.369..529F} {369, 529}

\bibitem[\protect\citeauthoryear{{Fall} \& {Efstathiou}}{{Fall} \&
  {Efstathiou}}{1980}]{1980MNRAS.193..189F}
{Fall} S.~M.,  {Efstathiou} G.,  1980, \mn@doi [\mnras]
  {10.1093/mnras/193.2.189}, \href
  {http://adsabs.harvard.edu/abs/1980MNRAS.193..189F} {193, 189}

\bibitem[\protect\citeauthoryear{{Fang}, {Faber}, {Koo}  \& {Dekel}}{{Fang}
  et~al.}{2013}]{2013ApJ...776...63F}
{Fang} J.~J.,  {Faber} S.~M.,  {Koo} D.~C.,   {Dekel} A.,  2013, \mn@doi [\apj]
  {10.1088/0004-637X/776/1/63}, \href
  {http://adsabs.harvard.edu/abs/2013ApJ...776...63F} {776, 63}

\bibitem[\protect\citeauthoryear{{Fisher} \& {Drory}}{{Fisher} \&
  {Drory}}{2008}]{2008AJ....136..773F}
{Fisher} D.~B.,  {Drory} N.,  2008, \mn@doi [\aj]
  {10.1088/0004-6256/136/2/773}, \href
  {http://adsabs.harvard.edu/abs/2008AJ....136..773F} {136, 773}

\bibitem[\protect\citeauthoryear{{Fisher} \& {Drory}}{{Fisher} \&
  {Drory}}{2010}]{2010ApJ...716..942F}
{Fisher} D.~B.,  {Drory} N.,  2010, \mn@doi [\apj]
  {10.1088/0004-637X/716/2/942}, \href
  {http://adsabs.harvard.edu/abs/2010ApJ...716..942F} {716, 942}

\bibitem[\protect\citeauthoryear{{Fliri} \& {Trujillo}}{{Fliri} \&
  {Trujillo}}{2016}]{2016MNRAS.456.1359F}
{Fliri} J.,  {Trujillo} I.,  2016, \mn@doi [\mnras] {10.1093/mnras/stv2686},
  \href {http://adsabs.harvard.edu/abs/2016MNRAS.456.1359F} {456, 1359}

\bibitem[\protect\citeauthoryear{{Gadotti}}{{Gadotti}}{2009}]{2009MNRAS.393.1531G}
{Gadotti} D.~A.,  2009, \mn@doi [\mnras] {10.1111/j.1365-2966.2008.14257.x},
  \href {http://adsabs.harvard.edu/abs/2009MNRAS.393.1531G} {393, 1531}

\bibitem[\protect\citeauthoryear{{Graham}}{{Graham}}{2001}]{2001AJ....121..820G}
{Graham} A.~W.,  2001, \mn@doi [\aj] {10.1086/318767}, \href
  {http://adsabs.harvard.edu/abs/2001AJ....121..820G} {121, 820}

\bibitem[\protect\citeauthoryear{{Graham}}{{Graham}}{2013}]{2013pss6.book...91G}
{Graham} A.~W.,  2013, {Planets, Stars and Stellar Systems. Volume 6:
  Extragalactic Astronomy and Cosmology}.
Springer, p.~91, \mn@doi{10.1007/978-94-007-5609-0_2}

\bibitem[\protect\citeauthoryear{{Graham} \& {Guzm{\'a}n}}{{Graham} \&
  {Guzm{\'a}n}}{2003}]{2003AJ....125.2936G}
{Graham} A.~W.,  {Guzm{\'a}n} R.,  2003, \mn@doi [\aj] {10.1086/374992}, \href
  {http://adsabs.harvard.edu/abs/2003AJ....125.2936G} {125, 2936}

\bibitem[\protect\citeauthoryear{{Gunn} et~al.,}{{Gunn}
  et~al.}{2006}]{2006AJ....131.2332G}
{Gunn} J.~E.,  et~al., 2006, \mn@doi [\aj] {10.1086/500975}, \href
  {http://adsabs.harvard.edu/abs/2006AJ....131.2332G} {131, 2332}

\bibitem[\protect\citeauthoryear{{H{\"a}u{\ss}ler} et~al.,}{{H{\"a}u{\ss}ler}
  et~al.}{2013}]{2013MNRAS.430..330H}
{H{\"a}u{\ss}ler} B.,  et~al., 2013, \mn@doi [\mnras] {10.1093/mnras/sts633},
  \href {http://adsabs.harvard.edu/abs/2013MNRAS.430..330H} {430, 330}

\bibitem[\protect\citeauthoryear{{Hernquist}}{{Hernquist}}{1992}]{1992ApJ...400..460H}
{Hernquist} L.,  1992, \mn@doi [\apj] {10.1086/172009}, \href
  {http://adsabs.harvard.edu/abs/1992ApJ...400..460H} {400, 460}

\bibitem[\protect\citeauthoryear{{Hopkins}, {Hernquist}, {Cox}, {Dutta}  \&
  {Rothberg}}{{Hopkins} et~al.}{2008}]{2008ApJ...679..156H}
{Hopkins} P.~F.,  {Hernquist} L.,  {Cox} T.~J.,  {Dutta} S.~N.,   {Rothberg}
  B.,  2008, \mn@doi [\apj] {10.1086/587544}, \href
  {http://adsabs.harvard.edu/abs/2008ApJ...679..156H} {679, 156}

\bibitem[\protect\citeauthoryear{{Hopkins}, {Cox}, {Younger}  \&
  {Hernquist}}{{Hopkins} et~al.}{2009}]{2009ApJ...691.1168H}
{Hopkins} P.~F.,  {Cox} T.~J.,  {Younger} J.~D.,   {Hernquist} L.,  2009,
  \mn@doi [\apj] {10.1088/0004-637X/691/2/1168}, \href
  {http://adsabs.harvard.edu/abs/2009ApJ...691.1168H} {691, 1168}

\bibitem[\protect\citeauthoryear{{Hubble}}{{Hubble}}{1926}]{1926ApJ....64..321H}
{Hubble} E.~P.,  1926, \mn@doi [\apj] {10.1086/143018}, \href
  {http://adsabs.harvard.edu/abs/1926ApJ....64..321H} {64}

\bibitem[\protect\citeauthoryear{{Hubble}}{{Hubble}}{1936}]{1936rene.book.....H}
{Hubble} E.~P.,  1936, {Realm of the Nebulae}

\bibitem[\protect\citeauthoryear{{Jiang} et~al.,}{{Jiang}
  et~al.}{2014}]{2014ApJS..213...12J}
{Jiang} L.,  et~al., 2014, \mn@doi [\apjs] {10.1088/0067-0049/213/1/12}, \href
  {http://adsabs.harvard.edu/abs/2014ApJS..213...12J} {213, 12}

\bibitem[\protect\citeauthoryear{{Kauffmann} et~al.,}{{Kauffmann}
  et~al.}{2003a}]{2003MNRAS.341...33K}
{Kauffmann} G.,  et~al., 2003a, \mn@doi [\mnras]
  {10.1046/j.1365-8711.2003.06291.x}, \href
  {http://adsabs.harvard.edu/abs/2003MNRAS.341...33K} {341, 33}

\bibitem[\protect\citeauthoryear{{Kauffmann} et~al.,}{{Kauffmann}
  et~al.}{2003b}]{2003MNRAS.346.1055K}
{Kauffmann} G.,  et~al., 2003b, \mn@doi [\mnras]
  {10.1111/j.1365-2966.2003.07154.x}, \href
  {http://adsabs.harvard.edu/abs/2003MNRAS.346.1055K} {346, 1055}

\bibitem[\protect\citeauthoryear{{Kauffmann}, {Heckman}, {De Lucia},
  {Brinchmann}, {Charlot}, {Tremonti}, {White}  \& {Brinkmann}}{{Kauffmann}
  et~al.}{2006}]{2006MNRAS.367.1394K}
{Kauffmann} G.,  {Heckman} T.~M.,  {De Lucia} G.,  {Brinchmann} J.,  {Charlot}
  S.,  {Tremonti} C.,  {White} S.~D.~M.,   {Brinkmann} J.,  2006, \mn@doi
  [\mnras] {10.1111/j.1365-2966.2006.10061.x}, \href
  {http://adsabs.harvard.edu/abs/2006MNRAS.367.1394K} {367, 1394}

\bibitem[\protect\citeauthoryear{{Kaviraj}}{{Kaviraj}}{2010}]{2010MNRAS.406..382K}
{Kaviraj} S.,  2010, \mn@doi [\mnras] {10.1111/j.1365-2966.2010.16714.x}, \href
  {http://adsabs.harvard.edu/abs/2010MNRAS.406..382K} {406, 382}

\bibitem[\protect\citeauthoryear{{Kelvin} et~al.,}{{Kelvin}
  et~al.}{2012}]{2012MNRAS.421.1007K}
{Kelvin} L.~S.,  et~al., 2012, \mn@doi [\mnras]
  {10.1111/j.1365-2966.2012.20355.x}, \href
  {http://adsabs.harvard.edu/abs/2012MNRAS.421.1007K} {421, 1007}

\bibitem[\protect\citeauthoryear{{Kere{\v s}}, {Katz}, {Weinberg}  \&
  {Dav{\'e}}}{{Kere{\v s}} et~al.}{2005}]{2005MNRAS.363....2K}
{Kere{\v s}} D.,  {Katz} N.,  {Weinberg} D.~H.,   {Dav{\'e}} R.,  2005, \mn@doi
  [\mnras] {10.1111/j.1365-2966.2005.09451.x}, \href
  {http://adsabs.harvard.edu/abs/2005MNRAS.363....2K} {363, 2}

\bibitem[\protect\citeauthoryear{{Khosroshahi}, {Wadadekar}, {Kembhavi}  \&
  {Mobasher}}{{Khosroshahi} et~al.}{2000}]{2000ApJ...531L.103K}
{Khosroshahi} H.~G.,  {Wadadekar} Y.,  {Kembhavi} A.,   {Mobasher} B.,  2000,
  \mn@doi [\apjl] {10.1086/312547}, \href
  {http://adsabs.harvard.edu/abs/2000ApJ...531L.103K} {531, L103}

\bibitem[\protect\citeauthoryear{{Kormendy}}{{Kormendy}}{1977}]{1977ApJ...217..406K}
{Kormendy} J.,  1977, \mn@doi [\apj] {10.1086/155589}, \href
  {http://adsabs.harvard.edu/abs/1977ApJ...217..406K} {217, 406}

\bibitem[\protect\citeauthoryear{{Kormendy} \& {Bender}}{{Kormendy} \&
  {Bender}}{2012}]{2012ApJS..198....2K}
{Kormendy} J.,  {Bender} R.,  2012, \mn@doi [\apjs]
  {10.1088/0067-0049/198/1/2}, \href
  {http://adsabs.harvard.edu/abs/2012ApJS..198....2K} {198, 2}

\bibitem[\protect\citeauthoryear{{Kormendy} \& {Ho}}{{Kormendy} \&
  {Ho}}{2013}]{2013ARA&A..51..511K}
{Kormendy} J.,  {Ho} L.~C.,  2013, \mn@doi [\araa]
  {10.1146/annurev-astro-082708-101811}, \href
  {http://adsabs.harvard.edu/abs/2013ARA%26A..51..511K} {51, 511}

\bibitem[\protect\citeauthoryear{{Kormendy} \& {Kennicutt}}{{Kormendy} \&
  {Kennicutt}}{2004}]{2004ARA&A..42..603K}
{Kormendy} J.,  {Kennicutt} Jr. R.~C.,  2004, \mn@doi [\araa]
  {10.1146/annurev.astro.42.053102.134024}, \href
  {http://adsabs.harvard.edu/abs/2004ARA%26A..42..603K} {42, 603}

\bibitem[\protect\citeauthoryear{{Kormendy}, {Bender}  \& {Cornell}}{{Kormendy}
  et~al.}{2011}]{2011Natur.469..374K}
{Kormendy} J.,  {Bender} R.,   {Cornell} M.~E.,  2011, \mn@doi [\nat]
  {10.1038/nature09694}, \href
  {http://adsabs.harvard.edu/abs/2011Natur.469..374K} {469, 374}

\bibitem[\protect\citeauthoryear{{Kruk} et~al.,}{{Kruk}
  et~al.}{2018}]{2018MNRAS.473.4731K}
{Kruk} S.~J.,  et~al., 2018, \mn@doi [\mnras] {10.1093/mnras/stx2605}, \href
  {http://adsabs.harvard.edu/abs/2018MNRAS.473.4731K} {473, 4731}

\bibitem[\protect\citeauthoryear{{Lackner} \& {Gunn}}{{Lackner} \&
  {Gunn}}{2012}]{2012MNRAS.421.2277L}
{Lackner} C.~N.,  {Gunn} J.~E.,  2012, \mn@doi [\mnras]
  {10.1111/j.1365-2966.2012.20450.x}, \href
  {http://adsabs.harvard.edu/abs/2012MNRAS.421.2277L} {421, 2277}

\bibitem[\protect\citeauthoryear{{Lang} et~al.,}{{Lang}
  et~al.}{2014}]{2014ApJ...788...11L}
{Lang} P.,  et~al., 2014, \mn@doi [\apj] {10.1088/0004-637X/788/1/11}, \href
  {http://adsabs.harvard.edu/abs/2014ApJ...788...11L} {788, 11}

\bibitem[\protect\citeauthoryear{{Laurikainen}, {Salo}  \&
  {Buta}}{{Laurikainen} et~al.}{2005}]{2005MNRAS.362.1319L}
{Laurikainen} E.,  {Salo} H.,   {Buta} R.,  2005, \mn@doi [\mnras]
  {10.1111/j.1365-2966.2005.09404.x}, \href
  {http://adsabs.harvard.edu/abs/2005MNRAS.362.1319L} {362, 1319}

\bibitem[\protect\citeauthoryear{{Laurikainen}, {Salo}, {Buta}, {Knapen}  \&
  {Comer{\'o}n}}{{Laurikainen} et~al.}{2010}]{2010MNRAS.405.1089L}
{Laurikainen} E.,  {Salo} H.,  {Buta} R.,  {Knapen} J.~H.,   {Comer{\'o}n} S.,
  2010, \mn@doi [\mnras] {10.1111/j.1365-2966.2010.16521.x}, \href
  {http://adsabs.harvard.edu/abs/2010MNRAS.405.1089L} {405, 1089}

\bibitem[\protect\citeauthoryear{{Laurikainen}, {Salo}, {Athanassoula},
  {Bosma}, {Buta}  \& {Janz}}{{Laurikainen} et~al.}{2013}]{2013MNRAS.430.3489L}
{Laurikainen} E.,  {Salo} H.,  {Athanassoula} E.,  {Bosma} A.,  {Buta} R.,
  {Janz} J.,  2013, \mn@doi [\mnras] {10.1093/mnras/stt150}, \href
  {http://adsabs.harvard.edu/abs/2013MNRAS.430.3489L} {430, 3489}

\bibitem[\protect\citeauthoryear{{Lilly}, {Le Fevre}, {Crampton}, {Hammer}  \&
  {Tresse}}{{Lilly} et~al.}{1995}]{1995ApJ...455...50L}
{Lilly} S.~J.,  {Le Fevre} O.,  {Crampton} D.,  {Hammer} F.,   {Tresse} L.,
  1995, \mn@doi [\apj] {10.1086/176555}, \href
  {http://adsabs.harvard.edu/abs/1995ApJ...455...50L} {455, 50}

\bibitem[\protect\citeauthoryear{{Lintott} et~al.,}{{Lintott}
  et~al.}{2008}]{2008MNRAS.389.1179L}
{Lintott} C.~J.,  et~al., 2008, \mn@doi [\mnras]
  {10.1111/j.1365-2966.2008.13689.x}, \href
  {http://adsabs.harvard.edu/abs/2008MNRAS.389.1179L} {389, 1179}

\bibitem[\protect\citeauthoryear{{Lisker}}{{Lisker}}{2008}]{2008ApJS..179..319L}
{Lisker} T.,  2008, \mn@doi [\apjs] {10.1086/591795}, \href
  {http://adsabs.harvard.edu/abs/2008ApJS..179..319L} {179, 319}

\bibitem[\protect\citeauthoryear{{Lotz}, {Primack}  \& {Madau}}{{Lotz}
  et~al.}{2004}]{2004AJ....128..163L}
{Lotz} J.~M.,  {Primack} J.,   {Madau} P.,  2004, \mn@doi [\aj]
  {10.1086/421849}, \href {http://adsabs.harvard.edu/abs/2004AJ....128..163L}
  {128, 163}

\bibitem[\protect\citeauthoryear{{Lotz}, {Madau}, {Giavalisco}, {Primack}  \&
  {Ferguson}}{{Lotz} et~al.}{2006}]{2006ApJ...636..592L}
{Lotz} J.~M.,  {Madau} P.,  {Giavalisco} M.,  {Primack} J.,   {Ferguson} H.~C.,
   2006, \mn@doi [\apj] {10.1086/497950}, \href
  {http://adsabs.harvard.edu/abs/2006ApJ...636..592L} {636, 592}

\bibitem[\protect\citeauthoryear{{Lupton}, {Gunn}, {Ivezi{\'c}}, {Knapp}  \&
  {Kent}}{{Lupton} et~al.}{2001}]{2001ASPC..238..269L}
{Lupton} R.,  {Gunn} J.~E.,  {Ivezi{\'c}} Z.,  {Knapp} G.~R.,   {Kent} S.,
  2001, in {Harnden} Jr. F.~R.,  {Primini} F.~A.,   {Payne} H.~E.,  eds,
  Astronomical Society of the Pacific Conference Series Vol. 238, Astronomical
  Data Analysis Software and Systems X. p.~269 (\mn@eprint {}
  {astro-ph/0101420})

\bibitem[\protect\citeauthoryear{{Lupton}, {Ivezic}, {Gunn}, {Knapp}, {Strauss}
   \& {Yasuda}}{{Lupton} et~al.}{2002}]{2002SPIE.4836..350L}
{Lupton} R.~H.,  {Ivezic} Z.,  {Gunn} J.~E.,  {Knapp} G.,  {Strauss} M.~A.,
  {Yasuda} N.,  2002, in {Tyson} J.~A.,  {Wolff} S.,  eds,  \procspie Vol.
  4836, Survey and Other Telescope Technologies and Discoveries. pp 350--356,
  \mn@doi{10.1117/12.457307}

\bibitem[\protect\citeauthoryear{{Lupton}, {Ivezi{\'c}}, {Gunn}, {Knapp}  \&
  {Strauss}}{{Lupton} et~al.}{2012}]{2012photolite}
{Lupton} R.~H.,  {Ivezi{\'c}} {\v Z}.,  {Gunn} J.~E.,  {Knapp} G.~R.,
  {Strauss} M.~A.,  2012, The photo-lite draft, plus other notes at RHL's Web
  site \url{http://www. astro.princeton.edu/~rhl/photo-lite.pdf}

\bibitem[\protect\citeauthoryear{{Lynden-Bell}}{{Lynden-Bell}}{1967}]{1967MNRAS.136..101L}
{Lynden-Bell} D.,  1967, \mn@doi [\mnras] {10.1093/mnras/136.1.101}, \href
  {http://adsabs.harvard.edu/abs/1967MNRAS.136..101L} {136, 101}

\bibitem[\protect\citeauthoryear{{MacArthur}, {Courteau}  \&
  {Holtzman}}{{MacArthur} et~al.}{2003}]{2003ApJ...582..689M}
{MacArthur} L.~A.,  {Courteau} S.,   {Holtzman} J.~A.,  2003, \mn@doi [\apj]
  {10.1086/344506}, \href {http://adsabs.harvard.edu/abs/2003ApJ...582..689M}
  {582, 689}

\bibitem[\protect\citeauthoryear{{M{\'a}rquez}, {Masegosa}, {Durret},
  {Gonz{\'a}lez Delgado}, {Moles}, {Maza}, {P{\'e}rez}  \&
  {Roth}}{{M{\'a}rquez} et~al.}{2003}]{2003A&A...409..459M}
{M{\'a}rquez} I.,  {Masegosa} J.,  {Durret} F.,  {Gonz{\'a}lez Delgado} R.~M.,
  {Moles} M.,  {Maza} J.,  {P{\'e}rez} E.,   {Roth} M.,  2003, \mn@doi [\aap]
  {10.1051/0004-6361:20031059}, \href
  {http://adsabs.harvard.edu/abs/2003A%26A...409..459M} {409, 459}

\bibitem[\protect\citeauthoryear{{Martin}, {Kaviraj}, {Devriendt}, {Dubois}  \&
  {Pichon}}{{Martin} et~al.}{2018}]{2018MNRAS.480.2266M}
{Martin} G.,  {Kaviraj} S.,  {Devriendt} J.~E.~G.,  {Dubois} Y.,   {Pichon} C.,
   2018, \mn@doi [\mnras] {10.1093/mnras/sty1936}, \href
  {http://adsabs.harvard.edu/abs/2018MNRAS.480.2266M} {480, 2266}

\bibitem[\protect\citeauthoryear{{Meert}, {Vikram}  \& {Bernardi}}{{Meert}
  et~al.}{2013}]{2013MNRAS.433.1344M}
{Meert} A.,  {Vikram} V.,   {Bernardi} M.,  2013, \mn@doi [\mnras]
  {10.1093/mnras/stt822}, \href
  {http://adsabs.harvard.edu/abs/2013MNRAS.433.1344M} {433, 1344}

\bibitem[\protect\citeauthoryear{{Meert}, {Vikram}  \& {Bernardi}}{{Meert}
  et~al.}{2015}]{2015MNRAS.446.3943M}
{Meert} A.,  {Vikram} V.,   {Bernardi} M.,  2015, \mn@doi [\mnras]
  {10.1093/mnras/stu2333}, \href
  {http://adsabs.harvard.edu/abs/2015MNRAS.446.3943M} {446, 3943}

\bibitem[\protect\citeauthoryear{{Mendel}, {Simard}, {Palmer}, {Ellison}  \&
  {Patton}}{{Mendel} et~al.}{2014}]{2014ApJS..210....3M}
{Mendel} J.~T.,  {Simard} L.,  {Palmer} M.,  {Ellison} S.~L.,   {Patton} D.~R.,
   2014, \mn@doi [\apjs] {10.1088/0067-0049/210/1/3}, \href
  {http://adsabs.harvard.edu/abs/2014ApJS..210....3M} {210, 3}

\bibitem[\protect\citeauthoryear{{M{\'e}ndez-Abreu}, {Aguerri}, {Corsini}  \&
  {Simonneau}}{{M{\'e}ndez-Abreu} et~al.}{2008}]{2008A&A...478..353M}
{M{\'e}ndez-Abreu} J.,  {Aguerri} J.~A.~L.,  {Corsini} E.~M.,   {Simonneau} E.,
   2008, \mn@doi [\aap] {10.1051/0004-6361:20078089}, \href
  {http://adsabs.harvard.edu/abs/2008A%26A...478..353M} {478, 353}

\bibitem[\protect\citeauthoryear{{M{\"o}llenhoff} \& {Heidt}}{{M{\"o}llenhoff}
  \& {Heidt}}{2001}]{2001A&A...368...16M}
{M{\"o}llenhoff} C.,  {Heidt} J.,  2001, \mn@doi [\aap]
  {10.1051/0004-6361:20000335}, \href
  {http://adsabs.harvard.edu/abs/2001A%26A...368...16M} {368, 16}

\bibitem[\protect\citeauthoryear{{Moorthy} \& {Holtzman}}{{Moorthy} \&
  {Holtzman}}{2006}]{2006MNRAS.371..583M}
{Moorthy} B.~K.,  {Holtzman} J.~A.,  2006, \mn@doi [\mnras]
  {10.1111/j.1365-2966.2006.10722.x}, \href
  {http://adsabs.harvard.edu/abs/2006MNRAS.371..583M} {371, 583}

\bibitem[\protect\citeauthoryear{{Negroponte} \& {White}}{{Negroponte} \&
  {White}}{1983}]{1983MNRAS.205.1009N}
{Negroponte} J.,  {White} S.~D.~M.,  1983, \mn@doi [\mnras]
  {10.1093/mnras/205.4.1009}, \href
  {http://adsabs.harvard.edu/abs/1983MNRAS.205.1009N} {205, 1009}

\bibitem[\protect\citeauthoryear{{Oke} \& {Gunn}}{{Oke} \&
  {Gunn}}{1983}]{1983ApJ...266..713O}
{Oke} J.~B.,  {Gunn} J.~E.,  1983, \mn@doi [\apj] {10.1086/160817}, \href
  {http://adsabs.harvard.edu/abs/1983ApJ...266..713O} {266, 713}

\bibitem[\protect\citeauthoryear{{Omand}, {Balogh}  \& {Poggianti}}{{Omand}
  et~al.}{2014}]{2014MNRAS.440..843O}
{Omand} C.~M.~B.,  {Balogh} M.~L.,   {Poggianti} B.~M.,  2014, \mn@doi [\mnras]
  {10.1093/mnras/stu331}, \href
  {http://adsabs.harvard.edu/abs/2014MNRAS.440..843O} {440, 843}

\bibitem[\protect\citeauthoryear{{Patton}, {Ellison}, {Simard}, {McConnachie}
  \& {Mendel}}{{Patton} et~al.}{2011}]{2011MNRAS.412..591P}
{Patton} D.~R.,  {Ellison} S.~L.,  {Simard} L.,  {McConnachie} A.~W.,
  {Mendel} J.~T.,  2011, \mn@doi [\mnras] {10.1111/j.1365-2966.2010.17932.x},
  \href {http://adsabs.harvard.edu/abs/2011MNRAS.412..591P} {412, 591}

\bibitem[\protect\citeauthoryear{{Pawlik}, {Wild}, {Walcher}, {Johansson},
  {Villforth}, {Rowlands}, {Mendez-Abreu}  \& {Hewlett}}{{Pawlik}
  et~al.}{2016}]{2016MNRAS.456.3032P}
{Pawlik} M.~M.,  {Wild} V.,  {Walcher} C.~J.,  {Johansson} P.~H.,  {Villforth}
  C.,  {Rowlands} K.,  {Mendez-Abreu} J.,   {Hewlett} T.,  2016, \mn@doi
  [\mnras] {10.1093/mnras/stv2878}, \href
  {http://adsabs.harvard.edu/abs/2016MNRAS.456.3032P} {456, 3032}

\bibitem[\protect\citeauthoryear{{Peletier} et~al.,}{{Peletier}
  et~al.}{2007}]{2007MNRAS.379..445P}
{Peletier} R.~F.,  et~al., 2007, \mn@doi [\mnras]
  {10.1111/j.1365-2966.2007.11860.x}, \href
  {http://adsabs.harvard.edu/abs/2007MNRAS.379..445P} {379, 445}

\bibitem[\protect\citeauthoryear{{Peng}, {Ho}, {Impey}  \& {Rix}}{{Peng}
  et~al.}{2002}]{2002AJ....124..266P}
{Peng} C.~Y.,  {Ho} L.~C.,  {Impey} C.~D.,   {Rix} H.-W.,  2002, \mn@doi [\aj]
  {10.1086/340952}, \href {http://adsabs.harvard.edu/abs/2002AJ....124..266P}
  {124, 266}

\bibitem[\protect\citeauthoryear{{Peng}, {Ho}, {Impey}  \& {Rix}}{{Peng}
  et~al.}{2010}]{2010AJ....139.2097P}
{Peng} C.~Y.,  {Ho} L.~C.,  {Impey} C.~D.,   {Rix} H.-W.,  2010, \mn@doi [\aj]
  {10.1088/0004-6256/139/6/2097}, \href
  {http://adsabs.harvard.edu/abs/2010AJ....139.2097P} {139, 2097}

\bibitem[\protect\citeauthoryear{{Pillepich} et~al.,}{{Pillepich}
  et~al.}{2018}]{2018MNRAS.473.4077P}
{Pillepich} A.,  et~al., 2018, \mn@doi [\mnras] {10.1093/mnras/stx2656}, \href
  {http://adsabs.harvard.edu/abs/2018MNRAS.473.4077P} {473, 4077}

\bibitem[\protect\citeauthoryear{{Robotham}, {Taranu}, {Tobar}, {Moffett}  \&
  {Driver}}{{Robotham} et~al.}{2017}]{2017MNRAS.466.1513R}
{Robotham} A.~S.~G.,  {Taranu} D.~S.,  {Tobar} R.,  {Moffett} A.,   {Driver}
  S.~P.,  2017, \mn@doi [\mnras] {10.1093/mnras/stw3039}, \href
  {http://adsabs.harvard.edu/abs/2017MNRAS.466.1513R} {466, 1513}

\bibitem[\protect\citeauthoryear{{Rodriguez-Gomez} et~al.,}{{Rodriguez-Gomez}
  et~al.}{2017}]{2017MNRAS.467.3083R}
{Rodriguez-Gomez} V.,  et~al., 2017, \mn@doi [\mnras] {10.1093/mnras/stx305},
  \href {http://adsabs.harvard.edu/abs/2017MNRAS.467.3083R} {467, 3083}

\bibitem[\protect\citeauthoryear{{Rodriguez-Gomez} et~al.,}{{Rodriguez-Gomez}
  et~al.}{2019}]{2019MNRAS.483.4140R}
{Rodriguez-Gomez} V.,  et~al., 2019, \mn@doi [\mnras] {10.1093/mnras/sty3345},
  \href {http://adsabs.harvard.edu/abs/2019MNRAS.483.4140R} {483, 4140}

\bibitem[\protect\citeauthoryear{{Sales}, {Navarro}, {Theuns}, {Schaye},
  {White}, {Frenk}, {Crain}  \& {Dalla Vecchia}}{{Sales}
  et~al.}{2012}]{2012MNRAS.423.1544S}
{Sales} L.~V.,  {Navarro} J.~F.,  {Theuns} T.,  {Schaye} J.,  {White} S.~D.~M.,
   {Frenk} C.~S.,  {Crain} R.~A.,   {Dalla Vecchia} C.,  2012, \mn@doi [\mnras]
  {10.1111/j.1365-2966.2012.20975.x}, \href
  {http://adsabs.harvard.edu/abs/2012MNRAS.423.1544S} {423, 1544}

\bibitem[\protect\citeauthoryear{{Sandage}}{{Sandage}}{1961}]{1961hag..book.....S}
{Sandage} A.,  1961, {The Hubble atlas of galaxies}

\bibitem[\protect\citeauthoryear{{Schade}, {Lilly}, {Crampton}, {Hammer}, {Le
  Fevre}  \& {Tresse}}{{Schade} et~al.}{1995}]{1995ApJ...451L...1S}
{Schade} D.,  {Lilly} S.~J.,  {Crampton} D.,  {Hammer} F.,  {Le Fevre} O.,
  {Tresse} L.,  1995, \mn@doi [\apjl] {10.1086/309677}, \href
  {http://adsabs.harvard.edu/abs/1995ApJ...451L...1S} {451, L1}

\bibitem[\protect\citeauthoryear{{Schaye} et~al.,}{{Schaye}
  et~al.}{2015}]{2015MNRAS.446..521S}
{Schaye} J.,  et~al., 2015, \mn@doi [\mnras] {10.1093/mnras/stu2058}, \href
  {http://adsabs.harvard.edu/abs/2015MNRAS.446..521S} {446, 521}

\bibitem[\protect\citeauthoryear{{Schiminovich} et~al.,}{{Schiminovich}
  et~al.}{2007}]{2007ApJS..173..315S}
{Schiminovich} D.,  et~al., 2007, \mn@doi [\apjs] {10.1086/524659}, \href
  {http://adsabs.harvard.edu/abs/2007ApJS..173..315S} {173, 315}

\bibitem[\protect\citeauthoryear{{Schlegel}, {Finkbeiner}  \&
  {Davis}}{{Schlegel} et~al.}{1998}]{1998ApJ...500..525S}
{Schlegel} D.~J.,  {Finkbeiner} D.~P.,   {Davis} M.,  1998, \mn@doi [\apj]
  {10.1086/305772}, \href {http://adsabs.harvard.edu/abs/1998ApJ...500..525S}
  {500, 525}

\bibitem[\protect\citeauthoryear{{Simard} et~al.,}{{Simard}
  et~al.}{2002}]{2002ApJS..142....1S}
{Simard} L.,  et~al., 2002, \mn@doi [\apjs] {10.1086/341399}, \href
  {http://adsabs.harvard.edu/abs/2002ApJS..142....1S} {142, 1}

\bibitem[\protect\citeauthoryear{{Simard}, {Mendel}, {Patton}, {Ellison}  \&
  {McConnachie}}{{Simard} et~al.}{2011}]{2011ApJS..196...11S}
{Simard} L.,  {Mendel} J.~T.,  {Patton} D.~R.,  {Ellison} S.~L.,
  {McConnachie} A.~W.,  2011, \mn@doi [\apjs] {10.1088/0067-0049/196/1/11},
  \href {http://adsabs.harvard.edu/abs/2011ApJS..196...11S} {196, 11}

\bibitem[\protect\citeauthoryear{{Snyder} et~al.,}{{Snyder}
  et~al.}{2015}]{2015MNRAS.454.1886S}
{Snyder} G.~F.,  et~al., 2015, \mn@doi [\mnras] {10.1093/mnras/stv2078}, \href
  {http://adsabs.harvard.edu/abs/2015MNRAS.454.1886S} {454, 1886}

\bibitem[\protect\citeauthoryear{{Stewart}, {Brooks}, {Bullock}, {Maller},
  {Diemand}, {Wadsley}  \& {Moustakas}}{{Stewart}
  et~al.}{2013}]{2013ApJ...769...74S}
{Stewart} K.~R.,  {Brooks} A.~M.,  {Bullock} J.~S.,  {Maller} A.~H.,  {Diemand}
  J.,  {Wadsley} J.,   {Moustakas} L.~A.,  2013, \mn@doi [\apj]
  {10.1088/0004-637X/769/1/74}, \href
  {http://adsabs.harvard.edu/abs/2013ApJ...769...74S} {769, 74}

\bibitem[\protect\citeauthoryear{{Stoughton} et~al.,}{{Stoughton}
  et~al.}{2002}]{2002AJ....123..485S}
{Stoughton} C.,  et~al., 2002, \mn@doi [\aj] {10.1086/324741}, \href
  {http://adsabs.harvard.edu/abs/2002AJ....123..485S} {123, 485}

\bibitem[\protect\citeauthoryear{{Strateva} et~al.,}{{Strateva}
  et~al.}{2001}]{2001AJ....122.1861S}
{Strateva} I.,  et~al., 2001, \mn@doi [\aj] {10.1086/323301}, \href
  {http://adsabs.harvard.edu/abs/2001AJ....122.1861S} {122, 1861}

\bibitem[\protect\citeauthoryear{{Strauss} et~al.,}{{Strauss}
  et~al.}{2002}]{2002AJ....124.1810S}
{Strauss} M.~A.,  et~al., 2002, \mn@doi [\aj] {10.1086/342343}, \href
  {http://adsabs.harvard.edu/abs/2002AJ....124.1810S} {124, 1810}

\bibitem[\protect\citeauthoryear{{Tal} \& {van Dokkum}}{{Tal} \& {van
  Dokkum}}{2011}]{2011ApJ...731...89T}
{Tal} T.,  {van Dokkum} P.~G.,  2011, \mn@doi [\apj]
  {10.1088/0004-637X/731/2/89}, \href
  {http://adsabs.harvard.edu/abs/2011ApJ...731...89T} {731, 89}

\bibitem[\protect\citeauthoryear{{Taranu}, {Dubinski}  \& {Yee}}{{Taranu}
  et~al.}{2013}]{2013ApJ...778...61T}
{Taranu} D.~S.,  {Dubinski} J.,   {Yee} H.~K.~C.,  2013, \mn@doi [\apj]
  {10.1088/0004-637X/778/1/61}, \href
  {http://adsabs.harvard.edu/abs/2013ApJ...778...61T} {778, 61}

\bibitem[\protect\citeauthoryear{{Teimoorinia}, {Bluck}  \&
  {Ellison}}{{Teimoorinia} et~al.}{2016}]{2016MNRAS.457.2086T}
{Teimoorinia} H.,  {Bluck} A.~F.~L.,   {Ellison} S.~L.,  2016, \mn@doi [\mnras]
  {10.1093/mnras/stw036}, \href
  {http://adsabs.harvard.edu/abs/2016MNRAS.457.2086T} {457, 2086}

\bibitem[\protect\citeauthoryear{{Toomre}}{{Toomre}}{1977}]{1977egsp.conf..401T}
{Toomre} A.,  1977, in {Tinsley} B.~M.,  {Larson} D.~Campbell R.~B.~G.,  eds,
  Evolution of Galaxies and Stellar Populations. p.~401

\bibitem[\protect\citeauthoryear{{Toomre} \& {Toomre}}{{Toomre} \&
  {Toomre}}{1972}]{1972ApJ...178..623T}
{Toomre} A.,  {Toomre} J.,  1972, \mn@doi [\apj] {10.1086/151823}, \href
  {http://adsabs.harvard.edu/abs/1972ApJ...178..623T} {178, 623}

\bibitem[\protect\citeauthoryear{{Vika}, {Bamford}, {H{\"a}u{\ss}ler}, {Rojas},
  {Borch}  \& {Nichol}}{{Vika} et~al.}{2013}]{2013MNRAS.435..623V}
{Vika} M.,  {Bamford} S.~P.,  {H{\"a}u{\ss}ler} B.,  {Rojas} A.~L.,  {Borch}
  A.,   {Nichol} R.~C.,  2013, \mn@doi [\mnras] {10.1093/mnras/stt1320}, \href
  {http://adsabs.harvard.edu/abs/2013MNRAS.435..623V} {435, 623}

\bibitem[\protect\citeauthoryear{{Vogelsberger} et~al.,}{{Vogelsberger}
  et~al.}{2014}]{2014MNRAS.444.1518V}
{Vogelsberger} M.,  et~al., 2014, \mn@doi [\mnras] {10.1093/mnras/stu1536},
  \href {http://adsabs.harvard.edu/abs/2014MNRAS.444.1518V} {444, 1518}

\bibitem[\protect\citeauthoryear{{White} \& {Rees}}{{White} \&
  {Rees}}{1978}]{1978MNRAS.183..341W}
{White} S.~D.~M.,  {Rees} M.~J.,  1978, \mn@doi [\mnras]
  {10.1093/mnras/183.3.341}, \href
  {http://adsabs.harvard.edu/abs/1978MNRAS.183..341W} {183, 341}

\bibitem[\protect\citeauthoryear{{Willett} et~al.,}{{Willett}
  et~al.}{2015}]{2015MNRAS.449..820W}
{Willett} K.~W.,  et~al., 2015, \mn@doi [\mnras] {10.1093/mnras/stv307}, \href
  {http://adsabs.harvard.edu/abs/2015MNRAS.449..820W} {449, 820}

\bibitem[\protect\citeauthoryear{{Woo}, {Dekel}, {Faber}  \& {Koo}}{{Woo}
  et~al.}{2015}]{2015MNRAS.448..237W}
{Woo} J.,  {Dekel} A.,  {Faber} S.~M.,   {Koo} D.~C.,  2015, \mn@doi [\mnras]
  {10.1093/mnras/stu2755}, \href
  {http://adsabs.harvard.edu/abs/2015MNRAS.448..237W} {448, 237}

\bibitem[\protect\citeauthoryear{{York} et~al.,}{{York}
  et~al.}{2000}]{2000AJ....120.1579Y}
{York} D.~G.,  et~al., 2000, \mn@doi [\aj] {10.1086/301513}, \href
  {http://adsabs.harvard.edu/abs/2000AJ....120.1579Y} {120, 1579}

\bibitem[\protect\citeauthoryear{{de Jong}}{{de
  Jong}}{1996}]{1996A&A...313...45D}
{de Jong} R.~S.,  1996, \aap, \href
  {http://adsabs.harvard.edu/abs/1996A%26A...313...45D} {313, 45}

\bibitem[\protect\citeauthoryear{{de Souza}, {Gadotti}  \& {dos Anjos}}{{de
  Souza} et~al.}{2004}]{2004ApJS..153..411D}
{de Souza} R.~E.,  {Gadotti} D.~A.,   {dos Anjos} S.,  2004, \mn@doi [\apjs]
  {10.1086/421554}, \href {http://adsabs.harvard.edu/abs/2004ApJS..153..411D}
  {153, 411}

\bibitem[\protect\citeauthoryear{{de Vaucouleurs}}{{de
  Vaucouleurs}}{1959}]{1959HDP....53..275D}
{de Vaucouleurs} G.,  1959, Handbuch der Physik, \href
  {http://adsabs.harvard.edu/abs/1959HDP....53..275D} {53, 275}

\makeatother
\end{thebibliography}


\appendix

\section{Measurement uncertainties}\label{sec:appA}

We compute the photometric uncertainties on several model parameters both as a reference and to illustrate the superior constraints on measurements using Stripe 82 images with respect to Legacy. The uncertainties we report are propagated uncertainties from the local mean sky measurement, $\sigma_{\langle \mathrm{sky} \rangle}$, and the \textsc{gim2d} parametric uncertainties computed by marginally sampling the posterior probability distribution for each parameter about the best-fitting \texttt{n4} model. Uncertainties from covariances between model parameters are not accounted for, however these should be more or less similar in each dataset given that the same decomposition routines and models are used in the comparison. Our catalogs include all of the necessary information needed to reproduce these uncertainty estimates. 

\begin{figure*}
  \includegraphics[width=\linewidth]{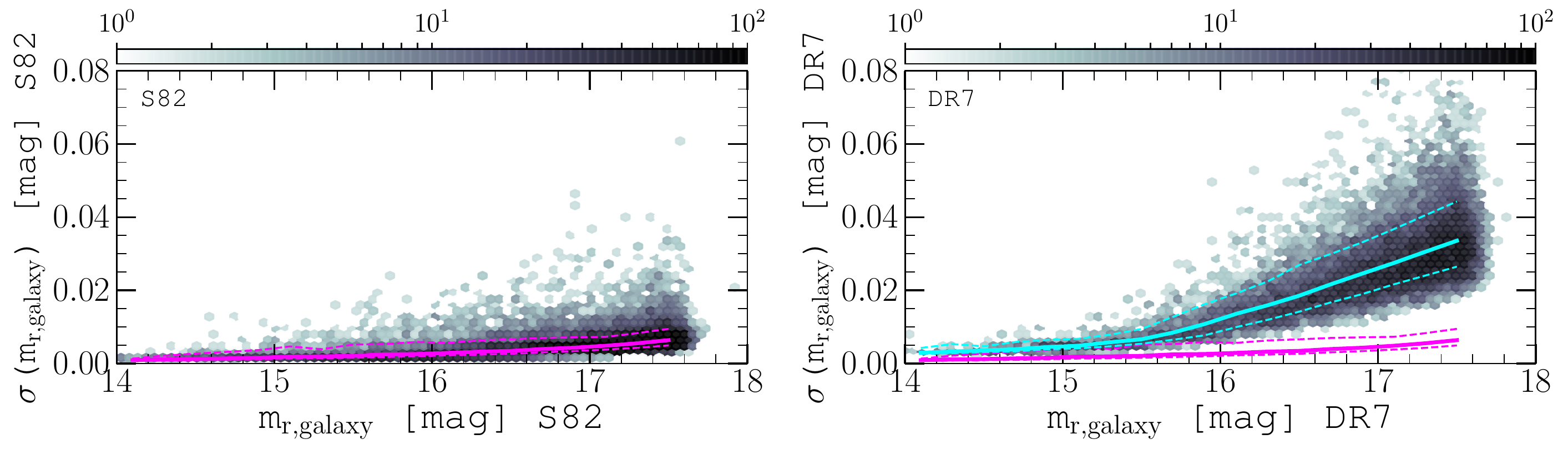}
  \includegraphics[width=\linewidth]{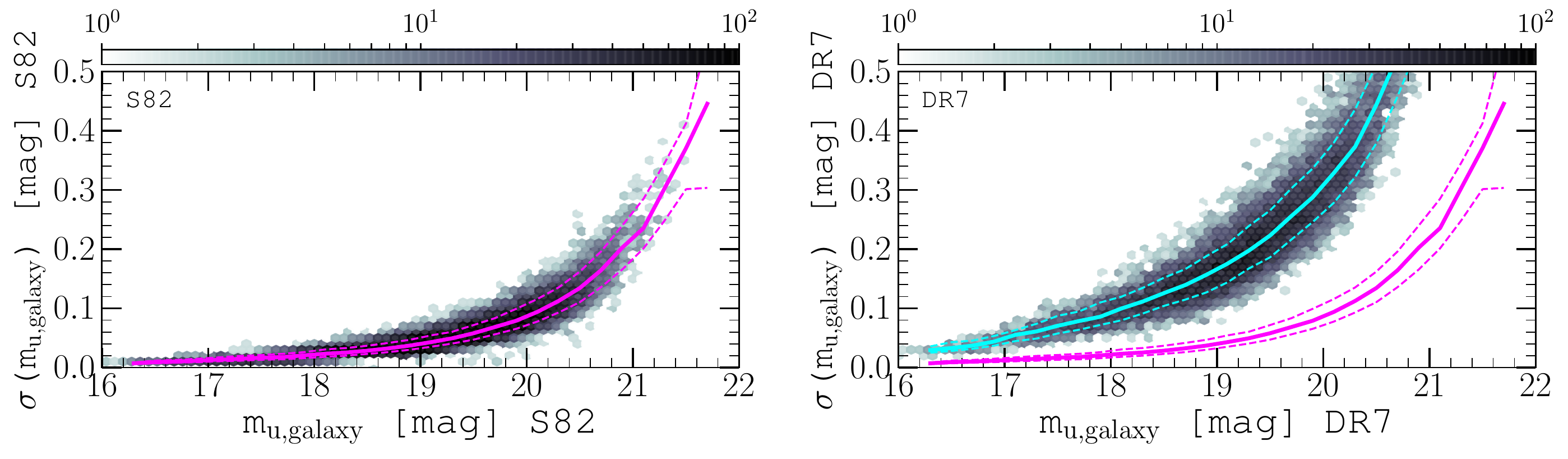}

\caption[Uncertainties]{Comparison of photometric total magnitude uncertainties, $\sigma^2($m$_{x,\mathrm{galaxy}}) = \sigma^2_{\langle\mathrm{sky}\rangle}+\sigma^2_{\textsc{gim2d}}$, for Stripe 82 (left panels, magenta curves) and Legacy (right panels, cyan curves). Upper and lower rows show the $r$ and $u$ band magnitude uncertainties. Stripe 82 curves are redrawn on the Legacy panels for visual impression. It is important to note the difference in scale between the upper and lower rows.}
\label{fig:mag_unc}
\end{figure*}

Figure \ref{fig:mag_unc} compares the $u$ (lower row) and $r$ (upper row) band total magnitude uncertainties, $\sigma($m$_{x,\mathrm{galaxy}})$, in Stripe 82 (left panels, magenta curves) and Legacy (right panels, cyan curves). For visual impression, the magenta curves for Stripe 82 are redrawn on the right panels. Background greyscale shows the 2D histogram of computed uncertainties. Curves show the median (solid, thick) and 16-84 percentile range (dashed). From Figure \ref{fig:mag_unc} it is clear that the constraints on total magnitudes are substantially better in the deep co-adds. In the $u$-band, for example, we can see that a galaxy with $m_{u\mathrm{galaxy}}\approx19$ in Stripe 82 has the same uncertainty (around 0.03 mag) as one at $m_{u\mathrm{galaxy}}\approx16$ in the single-epoch Legacy images.

\begin{figure*}
  \includegraphics[width=\linewidth]{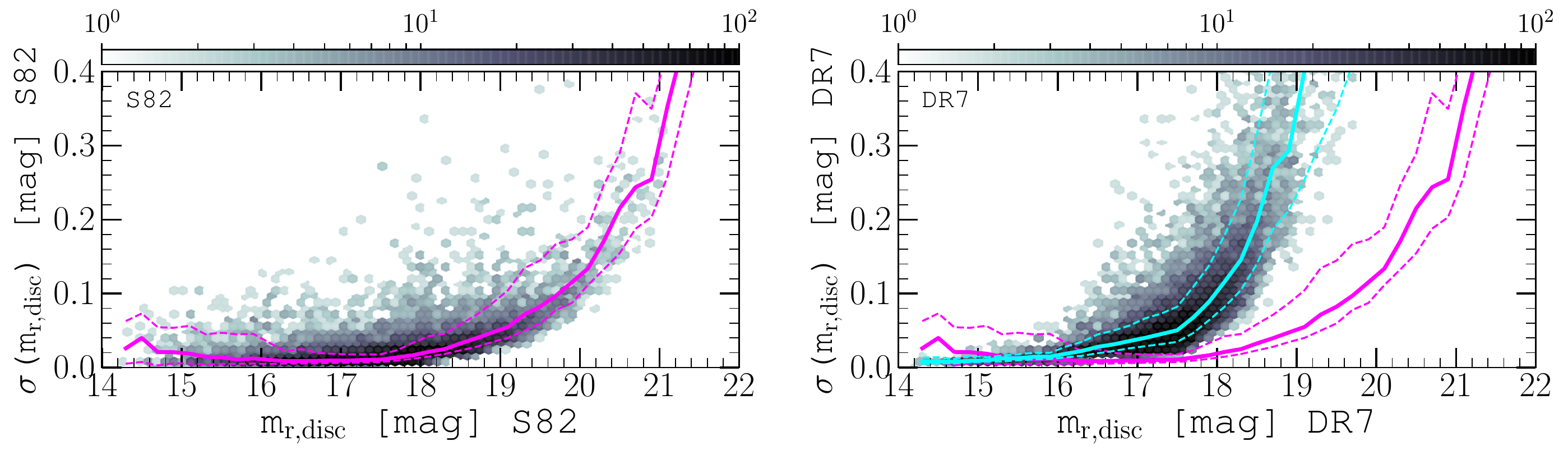}
  \includegraphics[width=\linewidth]{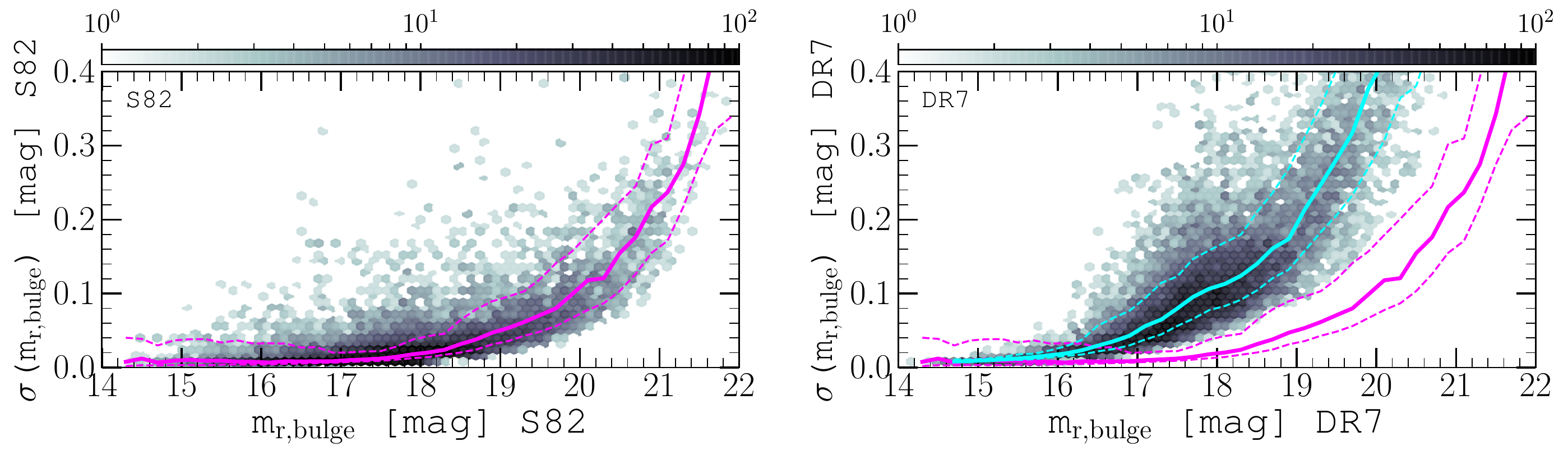}
  \includegraphics[width=\linewidth]{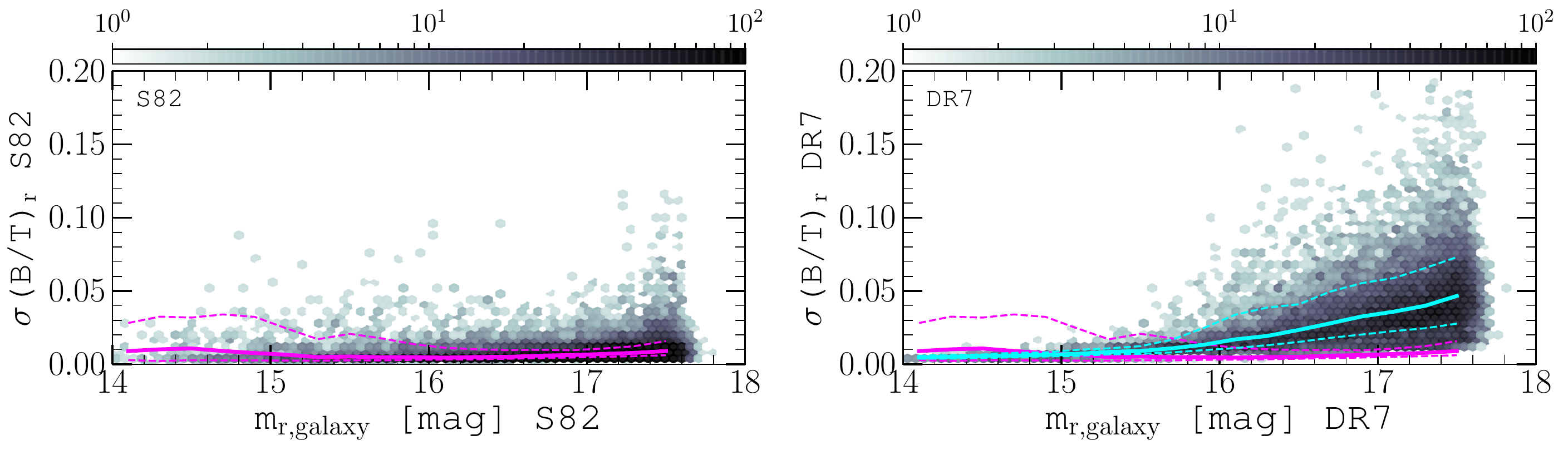}

\caption[Uncertainties]{Comparison of photometric component magnitude and bulge-to-total fraction uncertainties for Stripe 82 (left panels, magenta curves) and Legacy (right panels, cyan curves) in the $r$-band. Stripe 82 curves are redrawn on the Legacy panels for visual impression.}
\label{fig:comp_unc}
\end{figure*}

Figure \ref{fig:comp_unc} shows the photometric uncertainties for the bulge, $\sigma(m_{r,\mathrm{bulge}})$, and disc, $\sigma(m_{r,\mathrm{disc}})$, component magnitudes and the bulge-to-total ratio, $\sigma(B/T)_r$ in the $r$-band. The bottom right panel shows that Legacy $B/T$ uncertainties are small out to $m_{r,\mathrm{galaxy}} \approx 15.5$ mag -- after which the uncertainties begin to rise steadily. Stripe 82, however, has median $B/T$ uncertainties that are typically confined to $\sigma(B/T)_r\lesssim0.01$ over the full magnitude range. This result is consistent with our assertion that the improved depth in Stripe 82 enables characterization of $B/T$ in galaxies for which the photometry was insufficient to discriminate between bulge and disc light in Legacy. It also supports the claim that Stripe 82 should be more discriminating in statistical comparisons designed to show whether a two-component bulge+disc decomposition is favoured over a single-component fit -- particularly for faint targets. However, there is an interesting offset in the median Stripe 82 $B/T$ uncertainties relative to Legacy where $m_{r,\mathrm{galaxy}} \lesssim 15.5$ mag and which appears to increase with brightness. To test whether these high $B/T$ uncertainties came from the Stripe 82 sky or \textsc{gim2d} uncertainties we turned off the sky uncertainties in our calculations and found that the systematic persisted. The large scatter in the Stripe 82 $B/T$ uncertainties of bright galaxies therefore comes from parametric \textsc{gim2d} uncertainty in $B/T$. 

To understand what might be driving the inflated scatter and corresponding offset in Stripe 82 $B/T$ parametric uncertainties for bright sources, we inspected the Stripe 82 and Legacy decomposition mosaics of objects with $\sigma(B/T)_{r,\mathrm{S82}}>0.025$ and $m_{r,\mathrm{S82}}<14.5$ mag -- finding 41 in Stripe 82 and 2 in Legacy (both targets from Legacy were also satisfied the cut in Stripe 82). The uncertainties among the Stripe 82 targets that met this criterion were also generally larger than the two from Legacy. The mosaics of these sources revealed that they all had exceptionally asymmetric isophotes. Many were barred. Some were edge-on discs with long dust lanes. Others were spiral galaxies with strong twists and whose inner structures differed greatly in axis ratio or isophotal position angle from their outer structures. But all also had visible bulges. It might be that for these bright, asymmetric objects the bulge model is bouncing between fitting the bulge itself and fitting what the disc cannot. In other words, the bulge model is trying to compensate for the inadequacy of the disc model to the task of fitting a highly asymmetric disc or other extended structure. 

Following this intuition, the reason that the uncertainty is higher in Stripe 82 is because the outer structure has higher $S/N$ and therefore greater weight in the fit than in Legacy. If there are then twists in this outer structure then the bulge has to try to find a saddle point in likelihood space between properly modelling the bulge and modelling part of that outer structure -- driving up its parametric uncertainty relative to the Legacy images. We have not designed an experiment to explicitly test this assertion. Whether it is correct or not, the inflated $B/T$ uncertainties at the bright end of Stripe 82 propagate into the corresponding bulge and disc uncertainty calculations shown in the upper two rows of Figure \ref{fig:comp_unc}. Ultimately, this systematic highlights the limitations of modelling highly asymmetric structures with symmetric models and the value of more sophisticated (but also more sensitive) models which can can both fit galaxies with complex isophotes \emph{and} return realistic structural estimates (e.g., \citealt{2015ApJ...810..120C}). By finding inventive ways to inject priors into these models and find reasonable initial conditions, they may soon be scalable to large surveys.

The upper two rows of Figure \ref{fig:comp_unc} show the bulge, $\sigma(m_{r,\mathrm{bulge}})$, and disc, $\sigma(m_{r,\mathrm{disc}})$, component magnitudes and the bulge-to-total ratio, $\sigma(B/T)_r$ in the $r$-band. Again, in each case, we see the mark made by imaging depth on the Stripe 82 uncertainties. In general, the uncertainties on component magnitudes are much larger than for total magnitudes -- as expected since each presents a fraction of the whole. Taking $\sigma_{r,\mathrm{tol}}=0.1$ magnitudes ($\sim10\%$ flux difference) as a toy tolerance, disc magnitudes in Stripe 82 are robust out to $m_{r,\mathrm{disc}}\approx20$ mag, whereas they exceed the tolerance at $m_{r,\mathrm{disc}}\approx18$ mag with Legacy photometry. Similarly, the limit for bulges is extended from $m_{r,\mathrm{bulge}}\approx17.8$ mag to $m_{r,\mathrm{bulge}}\approx19.8$ mag. The $\sim2$ magnitude improvement in the limiting magnitude for a given tolerance generally holds for tolerances less than $\sigma_{r,\mathrm{tol}}=0.2$ magnitudes.

We caution that one should be careful not to over-interpret these uncertainties. These uncertainties are the statistical uncertainties \emph{given} an assumed model. Systematics that may arise from assuming models that may not be suited to real galaxies may not be captured (as highlighted in our discussion of the $B/T$ uncertainties). One may interpret the uncertainties we report as encoding the probabilities that a new measurement made using the same model and optimization routine in an image of similar quality would achieve similar results. 

\section{Catalog Structure and Parameters}
\label{sec:appB}

Table \ref{tab:appendix_catalogs} is an example table schema for the decomposition catalogs. Twelve tables are released in total -- one for each unique combination of model and $Xr-$band pairing. Tables for fits with the same decomposition model have identical schema. Tables in which structural parameters are fixed to results from $gr$-band decompositions include those forced values. A user may refer to Table \ref{tab:catalogs} to determine whether structural parameters in a given table are measured independently or fixed. One must again recall that the decompositions in each band are always performed pairwise with the $r-$band. The structural measurements in a given table are either: (1) covariant with the $r$-band structures (as in the case of S\'ersic index in the \texttt{ps} fits); or (2) fixed to the results of the simultaneous $gr-$band decompositions and as such are never \emph{truly} independent in any given band. 

The \texttt{ps}, \texttt{n4}, and \texttt{fn} tables are identical in structure with a few exceptions. First, the \texttt{n4} tables contain a $P_{pS}$ parameter. Second, the \texttt{fn} tables contain both $P_{pS}$ and $P_{n4}$ parameters. Each catalog also includes an estimate of the local sky background (relative to the pre-subtracted \sextractor{} background pixel mode in the full frame) and its uncertainty around each galaxy in each band. As tabulated, the sky uncertainties are normalized by the square root of the number of pixels used to estimate the sky level. These can be converted to calibrated sky surface brightness uncertainties, $\sigma_{\mathrm{sky}}$ (in magnitudes/arcsecond$^2$), as follows:
\begin{align}
\sigma_{\mathrm{sky}} =& -2.5\log_{10}\left(\frac{\texttt{skySigNorm}\times\sqrt{\mathrm{\texttt{skyNpix}}}}{\texttt{pixel\_scale}^2}\right) + 30.0
\label{eq:sky}
\end{align}

where \texttt{pixel\_scale} is the pixel scale, 0.396127 arcsec/pixel, of the SDSS camera. The local sky backgrounds themselves (\texttt{db} in the tables) are the relative offset between the global sky in the frame measured by \sextractor{} (which is then subtracted from the image) and the \textsc{gim2d} local estimate around the target galaxy. The example schema in Table \ref{tab:appendix_catalogs} is for a $gr$ decomposition table with the \texttt{n4} model. 

\onecolumn

\begin{longtable}{llll}
\captionsetup{width=\linewidth}
\caption{Stripe 82 galaxy morphology catalog example schema for the \texttt{n4} simultaneous $gr$ band decompositions (online supplementary information). The \texttt{fn} tables are identical to the \texttt{n4} tables but include the additional $P_{n4}$ statistic. The single-component \texttt{ps} model tables contain neither $P_{pS}$ nor $P_{n4}$ statistics and columns which pertained to the discs in the two-component tables are either NULL or are filled with nonsensical values in the \texttt{ps} tables: disc position angle, scale length, inclination and their associated uncertainties. Bulge-to-total fractions are also meaningless in the \texttt{ps} tables. Rest-frame quantities assume $(H_0,\Omega_m,\Omega_{\Lambda}) = (70$ km/s/Mpc, 0.3, 0.7). Absolute magnitudes incorporate $k-$corrections computed using \texttt{kcorrect} version 4.2 \citep{2007AJ....133..734B}. }
\label{tab:morph}. \\
\hline
 \textsc{Column} & \textsc{Type} & \textsc{Units} & \textsc{Description} \\
\hline
\endfirsthead
\multicolumn{4}{c}%
{\tablename\ \thetable\ -- \textit{Continued from previous page}} \\
\hline
 \textsc{Column} & \textsc{Type} & \textsc{Units} & \textsc{Description} \\
\hline
\endhead
\hline \multicolumn{4}{r}{\textit{Continued on next page}} \\
\endfoot
\hline
\endlastfoot
s82\_objID    & varchar(25) & n/a & Unique Stripe 82 catalog object ID \\
objID         & varchar(25) & n/a & Matched unique SDSS DR7 object ID \\
ra            & float       & degrees & Right ascension \\
decl          & float       & degrees & Declination \\
run           & int(11)     & n/a & Stripe 82 run: South-106, North-206 \\
rerun         & int(11)     & n/a & Processing/calibration identifier \\
camcol        & int(11)     & n/a & CCD camera column \\
field         & int(11)     & n/a & Stripe 82 field  \\
ID            & int(11)     & n/a & Object ID within given field \\
petroMag\_r   & float       & mag & SDSS Petrosian apparent magnitude \\
extinction\_r & float       & mag & Galactic extinction \\
z             & float       & n/a & Spectroscopic redshift \\
sciim\_nx     & int(11)     & pixels & Size of science cut-out x  \\
sciim\_ny     & int(11)     & pixels & Size of science cut-out y  \\
prchost       & varchar(15) & n/a & Processing host (decomposition) \\
start\_time   & datetime    & n/a & Processing start-time \\
end\_time     & datetime    & n/a & Processing end-tim \\
dfm\_r        & float       & counts/s  & (16 - 50)\% range f \\
f\_r          & float       & counts/s  & Galaxy total flux \\
dfp\_r        & float       & counts/s  & (50 - 84)\% range f \\
dg2dmagm\_r   & float       & mag & (16 - 50)\% range g2dmag \\
g2dmag\_r     & float       & mag & Galaxy apparent magnitude \\
dg2dmagp\_r   & float       & mag & (50 - 84)\% range g2dmag \\
dbtm\_r       & float       & n/a & (16 - 50)\% range bt \\
bt\_r         & float       & n/a & Bulge-to-total fraction \\
dbtp\_r       & float       & n/a & (50 - 84)\% range bt \\
drem          & float       & arcsec & (16 - 50)\% range re \\
re            & float       & arcsec & Bulge effective radius \\
drep          & float       & arcsec & (50 - 84)\% range re      \\
dem           & float       & n/a & (16 - 50)\% range e \\
e             & float       & n/a & Bulge ellipticity \\
dep           & float       & n/a & (50 - 84)\% range e \\
dphibm        & float       & degrees & (16 - 50)\% range phib \\
phib          & float       & degrees & Bulge position angle \\
dphibp        & float       & degrees & (50 - 84)\% range phib \\
drdm          & float       & arcsec & (16 - 50)\% range rd       \\
rd            & float       & arcsec & Disc scale length       \\
drdp          & float       & arcsec & (50 - 84)\% range rd       \\
didm          & float       & degrees & (16 - 50)\% range incd \\
incd          & float       & degrees & Disc inclination, q\_app = (b/a) = cos(incd) \\
didp          & float       & degrees & (50 - 84)\% range incd \\
dphidm        & float       & degrees & (16 - 50)\% range phib \\
phid          & float       & degrees & Disc position angle \\
dphidp        & float       & degrees & (50 - 84)\% range phid \\
ddxm\_g        & float       & pixels & (16 - 50)\% range dx       \\
dx\_g          & float       & pixels & Model centroid offset from \textsc{PHOTO} x     \\
ddxp\_g        & float       & pixels & (50 - 84)\% range dx      \\
ddym\_g        & float       & pixels & (16 - 50)\% range dy       \\
dy\_g          & float       & pixels & Model centroid offset from \textsc{PHOTO} y    \\
ddyp\_g        & float       & pixels & (50 - 84)\% range dy       \\
ddbm\_r        & float       & counts/s & (16 - 50)\% range db \\
db\_r          & float       & counts/s & (\textsc{gim2d} - \sextractor) \textit{residual} background \\
ddbp\_r        & float       & counts/s & (50 - 84)\% range db \\
dnm           & float       & n/a & (16 - 50)\% range n       \\
n             & float       & n/a & Sersic Index   \\
dnp           & float       & n/a & (50 - 84)\% range n       \\
rhalf\_r       & float       & arcsec & Galaxy half-light radius (HLR)      \\
rchi2\_r       & float       & n/a & Reduced CHI2 statistic       \\
c1\_r          & float       & n/a & Concentration parameter within 1 HLR       \\
c2\_r          & float       & n/a & Concentration parameter within 2 HLR       \\
c3\_r          & float       & n/a & Concentration parameter within 3 HLR       \\
c4\_r          & float       & n/a & Concentration parameter within 4 HLR       \\
a\_r           & float       & n/a & Asymmetry parameter       \\
da\_r          & float       & n/a & Uncertainty in asymmetry parameter       \\
az2\_1\_r       & float       & n/a & Simard et al. (2002) Az parameter for n=2 within 1 HLR        \\
az2\_2\_r       & float       & n/a & Simard et al. (2002) Az parameter for n=2 within 2 HLR        \\
az3\_1\_r       & float       & n/a & Simard et al. (2002) Az parameter for n=3 within 1 HLR        \\
az3\_2\_r       & float       & n/a & Simard et al. (2002) Az parameter for n=3 within 2 HLR        \\
az5\_1\_r       & float       & n/a & Simard et al. (2002) Az parameter for n=5 within 1 HLR        \\
az5\_2\_r       & float       & n/a & Simard et al. (2002) Az parameter for n=5 within 2 HLR        \\
dz\_r          & float       & n/a & Simard et al. (2002) Dz parameter       \\
rt1\_1\_r       & float       & n/a & Residual asymmetry R\_T within 1 HLR        \\
rt1\_2\_r       & float       & n/a & Residual asymmetry R\_T within 2 HLR        \\
rt1\_3\_r       & float       & n/a & Residual asymmetry R\_T within 3 HLR        \\
ra1\_1\_r       & float       & n/a & Residual asymmetry R\_A within 1 HLR        \\
ra1\_2\_r       & float       & n/a & Residual asymmetry R\_A within 2 HLR        \\
ra1\_3\_r       & float       & n/a & Residual asymmetry R\_A within 3 HLR        \\
prcflag       & int(11)       & n/a & Processing flag (0: converged) \\
rd\_kpc        & float       & kpc & Disc scale length [kpc]     \\
re\_kpc        & float       & kpc & Bulge effective radius [kpc]       \\
rhalf\_kpc\_r   & float       & kpc & Galaxy HLR [kpc]      \\
petroR50\_r    & float       & arcsec & Petrosian radius R50       \\
mu50\_r        & float       & mag/arcsec2 & Average surface brightness in R50 \\
Vmax          & float       & Mpc3 & Volume correction \\
Mr\_galaxy     & float       & mag & Galaxy absolute magnitude    \\
Mr\_bulge      & float       & mag & Bulge absolute magnitude    \\
Mr\_disk       & float       & mag & Disc absolute magnitude      \\
Mg\_galaxy     & float       & mag & Galaxy absolute magnitude    \\
Mg\_bulge      & float       & mag & Bulge absolute magnitude    \\
Mg\_disk       & float       & mag & Disc absolute magnitude      \\
extinction\_g  & float       & mag & Galactic extinction \\
petroMag\_g    & float       & mag & SDSS Petrosian apparent magnitude \\
dfm\_g         & float       & counts/s & (16 - 50)\% range f \\
f\_g           & float       & counts/s & Galaxy total flux \\
dfp\_g         & float       & counts/s & (50 - 84)\% range f \\
dbtm\_g        & float       & n/a & (16 - 50)\% range bt \\
bt\_g          & float       & n/a & Bulge-to-total fraction       \\
dbtp\_g        & float       & n/a & (50 - 84)\% range bt \\
dg2dmagm\_g    & float       & mag & (16 - 50)\% range g2dmag     \\
g2dmag\_g      & float       & mag & Galaxy apparent magnitude \\
dg2dmagp\_g    & float       & mag & (50 - 84)\% range g2dmag \\
rhalf\_g       & float       & arcsec & Galaxy half-light radius (HLR)      \\
rhalf\_kpc\_g   & float       & kpc & Galaxy HLR [kpc]      \\
rchi2\_g       & float       & n/a & Reduced CHI2 statistic       \\
ddbm\_g        & float       & counts/s & (16 - 50)\% range db \\
db\_g          & float       & counts/s & (\textsc{gim2d} - \sextractor) \textit{residual} background \\
ddbp\_g        & float       & counts/s & (50 - 84)\% range db \\
c1\_g          & float       & n/a & Concentration parameter within 1 HLR       \\
c2\_g          & float       & n/a & Concentration parameter within 2 HLR       \\
c3\_g          & float       & n/a & Concentration parameter within 3 HLR       \\
c4\_g          & float       & n/a & Concentration parameter within 4 HLR       \\
petroR50\_g    & float       & arcsec & Petrosian radius R50       \\
mu50\_g        & float       & mag/arcsec2 & Average surface brightness in R50 \\
a\_g           & float       & n/a & Asymmetry parameter       \\
da\_g          & float       & n/a & Uncertainty in asymmetry parameter       \\
az2\_1\_g       & float       & n/a & Simard et al. (2002) Az parameter for n=2 within 1 HLR        \\
az2\_2\_g       & float       & n/a & Simard et al. (2002) Az parameter for n=2 within 2 HLR        \\
az3\_1\_g       & float       & n/a & Simard et al. (2002) Az parameter for n=3 within 1 HLR        \\
az3\_2\_g       & float       & n/a & Simard et al. (2002) Az parameter for n=3 within 2 HLR        \\
az5\_1\_g       & float       & n/a & Simard et al. (2002) Az parameter for n=5 within 1 HLR        \\
az5\_2\_g       & float       & n/a & Simard et al. (2002) Az parameter for n=5 within 2 HLR        \\
dz\_g          & float       & n/a & Simard et al. (2002) Dz parameter       \\
rt1\_1\_g       & float       & n/a & Residual asymmetry R\_T within 1 HLR        \\
rt1\_2\_g       & float       & n/a & Residual asymmetry R\_T within 2 HLR        \\
rt1\_3\_g       & float       & n/a & Residual asymmetry R\_T within 3 HLR        \\
ra1\_1\_g       & float       & n/a & Residual asymmetry R\_A within 1 HLR        \\
ra1\_2\_g       & float       & n/a & Residual asymmetry R\_A within 2 HLR        \\
ra1\_3\_g       & float       & n/a & Residual asymmetry R\_A within 3 HLR        \\
specclass     & int(11)     & n/a    & Spectroscopic classification \\
ddxm\_r        & float       & pixels    & (16 - 50)\% range dx       \\
dx\_r          & float       & pixels    & Model centroid offset from \textsc{PHOTO} x     \\
ddxp\_r        & float       & pixels    & (50 - 84)\% range dx      \\
ddym\_r        & float       & pixels    & (16 - 50)\% range dy       \\
dy\_r          & float       & pixels    & Model centroid offset from \textsc{PHOTO} y    \\
ddyp\_r        & float       & pixels    & (50 - 84)\% range dy       \\
crhalf\_g      & float       & arcsec    & Circular aperture galaxy model half-light radius \\
crhalf\_r      & float       & arcsec    & Circular aperture galaxy model half-light radius \\
P\_pS          & float       & n/a    & F-test statistic PpS       \\
npxfit0       & int(11)     & n/a    & Number of sky pixels in science image used in fitting        \\
npxfit1       & int(11)     & n/a    & Number of target object pixels in science image used in fitting \\
skySigNorm\_r  & float       & counts/s/pixel   & Standard error in the sky background measurement \\
skySigNorm\_g  & float       & counts/s/pixel   & Standard error in the sky background measurement \\
skyNpix\_g     & float       & n/a    & Number of pixels used to evaluate local sky      \\
skyNpix\_r     & float       & n/a    & Number of pixels used to evaluate local sky \\
\hline
\label{tab:appendix_catalogs}
\end{longtable}
\twocolumn


\bsp	
\label{lastpage}
\end{document}